\documentstyle[twoside,epsf]{article}

\catcode`\@=11
\long\def\@makefntext#1{
\protect\noindent \hbox to 3.2pt {\hskip-.9pt  
$^{{\eightrm\@thefnmark}}$\hfil}#1\hfill}               

\def\@makefnmark{\hbox to 0pt{$^{\@thefnmark}$\hss}}    
        
\def\ps@myheadings{\let\@mkboth\@gobbletwo
\def\@oddhead{\hbox{}
\rightmark\hfil\eightrm\thepage}   
\def\@oddfoot{}\def\@evenhead{\eightrm\thepage\hfil
\leftmark\hbox{}}\def\@evenfoot{}
\def\sectionmark##1{}\def\subsectionmark##1{}}

\oddsidemargin=\evensidemargin
\newcounter{appendixc}
\newcounter{subappendixc}[appendixc]
\newcounter{subsubappendixc}[subappendixc]
\renewcommand{\thesubappendixc}{\Alph{appendixc}.\arabic{subappendixc}}
\renewcommand{\thesubsubappendixc}
        {\Alph{appendixc}.\arabic{subappendixc}.\arabic{subsubappendixc}}

\renewcommand{\appendix}[1] {\vspace{12pt}
        \refstepcounter{appendixc}
        \setcounter{figure}{0}
        \setcounter{table}{0}
        \setcounter{lemma}{0}
        \setcounter{theorem}{0}
        \setcounter{corollary}{0}
        \setcounter{definition}{0}
        \setcounter{equation}{0}
        \renewcommand{\thefigure}{\Alph{appendixc}.\arabic{figure}}
        \renewcommand{\thetable}{\Alph{appendixc}.\arabic{table}}
        \renewcommand{\theappendixc}{\Alph{appendixc}}
        \renewcommand{\thelemma}{\Alph{appendixc}.\arabic{lemma}}
        \renewcommand{\thetheorem}{\Alph{appendixc}.\arabic{theorem}}
        \renewcommand{\thedefinition}{\Alph{appendixc}.\arabic{definition}}
        \renewcommand{\thecorollary}{\Alph{appendixc}.\arabic{corollary}}
        \renewcommand{\theequation}{\Alph{appendixc}.\arabic{equation}}
        \noindent{\tenbf Appendix \theappendixc #1}\par\vspace{5pt}}
\newcommand{\subappendix}[1] {\vspace{12pt}
        \refstepcounter{subappendixc}
        \noindent{\bf Appendix \thesubappendixc. {\kern1pt \bfit #1}}
        \par\vspace{5pt}}
\newcommand{\subsubappendix}[1] {\vspace{12pt}
        \refstepcounter{subsubappendixc}
        \noindent{\rm Appendix \thesubsubappendixc. {\kern1pt \tenit #1}}
        \par\vspace{5pt}}

\newcommand{\textlineskip}{\baselineskip=13pt}
\newcommand{\smalllineskip}{\baselineskip=10pt}

\def\eightcirc{
\begin{picture}(0,0)
\put(4.4,1.8){\circle{6.5}}
\end{picture}}
\def\eightcopyright{\eightcirc\kern2.7pt\hbox{\eightrm c}} 

\newcommand{\copyrightheading}[1]
        {\vspace*{-2.5cm}\smalllineskip{\flushleft
        {\footnotesize International Journal of Modern Physics A, #1}\\
        {\footnotesize $\eightcopyright$\, World Scientific Publishing
         Company}\\
         }}

\newcommand{\publisher}[2]{{\begin{center}\footnotesize\smalllineskip 
        Received #1\\
        Revised #2
        \end{center}
        }}

\def\abstracts#1#2#3{{
        \centering{\begin{minipage}{4.5in}\baselineskip=10pt\footnotesize
        \parindent=0pt #1\par 
        \parindent=15pt #2\par
        \parindent=15pt #3
        \end{minipage}}\par}} 

\renewenvironment{thebibliography}[1]
        {\frenchspacing
         \ninerm\baselineskip=11pt
         \begin{list}{\arabic{enumi}.}
        {\usecounter{enumi}\setlength{\parsep}{0pt}
         \setlength{\leftmargin 12.7pt}{\rightmargin 0pt} 
         \setlength{\itemsep}{0pt} \settowidth
        {\labelwidth}{#1.}\sloppy}}{\end{list}}

\newcounter{itemlistc}
\newcounter{romanlistc}
\newcounter{alphlistc}
\newcounter{arabiclistc}

\newcommand{\fcaption}[1]{
        \refstepcounter{figure}
        \setbox\@tempboxa = \hbox{\footnotesize Fig.~\thefigure. #1}
        \ifdim \wd\@tempboxa > 5in 
           {\begin{center}
        \parbox{5in}{\footnotesize\smalllineskip Fig.~\thefigure. #1}
            \end{center}}
        \else
             {\begin{center}
             {\footnotesize Fig.~\thefigure. #1}
              \end{center}}
        \fi}

\newcommand{\tcaption}[1]{
        \refstepcounter{table}
        \setbox\@tempboxa = \hbox{\footnotesize Table~\thetable. #1}
        \ifdim \wd\@tempboxa > 5in
           {\begin{center}
        \parbox{5in}{\footnotesize\smalllineskip Table~\thetable. #1}
            \end{center}}
        \else
             {\begin{center}
             {\footnotesize Table~\thetable. #1}
              \end{center}}
        \fi}

\def\@citex[#1]#2{\if@filesw\immediate\write\@auxout
        {\string\citation{#2}}\fi
\def\@citea{}\@cite{\@for\@citeb:=#2\do
        {\@citea\def\@citea{,}\@ifundefined
        {b@\@citeb}{{\bf ?}\@warning
        {Citation `\@citeb' on page \thepage \space undefined}}
        {\csname b@\@citeb\endcsname}}}{#1}}

\newif\if@cghi
\def\cite{\@cghitrue\@ifnextchar [{\@tempswatrue
        \@citex}{\@tempswafalse\@citex[]}}
\def\citelow{\@cghifalse\@ifnextchar [{\@tempswatrue
        \@citex}{\@tempswafalse\@citex[]}}
\def\@cite#1#2{{$\null^{#1}$\if@tempswa\typeout
        {IJCGA warning: optional citation argument 
        ignored: `#2'} \fi}}

\def\pmb#1{\setbox0=\hbox{#1}
        \kern-.025em\copy0\kern-\wd0
        \kern.05em\copy0\kern-\wd0
        \kern-.025em\raise.0433em\box0}

\def\fnt#1#2{\footnotetext{\kern-.3em
        {$^{\mbox{\scriptsize #1}}$}{#2}}}

\def\fpage#1{\begingroup
\voffset=.3in
\thispagestyle{empty}\begin{table}[b]\centerline{\footnotesize #1}
        \end{table}\endgroup}

\def\runninghead#1#2{\pagestyle{myheadings}
\markboth{{\protect\footnotesize\it{\quad #1}}\hfill}
{\hfill{\protect\footnotesize\it{#2\quad}}}}
\font\tenit=cmti10 
\font\tenbf=cmbx10
\font\bfit=cmbxti10 at 10pt
\font\ninerm=cmr9

\font\eightrm=cmr8

\def\qed{\hbox{${\vcenter{\vbox{                        
   \hrule height 0.4pt\hbox{\vrule width 0.4pt height 6pt
   \kern5pt\vrule width 0.4pt}\hrule height 0.4pt}}}$}}

\newcommand{\Qtwo}{\mbox{$Q^2$}}

\newcommand{\etamax}{\mbox{$\eta_{\rm max}$}}

\newcommand{\pom}  {I\hspace{-0.2em}P} 
\newcommand{\xpom} {\mbox{$x_{_{\pom}}$}}

\newcommand{\pt}{\mbox{$p_{\mbox{\tiny{T}}}$}} 
\newcommand{\pts}{\mbox{$p_{\mbox{\tiny{T}}}^*$}} 
\newcommand{\ptsq}{\mbox{$p_{\mbox{\tiny{T}}}^{* 2}$}} 
\newcommand{\Et}{\mbox{$E_{\mbox{\tiny{T}}}$}} 
\newcommand{\xf}{\mbox{$x_{\mbox{\tiny{F}}}$}}

\newcommand{\MWtwo}{\mbox{$M^2_{\tiny{W}}$}} 
\newcommand{\MZtwo}{\mbox{$M^2_{\tiny{Z}}$}} 
\newcommand{\MZ}{\mbox{$M_{\tiny{Z}}$}} 
\newcommand{\eplus}{\mbox{$\rm e^+$}} 
\newcommand{\emin}{\mbox{$\rm e^-$}} 
\newcommand{\ee}{\eplus \emin} 
\newcommand{\Jpsi}{\mbox{$J/ \psi$}} 
\newcommand{\Ftwo}{\mbox{$F_2$}}

\newcommand{\Fl}{\mbox{$F_{\mbox{\tiny{L}}}$}} 
\newcommand{\degree}{\mbox{$^\circ$}} 
\newcommand{\Sigmapq} {\mbox{$\Sigma_{\pom q}$}} 
\newcommand{\Sigmap} {\mbox{$\Sigma_{\pom }$}} 
\newcommand{\beqn}{\begin{eqnarray}} 
\newcommand{\eeqn}{\end{eqnarray}} 
\def\ffig#1#2#3#4{ 
\begin{figure} [htbp]
\begin{center}  
\mbox{\hspace*{0.0in} \epsfysize=#2 \epsfxsize=#2 
\epsffile{#1} }  
\vskip 0.5cm 
{\fcaption {#3} 
\label{#4} } \end{center}
\end{figure}} 
\def\ffigp#1#2#3#4#5{ 
\begin{figure} [htbp]
\begin{center}  
\mbox{\hspace*{0.0in} \epsfysize=#2 \epsfxsize=#3 
\epsffile{#1} }  
\vskip 0.5cm 
{\fcaption {#4} 
\label{#5} } \end{center}
\end{figure}} 
\def\ffigq#1#2#3#4#5#6{ 
\begin{figure} [htbp]
\begin{center}  
\mbox{\hspace*{0.0in} \epsfysize=#2 \epsfxsize=#3 
\epsffile{#1} }  
\vskip #5 
{\fcaption {#4} 
\label{#6} } \end{center}
\end{figure}} 
 
\def\Eref#1{Equation~\ref{#1}} 
\def\eref#1{equation~\ref{#1}} 
\def\Fref#1{Figure~\ref{#1}} 
\def\fref#1{figure~\ref{#1}}

\begin{document} 
\pagestyle{plain} 
\runninghead{ Deep Inelastic Scattering at HERA $\ldots$}{ Deep Inelastic Scattering at HERA $\ldots$} 

\normalsize\textlineskip
\thispagestyle{empty}
\setcounter{page}{1}

\copyrightheading{}                     

\vspace*{0.88truein}

\fpage{1}
\centerline{\bf DEEP INELASTIC SCATTERING AT HERA}
\centerline{\footnotesize B. Foster\footnote{Supported by the UK
Particle Physics and Astronomy Research Council;\\ 
email: b.foster@bristol.ac.uk}}
\vspace*{0.015truein}
\centerline{\footnotesize\it H.H. Wills Physics Laboratory, University
of Bristol}
\baselineskip=10pt
\centerline{\footnotesize\it Bristol, BS8 1TL. U.K.}
\vspace*{10pt}
\publisher{(September 20th, 1997)}{}
\centerline{\footnotesize\it To be published in International Journal
of Modern Physics A}

\vspace*{0.21truein}
\abstracts{Results from the H1 and ZEUS experiments at HERA on deep inelastic  
scattering are reviewed. The data lead to a consistent 
picture of a steep rise in the $F_2$ structure function and in the gluon 
density within the proton. Important new information on the partonic 
structure of diffraction is emerging from H1 and ZEUS. The space-like 
region in which the weak and electromagnetic interactions become of 
equal strength is being explored for the first time. A possible excess of  
events at high $x$ and \Qtwo\ compared to the expectations of the  
Standard Model has been observed in both experiments.}{}{}
\section{Introduction} 
\label{sec-Intro} 
The study of the  deep inelastic scattering of leptons on protons has been of  
profound importance to the development of particle physics. The pioneering  
electron-proton scattering experiments at SLAC in the 1960s provided  
evidence for the quark structure of matter. Neutrino experiments at CERN in  
the 1970s discovered the weak neutral current and thereby the key  
experimental underpinning of the Standard Model. The observation of  
scaling violations laid the foundations for our modern understanding of the  
strong interaction in terms of QCD. The current programme of deep inelastic  
scattering experiments has been revolutionised by the advent of the HERA  
electron proton storage ring, which has improved the resolution with which  
the proton can be explored in deep inelastic scattering by a factor of 100.  
Some indication of the 
tremendous increase in the kinematic region which can be explored at 
HERA is given in \fref{fig-xq2-plane}. 
\ffig{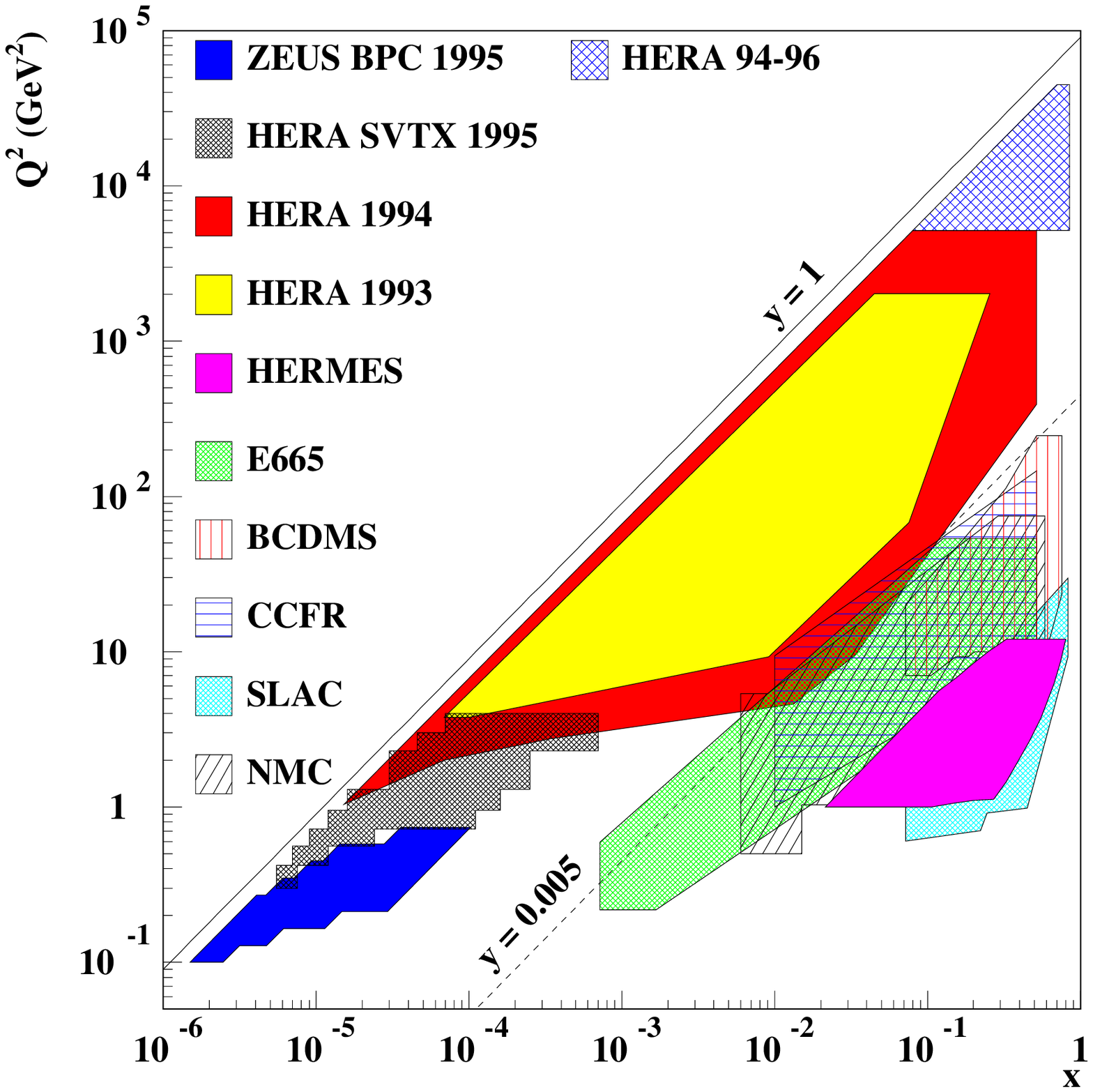} 
{80mm}{The $x$-\Qtwo\ plane which can be explored by the HERA  
experiments. The regions explored by ZEUS, H1 and HERMES in 
particular analyses are shown by the shaded regions. The enormous  
increase in phase space at HERA in comparison to the fixed target  
experiments (E665 etc.) is apparent.}  
{fig-xq2-plane}
 
A varied programme of physics has developed at HERA, for instance in the   
physics of quasi-real photon interactions. This article will however  
concentrate on the results achieved at HERA by the H1 and ZEUS
experiments in the area of deep inelastic  
scattering. These have had a major impact  
in several areas: the structure of  
the proton, the investigation of the mechanism of diffractive scattering, and  
the properties of the strong and electroweak interactions. 
\section{The HERA collider and the H1 and ZEUS experiments} 
\label{sec-HERA} 
HERA is  6.3 km  in circumference and collides electrons or positrons
of up to  
30 GeV with protons of up to 820 GeV at four interaction points around  
the ring. The bunches of electrons and protons cross at each interaction  
point  
every $96$~ns. The design luminosity in $1.5 \cdot 10^{31}$ cm$^{-2}$ 
s$^{-1}$.  
The electron storage ring is based on conventional magnets but the high  
energy of the protons requires superconducting magnets providing a bending  
field of approximately 4.7 T. Both electrons and protons are produced 
by a complex of pre-accelerators and 
accelerated up to 14 GeV and 40 GeV respectively before  
being extracted from the PETRA storage ring and injected into HERA. 
HERA began operation in 1992 and has run with electrons and positrons 
of up to  
27.7 GeV and protons of 820 GeV. The total luminosity 
delivered by HERA to the experiments is shown in  
figure~\ref{fig-HERA-del-lumi}. The maximum  
luminosity achieved by HERA up to spring 1997  
was $\sim 0.9  \cdot 10^{31}$~cm$^{-2}$ s$^{-1}$.    
The majority of data has been taken with positrons although a relatively  
small data sample with electrons was taken in runs from 1992-1994.  
\ffigq{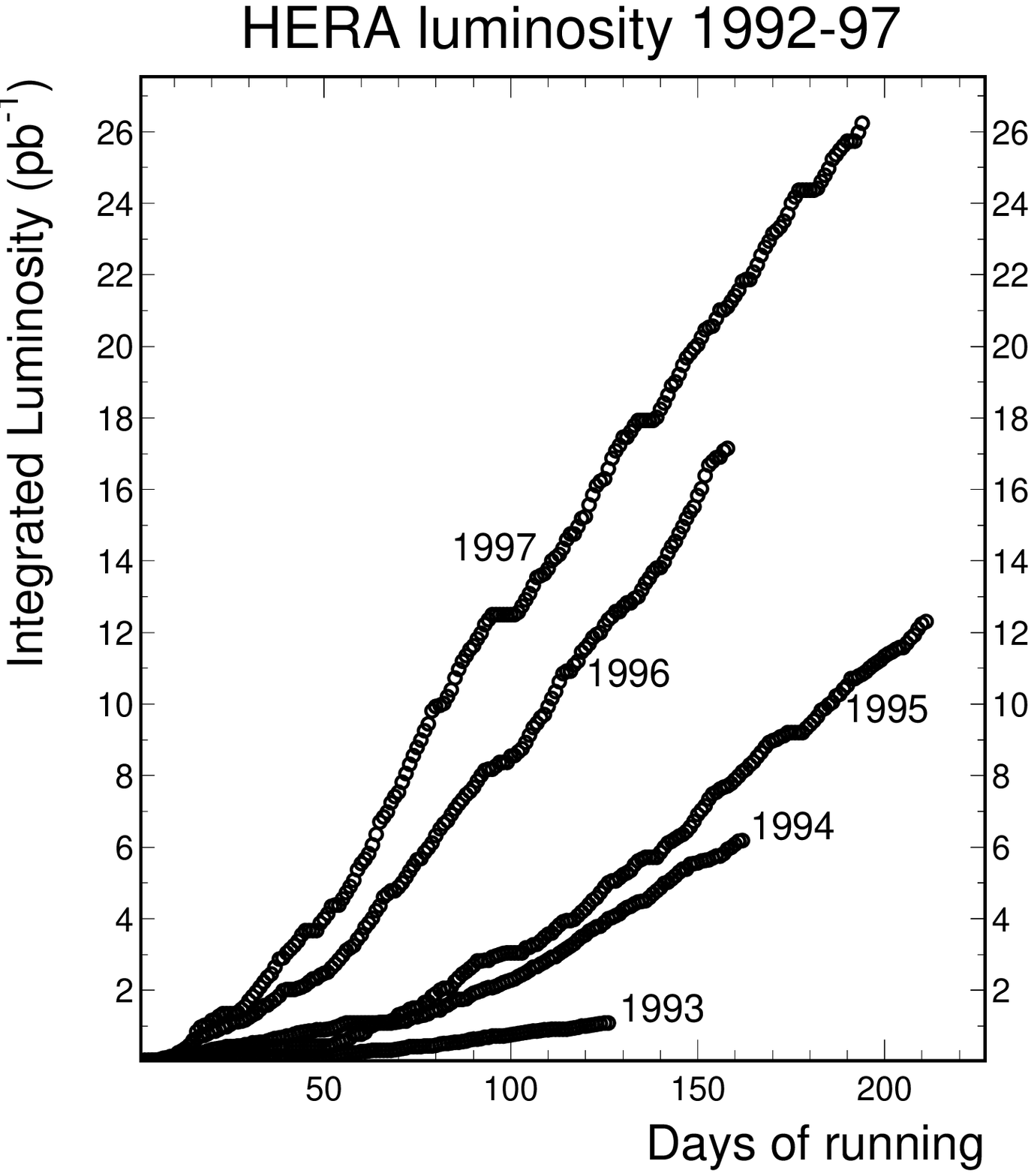} 
{100mm}{80mm}{The luminosity delivered by HERA between 1992 and mid-1997 as 
a function of day of the run.}{-0.5cm} 
{fig-HERA-del-lumi} 
 
The H1~\cite{H1-detector} and ZEUS~\cite{ZEUS-detector} detectors are  
large, general-purpose colliding beam detectors designed for maximum  
hermeticity. The most important characteristics of HERA which influenced  
the design of the experiments were: 
\begin{enumerate} 
\item The asymmetric nature of the collisions between the low energy  
electrons or positrons and the high energy protons, 
which results in a decided  
asymmetry in energy deposition, with most of the energy  in the proton or 
``forward'' direction. This is reflected in H1 and ZEUS, which have deeper  
calorimeters and more complex instrumentation in the forward compared to  
the rear direction. The large centre of mass energy together with the boost  
means that very high energy jets may be produced and thus there is an  
emphasis on precise calorimetry in both experiments.  
\item The very short beam crossing interval of $96$~ns required the  
development of complex pipelined electronics to store the signals from the  
detectors until such time as the fast logic at the first triggering
level was  
able to decide if the event was of interest. The trigger electronics
itself was also of unprecedented complexity.  
\end{enumerate} 
The H1 and ZEUS  collaborations eventually decided on rather different  
approaches to building an experiment to make accurate measurements in this  
difficult environment. However, both combine accurate and complex arrays  
of charged particle track detectors with large, high resolution  
calorimeters. 
 
The major components of the H1 detector can be summarised as  
follows. H1 has recently installed a silicon vertex 
detector~\cite{H1-Si} close  
to the interaction point, inside a ``jet chamber''~\cite{H1-CTD} drift
chamber with large drift cells and many radial
measurements of tracks  
producing drift times and energy loss measurements. The tracking system is  
supplemented in the forward and rear directions by dedicated planar  
chambers~\cite{H1-FDET}. Surrounding the tracking detectors is an array of  
calorimeters. In the forward and barrel parts of the detector 
the calorimetry   
uses the lead/liquid Argon technique~\cite{H1-LA}. 
In the rear direction H1 originally  
constructed a lead/scintillator sandwich calorimeter about one interaction  
length in depth. This was replaced for the 1995 data taking period with a  
so-called ``spaghetti'' calorimeter~\cite{H1-SPACAL} consisting of lead 
and scintillating fibres with both electromagnetic and hadronic sections .  
The quoted resolution $\sigma_E/E$ for the  
electromagnetic section of the liquid Argon calorimeter is 12\%/$\sqrt{ E}  
\oplus 0.01$, for the lead-scintillator calorimeter is 10\%/$\sqrt {E} \oplus  
0.42/E \oplus0.03$ and for the upgraded spaghetti calorimeter is 7.1/
$\sqrt{ E} \oplus 0.01$. The hadronic resolution of the liquid argon  
calorimeter is quoted as 0.50/$\sqrt{ E} \oplus 0.02$. 
Outside the calorimeter  
is a superconducting magnet producing a uniform magnetic field of 1.15 T.  
The iron return yoke is instrumented and acts as a ``tail-catcher'' to  
measure the leakage of hadronic showers as well as identify penetrating  
muons. The muon measurement is improved in the forward direction by  
a forward muon spectrometer.  
 
ZEUS also had a vertex detector~\cite{ZEUS-VXD}, a small drift chamber  
operating with DME which has a very slow drift velocity. This chamber 
was removed for the 1996 running period, and will be replaced by a 
silicon strip vertex detector~\cite{ZEUS-muVD}.  Surrounding this is  
the  Central Tracking Detector (CTD)~\cite{ZEUS-CTD} which is based on  
small drift cells with wire planes tilted at $45^{\circ}$ to the radial  
direction.  The CTD provides many drift-time and energy loss  
measurements and in addition provides a fast first level trigger by using  
information on the z  
position of tracks obtained from a measurement of the time difference of  
the arrival time of pulses at both ends of the  
CTD~\cite{ZEUS-CTD-Electronics}.  
The tracking is extended outside the central region by dedicated forward and  
rear tracking detectors. These are based on drift chamber  
layers with stereo measurements to give a space point.  Between the three  
tracking layers in the forward detector are two layers of 
transition radiation  
detector. The forward tracking detector was only fully instrumented from  
the start of the 1995 data taking period. Outside the CTD is a small  
superconducting solenoid producing a magnetic field of 1.47 T.  
Surrounding the coil and  
tracking detectors is a uranium-scintillator calorimeter~\cite{ZEUS-CAL}  
subdivided into a rear, barrel and forward detectors of differing thickness  
and tower size. The quoted resolution for electromagnetic showers is  
$\sigma/E =  18\%/\sqrt{E} \oplus 0.02$ and for hadrons is  
$\sigma/E =  35\%/\sqrt{E} \oplus 0.02$. The timing resolution of the  
calorimeter can be parameterised by the relation $1.5/\sqrt{E} \oplus 0.5 ns$. 
The iron return yoke is instrumented to act as a tail catcher and to
help in the  
identification of penetrating muons, which are also identified by dedicated  
muon chambers outside the iron. There is also a forward muon  
spectrometer which improves the measurement of high-momentum  
muons in the forward direction.  
 
Both experiments use the bremsstrahlung process, 
$ep \rightarrow ep\gamma$,  to measure the luminosity delivered by
HERA.  The cross-section for  this process
can be calculated in QED. Events are tagged by  
calorimetric detectors placed a substantial distance downstream of the main  
detectors in the electron direction. 
The luminosity is proportional to the rate  
of coincidences between an electron calorimeter, positioned in order to  
intercept electrons which have lost energy and are therefore bent out of the  
beam by the HERA magnets, and a photon calorimeter, placed in a straight  
line from the interaction point in order to intercept the photon from the  
bremsstrahlung process. In practice, since the backgrounds are low and 
well-understood, the singles rate in the photon calorimeter is 
normally used as the luminosity measurement. The electron calorimeter 
is also used to tag  
photoproduction events via the coincidence of a scattered electron in the  
electron calorimeter with activity in the main ZEUS detector.  
 
The accuracy of  
the luminosity determination has improved from typically 3.5\% for ZEUS and  
4.5\% for H1 in 1993, to around 1.5\%, although some difficulties with the  
bunch structure of HERA resulted in a somewhat larger uncertainty of  
about 2.3\% in the 1996 data taking.  
 
Both experiments have also installed detectors designed to tag ``leading''  
protons carrying  almost the full beam energy emanating from the interaction  
point. This is achieved via a ``leading proton spectrometer'', which has a  
number of stations 
of Roman pots containing silicon strip detectors which can be lowered into 
the neighbourhood of the beam. The magnetic dipole elements of HERA act as 
a magnetic spectrometer so that very high energy protons can be tracked 
through the detector stations to give a reconstructed proton with very good  
energy resolution. Forward neutron calorimeters have also been
installed. 
These 
are relatively small and crude devices intended to measure the energy of 
neutrons produced at the interaction point and carrying almost the full beam 
energy. More information on these detectors is given in  
sections~\ref{sec-diff-ZEUS-FNC} and \ref{sec-diff-concl}. 
 
Data from some of the major components of the H1 and ZEUS detectors can  
be seen in figures~\ref{fig-H1-NCev} and ~\ref{fig-ZEUS-CCev}, which  
show typical neutral current and charged current deep inelastic scattering  
events as visualised by the event display programs. 
\ffigp{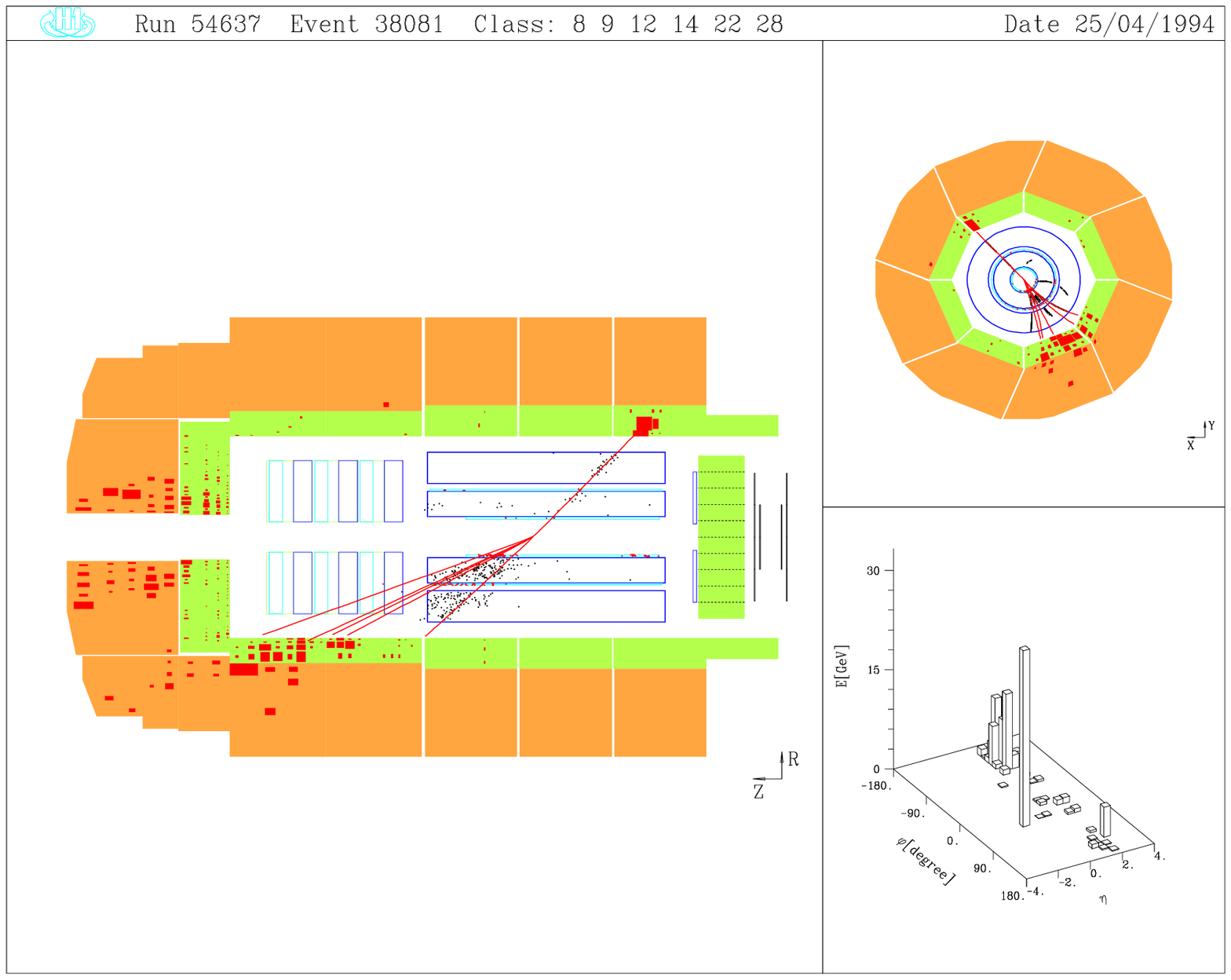}
{80mm}{110mm}{A typical neutral current
event as seen in the H1 experiment. The main view shows an $r-z$ section
through the main H1 apparatus. The electron is clearly isolated from
the other tracks in the event and extrapolates to a large electromagnetic
deposit in the electromagnetic liquid argon calorimeter. The large
energy deposit around the beam pipe due to the break-up of the incident
proton can also be seen, as can the jet of hadrons corresponding to
the current jet. The other two views show an $r-\phi$ section through
the
central detector and a two-dimensional map of the deposited
energy.} 
{fig-H1-NCev}
\ffigq{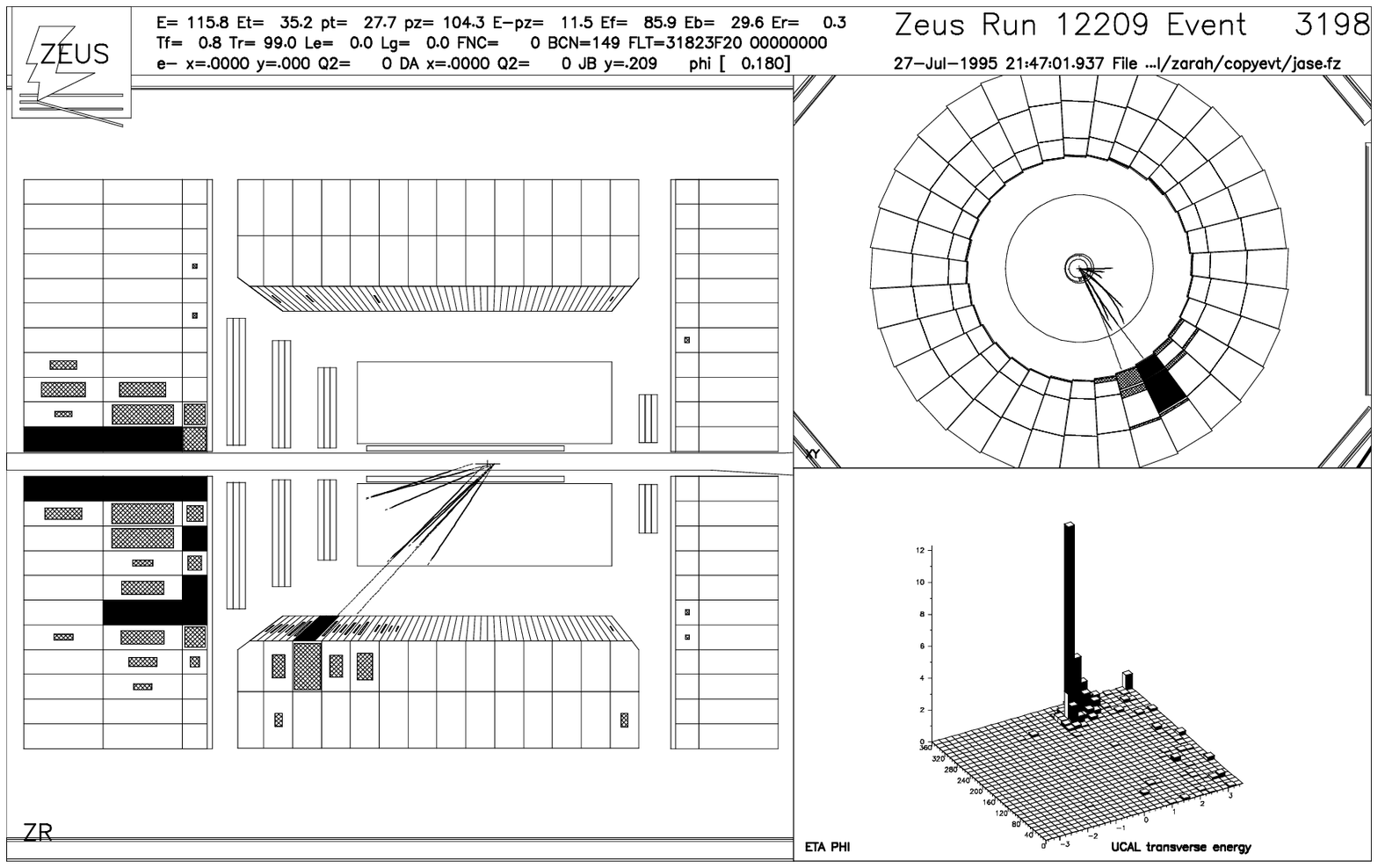} 
{80mm}{120mm}{\protect 
A typical charged current DIS event observed in the ZEUS detector.  
The topology of this event is very similar to that of
\fref{fig-H1-NCev},
except that there is no scattered electron visible.  The
reconstructed direction of the missing energy is such that it
would be fully contained
inside the detector, so that the absence of the electron cannot
correspond to a loss due to the fact that the detector does not have
perfect acceptance. The large
missing energy is therefore assumed to correspond to a scattered neutrino.} 
{-0.5cm}{fig-ZEUS-CCev}
\section{A brief theoretical introduction to Deep Inelastic Scattering} 
\label{sec-kinematics} 
\subsection{Relativistic invariants}  
\label{sec-kin-invariants} 
The scattering of a lepton from a proton at high \Qtwo\ can be viewed as the  
elastic scattering of the lepton from a quark or antiquark inside the proton.  
As such the process can be fully described by two relativistic invariants.  
If the  
initial (final) four-momentum of the lepton is $k (k')$, the initial 
four-momentum of the  proton is $p$, the fraction of the proton's  
momentum carried by the struck quark is $x$ and the final four-momentum  
of the hadronic system is $p'$ then the following invariants may be  
constructed: 
\begin{eqnarray} 
s &=& (p+k)^2\\ 
t &=& (p-p')^2\\
\Qtwo &=& -q^2 = -(k' - k)^2 \\ 
 y &=& \frac{p \cdot q}{p \cdot k} \\ 
W^2 &=& (p')^2 = (p+q)^2 
\end{eqnarray} 
where $s$ is the centre of mass energy squared, $t$ is the
four-momentum transfer squared between the proton and final state hadronic 
system, \Qtwo\ is the
four-momentum
transfer squared between the lepton and the proton, $y$ is the 
inelasticity of the scattered lepton and $W^2$ is the
invariant mass squared of the final state hadrons. 
It is easy to show that energy-momentum conservation implies that  
\beqn 
x = \frac{Q^2}{2 p \cdot q} 
\eeqn 
and that thus  
\beqn 
y &=& \frac{Q^2}{sx}
\label{eq-yq2sx}
\eeqn 
\beqn
W^2 &\sim& Q^2\frac{1-x}{x} 
\eeqn 
Since DIS at a given $s$ can be specified by any two of these 
invariants we are free to  
choose the most convenient; we will normally use $x$ and \Qtwo. 
\subsection{Parton evolution}
\label{sec-parton-ev}
One of the most important experimental observations of modern
particle physics has been the discovery of so-called ``scaling violations'',
in which the simple, three valence quark, picture of the proton
had to be
abandoned in favour of a dynamic, evolving parton content driven
by fundamental QCD processes.  The experimentally observed
evolution of the differential cross-sections as
a function of the kinematic variables implies a corresponding evolution of the
parton distributions. The description of parton evolution  
involves many different aspects of QCD depending on the kinematic
range of the measurements and thus is an important testing ground
of the theory. A very brief discussion of the QCD treatment of 
parton evolution
is included in this section in order to introduce the reader to any unfamiliar
notation used in the field. However, the interested reader is
strongly recommended to consult a standard text, such as Ellis, Stirling
and Webber~\cite{Ellis-Stirling-Webber} for a full exposition of this
complex subject.

The evolution of the parton density in the proton away from the familiar
static picture of three valence quarks bound by the exchange of gluons
is driven by fundamental processes of QCD. The important processes are
the tendency of quarks to radiate hard gluons, and of gluons to split
into quark-antiquark and gluon-gluon pairs. The probability of such
splittings depends on the gauge structure of QCD and the value
of the strong coupling constant and its variation with the kinematic
scale. Clearly the radiation of a hard gluon from a quark, and the 
splitting of a gluon into a quark-antiquark or gluon-gluon, 
decreases the four-momentum of
the radiating quark and contributes to a growth in the number of
low $x$ partons which can be found within the proton. Since these 
processes are virtual, their duration is controlled by the uncertainty 
principle; thus probes with small wavelength able to localise transient
phenomena are able to resolve more and more of these splittings. 
At large \Qtwo\ it is therefore to 
be expected that the parton distributions would
grow at small $x$ and shrink at large $x$, as is observed in
experiment.  The form of the
splitting functions given below in the limit
as $x \rightarrow 0$ shows that it is significantly more 
likely for a gluon to split into
a pair of gluons than a quark-antiquark. Since quarks
also radiate gluons, producing a softer gluon as well as a
softer quark, the sum of the quark and gluon
splitting processes causes the gluon density to increase
strongly as $x$ falls. This phenomenon of the
growth of the gluon density at low $x$ is one of the most notable
characteristics of deep inelastic scattering at HERA, and forms one of
the major threads in this review. 

The parton splitting functions can be considered
as the probability of a parton ``containing'' another type of parton
with a fraction $z$ of the initial momentum. 
The probability that a quark contains another quark of lower
fractional momentum can then be written as $P_{qq}(z)$, that
it contains a gluon as $P_{gq}(z)$, that a gluon contains as
quark as $P_{qg}(z)$ and finally that a gluon contains a lower
momentum gluon as $P_{gg}(z)$. 
These different processes are shown scematically in 
\fref{fig-jase-apsplit}. Clearly the existence of such
branchings couples the quark density in the proton to the
gluon density, and vice-versa, so that the evolution of the
parton densities follows a set of coupled integro-differential equations
as a function of the momentum scale, known as the Altarelli-Parisi,
or Dokshitzer-Gribov-Lipatov-Altarelli-Parisi (DGLAP) 
equations~\cite{DGLAP}. These can be written as
\ffigq{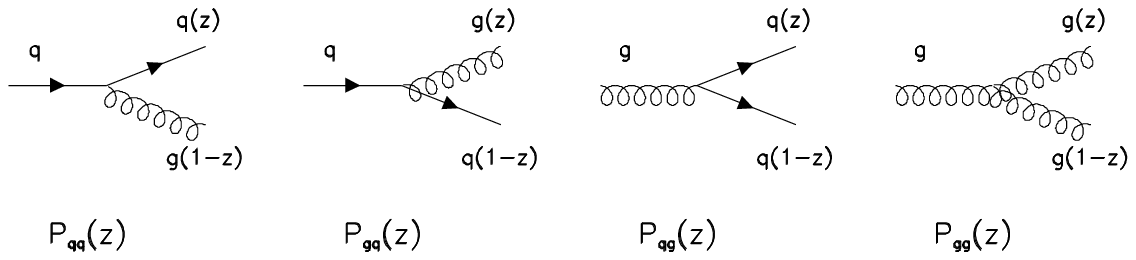} 
{80mm}{100mm}{\protect 
The parton branchings described by the four Altarelli-Parisi
splitting functions.}  
{-0.5cm}{fig-jase-apsplit}
\beqn
t \frac{\partial}{\partial t} \left( \begin{array}{c}
q_i(x,t) \\
 g(x,t) \end{array} \right) & = &  \frac{\alpha_s(t)}{2 \pi} 
\sum_{q_j, \overline{q}_j} \int^1_x 
\frac{d \xi}{\xi} \nonumber \\
& \cdot & \left( \begin{array}{cc}
P_{q_iq_j} \left(\frac{x}{\xi}, \alpha_s(t) \right) & P_{q_ig} 
\left(\frac{x}{\xi}, \alpha_s(t)\right) \\
P_{g q_j} \left(\frac{x}{\xi}, \alpha_s(t) \right) & P_{gg}
\left(\frac{x}{\xi}, \alpha_s(t)\right) \end{array} \right)
\left( \begin{array}{c}
q_j(\xi,t)\\
g(\xi,t) \end{array}
\right)  
\eeqn  
where $t$ is the squared momentum scale of interest and
$q_i(x,t),g(x,t)$ are the quark and gluon densities. Each of the splitting
functions has a perturbative expansion in powers of $\alpha_s$, and it
can be shown in general that
\beqn
xP(x,\alpha_s) =\sum_{n=0}^\infty \left(\frac{\alpha_s}{2\pi}\right)^n
\left[\sum_{m=0}^n A^{(n)}_m (\ln 1/x)^m + 
x\overline{P}^{(n)}(x)\right]
\label{eq-DGLAP-gen-exp}
\eeqn
where $P = P_{q_{i}q_{j}}, P_{q_ig}$ etc.\ and $A^{(n)}_m$ are calculable 
coefficients. The $\overline{P}(x)$ terms
correspond to expansion in terms of $\ln Q^2$. Both the leading order and
next-to-leading order forms of the DGLAP splitting functions have been
calculated; they depend upon the group structure of QCD. To illustrate
the rather simple form of the splitting functions, we show below the
leading order expressions:
\beqn
P^{(0)}_{qq}(x) & = & \frac{4}{3}\left\{\frac{1+x^2}{(1-x)_+} + \frac{3}{2}
\delta(1-x)\right\} \nonumber \\
P^{(0)}_{qg}(x) & = & \frac{1}{2}\left\{x^2 + (1-x)^2\right\} \nonumber \\
P^{(0)}_{gq}(x) & = & \frac{4}{3}\left\{\frac{1+(1-x)^2}{x}\right\} \\
\label{eq-LO-AP-splitters}
P^{(0)}_{gg}(x) & = & 6\left\{\frac{x}{(1-x)_+} + \frac{1-x}{x}
+x(1-x)\right\} \nonumber \\ 
&+& \delta(1-x)\frac{(33-2n_f)}{6} \nonumber 
\eeqn
where the $+$ subscript implies continuity from below as $x \rightarrow 1$.

The existence of both
$\ln Q^2$ and $\ln 1/x$ terms in the perturbative expansion of
\eref{eq-DGLAP-gen-exp} can be traced to
the fact that the transverse momentum, $k_t$, which is  developed in the 
quark-gluon splitting process\footnote{This is another symptom of the
breakdown of the quark-parton model, which assumes that partons
have zero $k_t$.} ~must be integrated over to produce an
observable quantity. This integral appears in the calculation of
observables in the form of an integral over the virtuality $\left|k^2\right|$
\beqn
\int^{k^2_{max}} \frac{d \left| k^2 \right|}{\left| k^2 \right|}
\eeqn
where $k^2_{max} = Q^2/x$. The expansion given in \eref{eq-DGLAP-gen-exp}
is in general truncated, depending on the kinematic conditions, to produce
several well-known approximations. The standard leading order (LO) 
approximation is given by ignoring all terms except for those with $n,m = 0$
and $\overline{P}^{(1)}$; the NLO approximation keeps terms in $n,m = 0,1$ 
plus $\overline{P}^{(0)}$,$\overline{P}^{(1)}$; 
the double-leading log. approximation (DLLA) has
$n, m =0$; the leading logarithm approximation in $1/x$ has $n \geq 0, m = n$,
etc. The latter approximation is resummed by the BFKL 
equation~\cite{BFKL}, and is appropriate
when \Qtwo\ is small and $1/x$ very large;
the DGLAP equation by contrast is appropriate for \Qtwo\ much larger
than $1/x$. Whereas the DGLAP equation
corresponds to considering the gluon emission to be strongly ordered
in $k_t$, the BFKL equation is not ordered in $k_t$, but rather in $x$.
It would be useful to have a formalism which was valid and soluble
in all kinematic ranges; work to this end is reported below in
section~\ref{sec-quarks-F2-theory}.

The limits of applicability of these various approximations is one of the
interesting areas of QCD which can in principle be studied at HERA, where
the increased kinematic region available compared to fixed
target experiments means that regions where \Qtwo\ is very much greater
than $1/x$, and {\it vice-versa}, can be investigated. 
It may be possible to establish the breakdown of ``normal''
DGLAP evolution and the applicability of BFKL evolution.
Such studies form another of the major threads of this review.   
\subsection{Differential cross-sections in DIS} 
\label{sec-dsigdo-dis} 
\Eref{eq-general-sigma} shows the 
general form for the spin-averaged neutral current differential cross-section 
in the one-boson exchange approximation 
in terms of the structure functions ${ F}_1$, ${ F}_2$ and  
${F}_3$.    
\beqn  
\frac{d^2 \sigma}{dx dQ^2}  
& = & \frac{2\pi \alpha^2}{x Q^4}  
\left[2xy^2{ F}_1 + 2(1-y)  { F}_2 \right. \nonumber \\ 
& + & \left. \{ 1-(1-y)^2 \} x { F}_3 \right]  
\label{eq-general-sigma} 
\eeqn 
The structure functions are products of quark distribution functions 
and the couplings of the current mediating the interaction. They are in  
general functions of the two invariants required to describe the interaction,  
$x$ and \Qtwo.  
 
In the naive quark-parton model in which quarks are massless, 
have spin $\frac{1}{2}$ and  
in which they develop no \pt,  the Callan-Gross  
relation~\cite{Callan-Gross} 
\begin{eqnarray} 
2 x F_1 (x) & = & F_2 (x)
\label{eq-CallanGross} 
\end{eqnarray} 
is satisfied. In the QCD improved parton model \pt\ must be taken into  
account  and this relation is violated. This is usually quantified by  
defining a  
longitudinal structure function, \Fl, such that   
\Fl\  $= F_2 - 2x F_1$. Substituting into \eref{eq-general-sigma} gives 
\beqn  
\frac{d^2 \sigma} 
{dx dQ^2} & = & \frac{2\pi \alpha^2}{x Q^4}  
 \left[   
\{ 1+(1-y)^2 \} 
 F_2(x, Q^2) \right. \nonumber \\
 & - & \left.  {y^2} \Fl(x, Q^2) + \{ 1-(1-y)^2 \} x F_3(x, Q^2) 
 \right] 
\label{eq-Fl-sigma} 
\eeqn 
\noindent 
In general, the form of the structure functions beyond 
zeroth order ({\it vis.} the quark-parton model) depends on 
the renormalisation and factorisation scheme used.  
In the so-called ``DIS" scheme~\cite{DIS-scheme}, 
the  logarithmic singularity  
produced by collinear gluon emission is absorbed into 
the definition of the quark distribution, so that 
the structure functions can be expressed as 
\beqn 
 F_2(x, Q^2) & = & \sum_{f=u,d,s,c,b} A_f(Q^2) \left[ xq_f(x,Q^2) + 
   x\overline{q}_f(x,Q^2) \right]  \label{eq-F2-qpm} \\ 
x F_3(x,Q^2) & = & \pm \sum_{f=u,d,s,c,b} B_f(Q^2) \left[ xq_f(x,Q^2) - 
   x\overline{q}_f(x,Q^2) \right]  \label{eq-F3-qpm} 
\eeqn 
where the positive sign in the expression for $xF_3$ is
appropriate to electron scattering, the 
negative sign to positron scattering. 
The parton distributions $q_f(x,Q^2)$ and $\overline{q}_f(x,Q^2)$ refer  
to quarks and anti-quarks of type 
$f$. The flavour dependent coefficients  
$A_f(Q^2)$ and $B_f(Q^2)$ contain 
the electric charge of the fermion, $e_f$,  
the electro-weak axial $(c_A)$ and vector $(c_V)$ coupling constants  and  
propagator terms $P_Z$, 
\beqn
  A_f(Q^2) & = & e_f^2 - 2e_fc_V^ec_V^fP_Z +  
                 ({c_V^e}^2+{c_A^e}^2)({c_V^f}^2+{c_A^f}^2){P_Z}^2, \\ 
  B_f(Q^2) & = & \ \ \ - 2e_fc_A^ec_A^fP_Z +  
                 4c_V^ec_V^fc_A^ec_A^f{P_Z}^2, \\ 
  {\rm where} \ \ \ \ \ \ \ P_Z & = & \frac{Q^2}{Q^2+M_Z^2}\\ 
c_V^f & = & T_3^f - 2 e_f \sin^2 \theta_W\\ 
c_A^f & = & T_3^f\\ 
{\rm and} \ \ \ \ \ \ \ \ T_3^f & = & 
+ \frac{1}{2} \ {\rm for} \ f = \nu, u, c, t\\ 
& = & - \frac{1}{2} \ {\rm for} \ f = e, d, s, b 
\eeqn 

In the $\overline{\mbox{MS}}$ scheme~\cite{MSbar-scheme},  
another scheme in common use, equations~\ref{eq-F2-qpm} 
and \ref{eq-F3-qpm} have a more complex
form. For example, $F_2$ is given in first order QCD by  
\beqn 
 F_2(x, Q^2) & = & \sum_{f=u,d,s,c,b} A_f(Q^2)  
\int_{x}^{1} 
   \frac{dy}{y} \left(\frac{x}{y}\right) \left[ 
\left\{ \delta(1 - \frac{x}{y})  
+ \frac{\alpha_s}{2 \pi} C^{\overline{\mbox{\tiny {MS}}}}_q 
\left(\frac{x}{y}\right)\right\} \right. \nonumber \\
&\cdot & \left.     
\left( yq_f(y,Q^2) + 
   y\overline{q}_f(y,Q^2) \right)  +  \left\{ \frac{\alpha_s}{2 \pi} 
C^{\overline{\mbox{\tiny{MS}}}}_g 
\left(\frac{x}{y}\right) \right\} yg(y,Q^2) \right] \label{eq-F2-MSbar} 
\eeqn 
where  $g(x,Q^2)$ is the gluon density in the proton, 
$\alpha_s(Q^2)$ is the QCD running coupling constant
and $C_q(x)$ and $C_g(x)$ are scheme-dependent ``coefficient 
functions", closely related to the integrals of the
DGLAP splitting functions discussed
in section~\ref{sec-parton-ev}. 

In contrast, the longitudinal structure function contains 
no collinear divergence at first-order in QCD so that  
\beqn
 \Fl (x, Q^2) & = &    
 \frac{ \alpha_{s}(Q^2)} {2\pi} \sum_{f=u,d,s,c,b} A_f(Q^2) 
\left\{ 2 \int_{x}^{1} \frac{dy}{y} \left(\frac{x}{y} \right)^2  
\left( 1 - \frac{x}{y} \right)yg(y,Q^2) \right.
   \nonumber \\ 
 & + & \left.  \frac{4}{3}  
\int_{x}^{1}  
   \frac{dy}{y} \left( \frac{x}{y} \right) ^2 \left[ yq_f(y,Q^2) + 
   y\overline{q}_f(y,Q^2) \right] \right\}  
\label{eq-Fl} 
\eeqn
independent of the factorisation scheme employed.  
This can also be expressed in terms of $F_2$ as 
\beqn
 \Fl (x, Q^2)     & = &    
\frac{\alpha_s (Q^2)}{2\pi} 
\left\{ 
\frac{4}{3} \int_{x}^{1}  
   \frac{dy}{y} \left(\frac{x}{y}\right)  F_2(y,Q^2) \hfill 
\right.   \nonumber \\ 
  & + & \left.  2 \sum_{f=u,d,s,c,b} A_f(Q^2) 
   \int_{x}^{1} \frac{dy}{y} \left(\frac{x}{y}\right)^2 \left( 1 - 
   \frac{x}{y} \right) yg(y,Q^2) \right\} 
\label{eq-F2Fl} 
\eeqn

These structure functions can be related to the differential
cross-section, taking  
account of electroweak radiative corrections, e.g.\ virtual 
loops and photon bremsstrahlung~\cite{rad-corr-refs} via the term $\delta_r$.  
Provided that \Qtwo $\ll M_Z^2$ the terms proportional to  $P_Z$ can 
be neglected, $xF_3$ vanishes and \eref{eq-F2-qpm} reduces to that for
photon exchange. \Eref{eq-Fl-sigma} can then be written as:
\beqn  
\frac{d^2 \sigma} 
{dx dQ^2} & =& \frac{2\pi \alpha^2}{x Q^4}  
\left[  
\{ 1+(1-y)^2 \} 
 F_2(x, Q^2) \right. \nonumber \\
 & - & \left. {y^2} \Fl(x, Q^2) 
 \right] (1 + \delta_r(x, Q^2))  
\label{eq-F2} 
\eeqn 
\noindent 
 
\Eref{eq-F2} shows that the differential cross-section  
for one-photon exchange is proportional to two  
structure functions, $F_2$ and \Fl. However, it is 
only possible to isolate the  
contribution from \Fl\ directly  by measuring the cross-section  
at different centre of mass energies for
fixed $x$ and \Qtwo. Since this is a difficult procedure  
at HERA this has not yet been attempted, although an indirect 
determination is reported in section~\ref{sec-quarks-fl}. 
In order to determine $F_2$ from the differential 
cross-section data reported in section~\ref{sec-quarks}, 
the value of \Fl\ has  
been assumed to be that given by QCD 
using parton densities from fits to many data sets including 
from fixed target DIS experiments. In the 
range of $x, Q^2$  used in the  
results presented here, \Fl\ is generally expected to be much 
smaller than $F_2$ and suppressed by the factor $y^2$
in \eref{eq-F2}. 
  
For charged current processes, the quark-parton model 
ignoring the $b$ quark contribution  
will be sufficiently accurate for our 
purposes. This gives 
the following expressions for the spin-averaged electron and positron
induced reactions respectively:  
\begin{eqnarray} 
\left. \frac{d^2 \sigma}{dx dQ^2} \right|_{\emin} & = &   
\frac{G_F^2}{2\pi} \left( \frac{\MWtwo}{\MWtwo+Q^2}\right)^2 
 \nonumber \\ 
 & & 2x \cdot \{u(x) + c(x) + (1-y)^2(\overline{d}(x) + \overline{s}(x))\} 
\label{eq-CCem}  
\end{eqnarray} 
\begin{eqnarray} 
\left. \frac{d^2 \sigma}{dx dQ^2} \right|_{\eplus} & = &   
\frac{G_F^2}{2\pi} \left(\frac{\MWtwo}{\MWtwo+Q^2}\right)^2 
\nonumber \\ 
 & & 2x \cdot \{\overline{u}(x) +\overline{c}(x) + (1-y)^2(d(x) + s(x))\} 
\label{eq-CCep} 
\end{eqnarray} 
 where $q(x), \overline{q}(x)$ etc. refer to the density distribution of the  
quark  
or antiquark of flavour $q$.   

\subsection{Reconstruction of the kinematic variables} 
\label{sec-kin-var-recons} 
As noted above, two invariants are required to specify inclusive deep  
inelastic scattering processes fully. 
Both the H1 and ZEUS detectors are sufficiently 
hermetic that the invariants can be reconstructed from measurements on the 
electron, on the hadronic final state corresponding to the struck quark, or 
on a mixture of the two. The optimal method depends on the kinematic  
region  
of interest and on the properties and resolutions of the detectors. The size 
of the radiative corrections is also very dependent on the reconstruction 
method employed. The basic measurements made are the angles and energies  
of the electron and current jet, where the angles are defined with respect to 
the proton beam direction, i.e. $\theta_e = 180 \degree$ corresponds to zero 
electron scattering angle. From these angles and energies the following  
quantities can be constructed: 
\begin{itemize} 
\item {\bf The Electron method}\\ 
\beqn 
y_e & = & 1 - \frac{E^{\prime}_e}{E_e}\sin^2\theta_e/2 \\ 
Q^2_e & = & 4 E_e E^{\prime}_e \cos^2\theta_e/2 
\eeqn 
\item {\bf The Jacquet-Blondel~\cite{Jacquet-Blondel} (JB) method} 
\beqn 
y_{JB} & = & \frac{\Sigma}{2E_e} \\ 
Q^2_{JB} & = & 2 E_e \frac{(\pt^h)^2}{2E_e - \Sigma} 
\eeqn 
where 
\beqn 
\Sigma & = & \sum_{i} (E_i - p_{z,i}) \\ 
(\pt^h)^2 & = &  \left( \sum_{i} p_{x,i} \right)^2 +  
               \left( \sum_{i} p_{y,i} \right)^2  
\eeqn 
where the summations exclude the scattered electron or positron.  
\item {\bf The Double Angle (DA)~\cite{DA} Method} \\ 
\beqn 
y_{DA} & = & \frac{\Sigma}{\pt^h \tan (\theta_e/2) + \Sigma} \\ 
Q^2_{DA} & = & 4 E_e^2 \pt^h \left[\frac{cot(\theta_e/2)}{\pt^h \tan  
(\theta_e/2) + \Sigma} \right]
\eeqn 
or alternatively
\beqn
y_{DA} & = & \frac{\sin \theta_e(1-\cos \gamma_h)}{\sin \gamma_h +
\sin \theta_e - \sin(\gamma_h + \theta_e)} 
\label{eq-DA-y-alt}
\eeqn
\beqn
Q^2_{DA} & = & 4E_e^2 \frac{\sin \gamma_h(1+\cos \theta_e)}{\sin \gamma_h +
\sin \theta_e - \sin(\gamma_h + \theta_e)}
\label{eq-DA-Q2-alt}
\eeqn
\item {\bf The Sigma~\cite{Sigma} Method} \\ 
\beqn 
y_{\Sigma} & = & \frac{\Sigma}{\Sigma + E^{\prime}_e(1 - \cos  
\theta_e)}  \\ 
Q^2_{\Sigma} & = & \frac{E^{\prime 2}_e \sin^2 \theta_e}{1 -  
y_{\Sigma}} 
\eeqn 
\item {\bf The PT~\cite{PT} Method} \\
This method is rather more complex than the others, and proceeds in
two steps. In the first step, the event-by-event \pt\ balance is used
to correct the hadronic energy estimate of $y$ using a functional form
derived from Monte Carlo studies. Then the electron,
JB and DA methods using the corrected hadronic energy 
are combined to give the best reconstruction over
the full kinematic range. In the ideal case
\beqn
y = y_{JB}/(\frac{\pt^h}{\pt^e})
\eeqn
However, hadronic energy flow between the proton remnant and the
current jet causes a correction to this formula
\beqn
y = y_{JB}/C
\eeqn
where $C$ is determined from Monte Carlo as a function of 
$\pt^h/\pt^e, \pt^h$ and $\gamma_h$. At high $y$ the balance between
$\pt^h$ and $\pt^e$ is less efficient at correcting $y$, so that the
Sigma method is used in this region. 
This corrected $y$ is then used to
calculate
the hadronic angle $\gamma_{PT}$ using
\beqn
\cos \gamma_{PT}   = \frac{(\pt^e)^2 - 4E^2_e y^2}{(\pt^e)^2 + 
4E^2_ey^2}
\eeqn 
which can then be used in equations~\ref{eq-DA-y-alt} and 
\ref{eq-DA-Q2-alt}.
\end {itemize}
In all cases $x$ can be derived from any of the above methods using  
\eref{eq-yq2sx}. 
\section{The quark substructure of the proton} 
\label{sec-quarks} 
\subsection{Measurement of $F_2$} 
\label{sec-quarks-F2} 
Both ZEUS and H1 published \Ftwo\ measurements from the data taken in  
1992 and 1993~\cite{H1-F2-1993,ZEUS-F2-1993}. 
These results showed a rapid rise in  
\Ftwo\ as $x$ decreases below $10^{-2}$. Subsequent data taking, together with 
upgrades to the experimental apparatus in the two experiments, have 
resulted in a major improvement in both the kinematic 
range and the accuracy of the \Ftwo\  
measurement. 
The most important factor here has been a decrease in the minimum  
scattering angle of the positron which can be used in the analysis 
due to improvements in instrumentation near to the beam pipe in the rear  
direction, i.e. the direction of the positron beam.  
In addition, some special runs were taken 
with the interaction vertex 
displaced by about 67 cm in the proton direction. This means that 
positrons scattered at even smaller angles, and hence lower \Qtwo , emerge 
into the well instrumented and understood parts of the detector and hence 
can be accepted for the \Ftwo\ analysis. The experiments accumulated some  
58 nb$^{-1}$ of shifted vertex data in the 1994 data taking period. In  
addition H1 utilised data accumulated  
from the small ``proton satellite" bunches which accompany the main proton  
bunch in adjacent buckets. This data corresponds to an effective interaction  
point of 68 cm in the proton direction and gives an effective luminosity  
contribution of about 68 nb$^{-1}$.  
 
In order to determine $F_2$ 
the following steps are carried out.  
Firstly a sample of DIS events is selected,  
basically by requiring an identified electron or positron 
in the main detector.  
The data are binned in $x, Q^2$  
with bin sizes determined by detector resolution,  
statistics, migration in and out of the 
bin due to the finite resolution of the experiments, etc.   
Estimates of the background in each bin are made and the background  
statistically subtracted. 
The data are corrected for acceptance, radiative effects and 
migration via Monte 
Carlo simulation.   
Multiplying the corrected number of events by the 
appropriate kinematic factors shown in  
\eref{eq-F2} and subtracting an estimate for \Fl\ 
gives values of \Ftwo\ and hence the quark and antiquark 
densities in the chosen bins.      
 
Several independent positron finding algorithms are used. ZEUS  
requires a minimum scattered positron energy of 10 GeV, whereas H1  
requires 11 GeV. As discussed in section~\ref{sec-kin-var-recons}, the
method of  
kinematic variable reconstruction employed varies between the experiments  
and in different kinematic regions. ZEUS uses the Electron Method for the  
low \Qtwo\ data and the PT Method for \Qtwo $ > $3 GeV$^2$,  
whereas H1 uses the Electron Method for  $y > 0.15 $  and the Sigma Method  
for  $y< 0.15$. Both experiments use data from unpaired positron and proton  
bunches to show that the background from non-$ep$ processes such as  
beam-gas interactions is very small over the entire kinematic range. The  
comparison of Monte Carlo simulation of DIS processes with the data shows  
excellent agreement except for kinematic regions where photoproduction  
background would be expected to be important. Photoproduction has a much  
larger total cross-section than DIS, and can form a background when the  
scattered positron escapes undetected down the beam pipe, but a fake  
positron is produced by a $\pi^0$ or $\gamma$ produced in the hadronic  
final state. Since it is easier to fake low energy positrons 
by this mechanism  
than high energy positrons, this background becomes progressively more  
important at high $y$. Both experiments use MC simulation of  
photoproduction processes to estimate this background. These estimates can  
be cross-checked by  
examining the subset of photoproduction events which have the scattered  
electron detected in the luminosity monitor. In addition ZEUS uses fits to  
the data plotted as a function of the total $E-p_z$ 
in the event to estimate the photoproduction  
background, which occurs predominantly at $E-p_z \ll 2E_e$ in contrast to  
DIS which peaks at $E-p_z \sim 2E_e$. All of these methods give  
consistent results. The highest levels of photoproduction background  
occur at high-$y$ and are typically $\sim$ 2\% in ZEUS and $\sim$ 3\%  
in  H1. The \Fl\ value is calculated from \eref{eq-Fl} using the  
GRV~\cite{GRV}   
(H1), or  MRSA~\cite{MRSA} (ZEUS) parton distributions (see 
section~\ref{sec-quarks-F2-theory}). 
 
A careful estimation of the systematic error on the \Ftwo\ measurement has  
been carried out by both experiments. A great many systematic checks are  
necessary, and space limitations permit only a cursory summary here.   
The most important sources for ZEUS are the uncertainty on the efficiency of  
the electron finding algorithm at high $y$ and the energy scale at low
$y$. For  
H1 the most important uncertainties are also the energy scale at low $y$ and  
the photoproduction background subtraction at high $y$ and low \Qtwo.  
 
In addition to the runs taken with shifted event vertex, the large
increase in  
data taken allowed the experiments  to analyse events with an energetic  
photon produced by initial state radiation (ISR). Since such radiation  
produces an effective reduction in the energy of the deep inelastic  
scattering process, a positron scattered at a given angle corresponds to a  
lower \Qtwo\ than would have been the case for an event at the nominal  
centre-of-mass energy.  Thus this type of event extends the \Qtwo\ reach of  
the experiment in a similar and complementary way to the shifted vertex  
running. In H1, these events are isolated by the detection of a photon with  
energy greater than 4 GeV in the photon calorimeter of the luminosity  
monitor together with a scattered positron detected in the main detector with  
energy greater than 8  GeV. In ZEUS the photon energy requirement is  
between 6 and 18 GeV and additionally less than 3 GeV in the luminosity  
electron calorimeter,  together with a scattered positron in the 
main detector  
with energy greater than 10 GeV. Many of the sources of systematic error 
associated with ISR  
events are similar to those associated with the nominal and shifted vertex 
data, although 
there are important and sizeable additional sources of uncertainty. These 
include the energy scale of the photon calorimeter of the  
luminosity monitor, the maximum energy allowed to be detected in the  
luminosity electron calorimeter (for ZEUS), the effect of \Fl\ and higher  
order radiative corrections and the subtraction of the Bethe-Heitler  
background.  
 
\Fref{fig-ZEUSH1F2x} shows the 1994 \Ftwo\ data from  
H1~\cite{H1-1996-F2} 
and ZEUS~\cite{ZEUS-1996-F2} together with the fixed target data at higher 
$x$ from the BCDMS~\cite{BCDMS},  
NMC~\cite{NMC-F2} and E665~\cite{E665-F2} experiments. The  
improvements in the detectors, the shifted vertex running and the 
increase in  
integrated luminosity has resulted in a major extension of both the accuracy  
and of the $x$, \Qtwo\ reach of the measurements.  
\ffigp{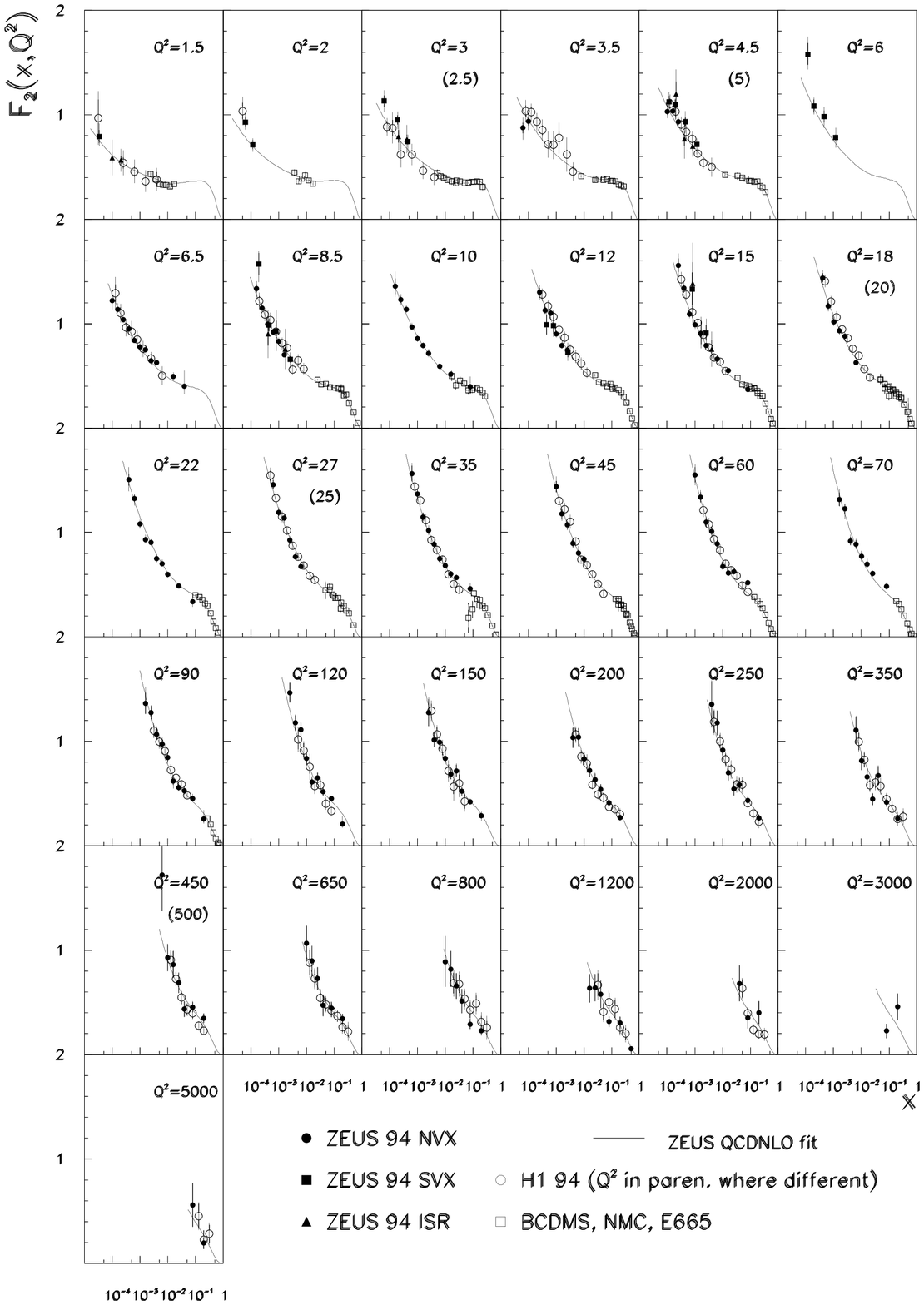} 
{140mm}{115mm}{Values of \Ftwo\ in selected \Qtwo\ intervals  
as determined from the 1994 ZEUS (filled circles,
squares, and triangles) and H1 (open circles)  
data plotted as a function of $x$. The error bars correspond to the 
statistical and  
systematic error added in quadrature, not including the uncertainty on the  
overall luminosity determination. In those bins with only one \Qtwo\  
value,  
both experiments have used identical binning, otherwise the data from the  
nearest equivalent \Qtwo\ value have been used and the H1 and ZEUS values  
are denoted separately. The BCDMS, E665 
and NMC data (open squares) are shown in those bins for which measurements  
have been published, predominantly at high $x$ and low \Qtwo. Also shown  
is a parameterisation from the ZEUS QCD fit to the data.} 
{fig-ZEUSH1F2x} 
It can be seen that there is good agreement between the ZEUS and H1 data  
points. The strong rise in \Ftwo\ as $x$ 
decreases persists to the lowest values  
of \Qtwo\ measured in 1994 data, \Qtwo\ $\sim 1.5$ GeV$^2$.  
 
As will be discussed in the next section, there is considerable theoretical interest in measuring \Ftwo\ at very low \Qtwo\ and $x$. Both H1 and ZEUS have been able to extend the range of $F_2$  
measurement to very low $x$ and \Qtwo\ by improvements to their 
detectors. H1 replaced their rear calorimeter with a lead/scintillating fibre 
calorimeter~\cite{H1-SPACAL} with an energy resolution of 
 $7.5\%/\sqrt{E{\rm( GeV)}} \oplus 2.5\%$ and a spatial resolution of 
about 4 mm. They have also added an 8-layer drift chamber  
which gives 
a precision of about 0.5 mm and 2.5 mm in the radial and azimuthal  
directions respectively. ZEUS added a 
``Beam pipe calorimeter", consisting of two tungsten-scintillator  
sandwich calorimeters placed in front of thin windows in the 
beam pipe at a distance of about 2.94 m from the interaction point. The 
detectors cover the positron scattering angle range from 17 to 34 mrad.  
The energy resolution of these devices is estimated to be  
$17\%/\sqrt{E{\rm (GeV)}}$. These upgrades greatly improve the electron 
identification capabilities of both experiments down to very small angles. 
 
The kinematic variables are reconstructed by both experiments using the  
electron method. In addition, depending on the region of 
phase space considered, H1 uses the Sigma method while ZEUS  
uses the Jacquet-Blondel method. The data from H1~\cite{H1-lowq2} 
 are shown in 
\fref{fig-H1-lowq2} and extend down to a \Qtwo\ of 0.35 GeV$^2$. 
The ZEUS data~\cite{ZEUS-BPC} extend to even lower  
values of \Qtwo\ and $x$, reaching \Qtwo\ of 0.11 GeV$^2$.  
\ffigp{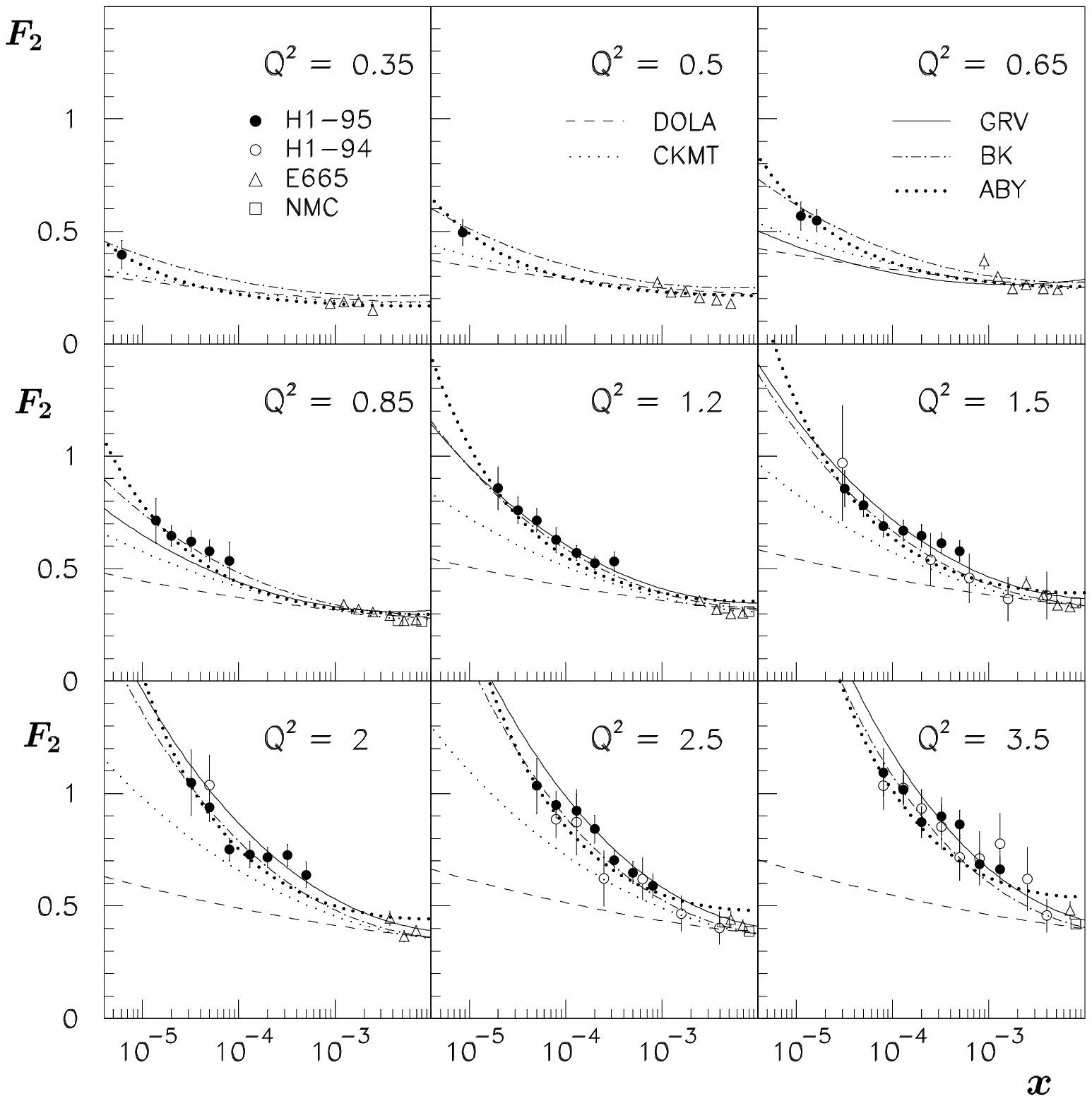} 
{120mm}{120mm}{H1 1995 data at low \Qtwo\ together with previously 
published data from 1994~\protect\cite{H1-1996-F2} and data from the 
E665~\protect\cite{E665-F2} and NMC~\protect\cite{NMC-F2} experiments. 
The predictions of Donnachie and Landshoff~\protect\cite{F2-DL} 
(DOLA) are shown 
by a dashed line, those of Capella {\it et al.}~\protect\cite{Capella} 
(CKMT) by a 
line with small 
dots, those of Badelek and Kwiecinski~\protect\cite{BK} (BK) by a  
dashed-dotted line, those of Gl\"{u}ck {\it et al.}~\protect\cite{GRV-various}  
(GRV) by a full line and those of Adel {\it et al.}~\protect\cite{Adel} (ABY) by a 
line with large dots. 
Overall normalisation uncertainties are not included in the error bars. } 
{fig-H1-lowq2}  
\subsection{Theoretical implications of $F_2$ measurement} 
\label{sec-quarks-F2-theory} 
 
We have seen above that the ZEUS and H1 collaborations have been able 
to make precision measurements of $F_2$ over an unprecedented range 
in both $x$ and \Qtwo. This has acted as a tremendous stimulant to 
theoretical activity, so that it is appropriate at this point to give an 
overview of some of this theoretical work.  
Within the framework of perturbative QCD there are several possible  
explanations for the strong rise of \Ftwo\ as $x$ decreases.  
There are essentially three relevant issues. The first is the appropriate 
form of the evolution equations within the relevant kinematic range, 
i.e.\ whether Altarelli-Parisi (DGLAP)~\cite{DGLAP}, BFKL~\cite{BFKL}, 
or indeed other forms are most  
appropriate. Secondly, the form of the non-perturbative input which is 
required in all models must be specified. Finally, the starting scale 
from which  
a perturbative evolution can be carried out must be decided upon. It 
is possible to get a good description of the data with various combinations 
of assumptions about these three factors. The experimental challenge is 
to measure accurately as large a number of independent quantities, such 
as $F_2$, \Fl, $F_2^{charm}$, etc., in as large a kinematic range as  
practicable,  
so that the theoretical implications are unambiguous.  
 
Qualitatively,  
the rise of the structure functions at low $x$ was predicted in the double  
leading log limit of QCD some  
twenty years ago~\cite{DLL-Rujula}. It seems possible to generate the  
strong rise of \Ftwo\ as $x$ decreases from either the conventional  
DGLAP evolution in $\ln Q^2$, or from BFKL  
evolution in $\ln (1/x)$. At some point both these  
evolutions must be modified by the very large parton densities  
encountered at low $x$, where non-linear parton recombination  
effects~\cite{Parton-recomb} must become important, although this 
is not expected to take place within the kinematic range accessible to 
HERA. Attempts have been made to embody both DGLAP and  
BFKL  
behaviour in a single equation based on angular ordering of gluon 
emission~\cite{CCFM}, in the so-called CCFM equation, explicit solutions  
of which have been obtained for  
the small $x$ region~\cite{Kwiecinski-MS}. In contrast to the above 
approaches, which are based on perturbative QCD and parton dynamics, 
other models based on Regge considerations, in which the virtual photon 
can be considered as a hadron, have been proposed. Such models, which 
predict a much softer rise with decreasing $x$ were  
championed by Donnachie and Landshoff~\cite{F2-DL}. However, the  
Donnachie and Landshoff approach, which generically has been so  
successful in describing the total cross-section data for a wide variety of  
``soft'' hadronic processes, including real photoproduction, falls  
significantly below the \Ftwo\ data even at  \Qtwo\ $\sim 1$~GeV$^2$. 
   
Several groups have developed very sophisticated procedures to combine  
all data relevant to the determination of proton parton distributions. This  
data comes from the HERA experiments, fixed target DIS experiments,  
$W^{\pm}$ data from $p\overline{p}$, Drell-Yan processes and direct  
photon production. Since parton distributions at very low \Qtwo\ are not  
calculable perturbatively the procedure is to introduce simple functional  
forms for the parton distributions motivated by QCD at a sufficiently large  
\Qtwo\ (typically \Qtwo\ $\sim 4$ GeV$^2$) that next to leading order  
Altarelli-Parisi evolution equations can be used to evolve to higher  
\Qtwo. The data are then compared to predictions based on the 
parton distributions evolved to the relevant scale in order to 
obtain the best values for the free parameters in the assumed 
functional forms. The 
groups most active in performing such fits are Martin, Roberts and  
Stirling~\cite{MRS} and the CTEQ group~\cite{CTEQ}. A  
somewhat different approach is adopted by Gl\"{u}ck, Reya and  
Vogt~\cite{GRV}. They utilise the fact that  as \Qtwo\  
$\rightarrow 0$ parton distributions are fully constrained by the charge  
and momentum sum rules. By assuming valence-like distributions for the  
quarks at a very low starting \Qtwo, in principle the gluon and sea  
distributions can be generated purely dynamically. However, it is 
found that such a procedure generates parton distributions which are too  
steep as $x$ decreases~\cite{GRV-90}, and in addition the theoretical  
validity of such a procedure is open to doubt~\cite{Brodsky-Schmidt}.  
More theoretically respectable, although perhaps not so intuitively  
attractive, is to input ``valence-like" distributions for both quarks and  
gluons fixed by high-$x$ data at a larger though still very small \Qtwo. 
The starting \Qtwo\ value, $Q^2_0$, is determined by the point at which  
the input gluon distribution is of the same order as the input $u$ valence  
quark distribution. $Q_0^2$ is of order $0.3 -- 0.5$ GeV$^2$ in NLO  
QCD~\cite{GRV-92}. Although there are quite large uncertainties on the  
value of $Q^2_0$ and on the valence-like distributions assumed at  
$Q^2_0$, the effect of these is suppressed in the comparison with the  
HERA data by the long evolution distance. \Fref{fig-ZEUSH1F2x} shows a  
comparison of the ZEUS QCD fit with the ZEUS and H1 $F_2$ data. 
Similarly good agreement is obtained with the  
parameterisations of GRV, MRS and CTEQ; all of these approaches have a rather 
similar number of free parameters.  
 
It is instructive to plot \Ftwo\ converted to a total  
virtual photon-proton cross-section using the approximate relations 
\beqn 
W^2  & \sim &  \frac{Q^2}{x} \\ 
\sigma^{\gamma^*p}_{tot}(W^2,Q^2)  & \sim  & \frac{4 \pi^2 \alpha}{Q^2}  
F_2(x, Q^2) 
\eeqn 
based on the assumption that the $ep$ cross-section can be factorised
into the product of a photon flux factor and a virtual photon-proton 
cross-section. The $\sigma^{\gamma^*p}$ thus obtained is 
plotted against $W^2$ for various 
\Qtwo\ values in \fref{fig-F2-sig-W2}.  
\ffig{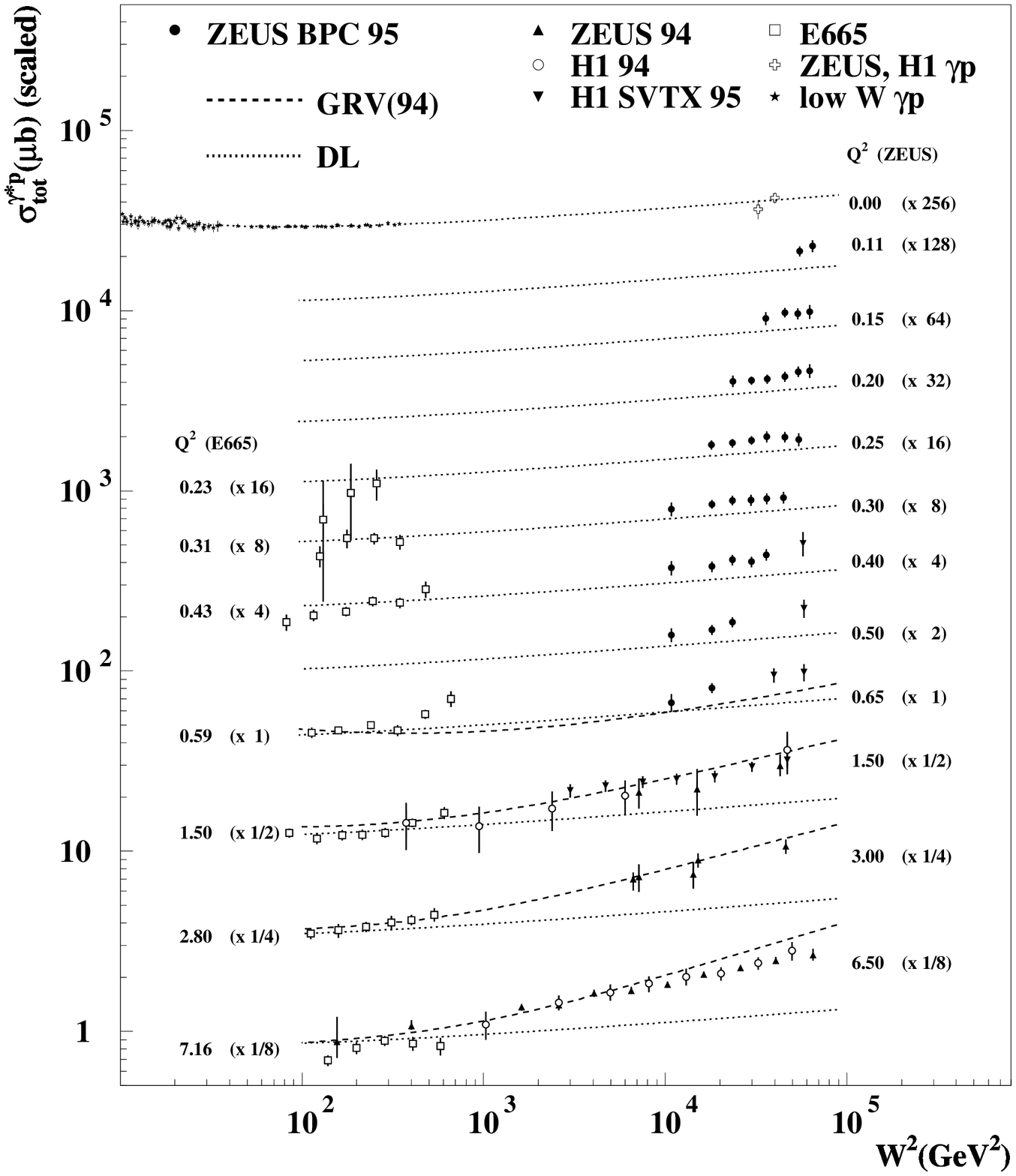} 
{120mm}{ Values of the total virtual photon-proton  
cross-section for various \Qtwo\ (in GeV$^2$). The
ZEUS and H1 points are obtained from a variety of analyses
and run periods, including shifted vertex running (SVTX) and
using the ZEUS Beam Pipe Calorimeter (BPC).   
The points at high $W^2$ and \Qtwo\ = 0 are obtained by 
averaging the total  
photoproduction cross-section measurement of 
ZEUS~\protect \cite{ZEUS-photoprod-SIGMA} and
H1~\protect\cite{H1-photoprod-SIGMA}.  
The remaining points at low $W^2$ are taken from a variety of fixed 
target experiments, in particular E665~\protect\cite{E665-F2}.   
The dotted lines are parameterisations from Donnachie and 
Landshoff~\protect\cite{F2-DL} while the dashed lines are
parameterisations from GRV~\protect\cite{GRV-F2}.} 
{fig-F2-sig-W2} 
Also shown is the total photoproduction cross-section 
as measured by ZEUS~\cite{ZEUS-photoprod-SIGMA} and
H1~\cite{H1-photoprod-SIGMA}, together with data from  
fixed target experiments. The striking feature of this plot is that the 
steeply rising cross-section as a function of $W^2$ 
seen at high values of \Qtwo\ persists down to the lowest 
measured values of  
\Qtwo. However, as shown by the dotted line in 
\fref{fig-F2-sig-W2}, at $Q^2 = 0$ the rise is less steep and 
is well represented by Donnachie-Landshoff (DL)-type models using a ``soft''  
prescription based on 
Regge theory. There must therefore be a transition region between 
the ``high" \Qtwo\ region in which perturbative QCD gives a good 
 description of the data and the photoproduction region which is well 
described by ``soft" physics. This 
transition can be seen in \fref{fig-F2-sig-W2} and also
in the data 
from H1 shown in \fref{fig-H1-lowq2}. Other Regge-inspired models  
shown in this figure, such as that of Capella {\it et al.}~\cite{Capella},
although nearer to the data  
than DL, still fail to give a good representation. Models based  
on perturbative QCD evolution, such as GRV, typically work well for higher 
\Qtwo, \Qtwo\ $\geq 1$ GeV$^2$, but do not fit the data at lower 
\Qtwo. The model of Badelek and Kwiecinski~\cite{BK} combines 
vector meson dominance with perturbative QCD in a smooth way. This 
model gives a good fit to the data. The model of 
Adel {\it et al.}~\cite{Adel}
assumes that perturbative QCD is applicable to very low 
\Qtwo\ and modifies this behaviour by adding a singular ``hard" 
contribution and  
assuming that the strong coupling ``saturates" below 
1 GeV$^2$. This also leads to good agreement with the data. 
 
It is interesting to take a very-simple minded approach to the data,  
determining a slope in $W$ for each value of \Qtwo. Such a plot 
for the ZEUS data shows a monotonic decrease in the slope until 
\Qtwo\ $\sim$ 0.65 GeV$^2$, after which the slope is consistent 
with being flat or even slightly rising. Fits to a simple vector
dominance model (VDM), in which the virtual photon is considered to
consist of a superposition of light vector meson states and therefore to
behave essentially as a hadron, 
give an acceptable $\chi^2$ until \Qtwo\ reaches 1 GeV$^2$, 
after which the $\chi^2$ becomes rapidly unacceptable. Similarly 
the QCD fit has a strong increase in $\chi^2$ above about 0.65 GeV$^2$. 
The transition between acceptable fits for these two different models 
of the interaction seems rather rapid~\cite{Tickner}. Clearly some 
variation of models which engineer a transition between ``soft" and  
``hard" behaviour ought to be able to describe the data. 
In particular, Schildkneckt and 
Spiesberger~\cite{Sch-Spi} revive the old idea of
generalised vector dominance~\cite{Sak-Sch} (GVD) and apply it
to the HERA data. Schildkneckt and Spiesberger make a couple of simple
assumptions on the form
of the energy dependence of the forward scattering amplitude in
the GVD formalism and obtain a remarkably good fit to the
ZEUS and H1 data even below \Qtwo\ $< 1$ GeV$^2$, as shown in
\fref{fig-SchSpi-sigma-Qtwo}. Further more 
accurate measurements at even lower \Qtwo\ can be expected in the  
future, which should shed further light on this very interesting region 
at the boundaries of applicability of perturbative QCD. 
\ffig{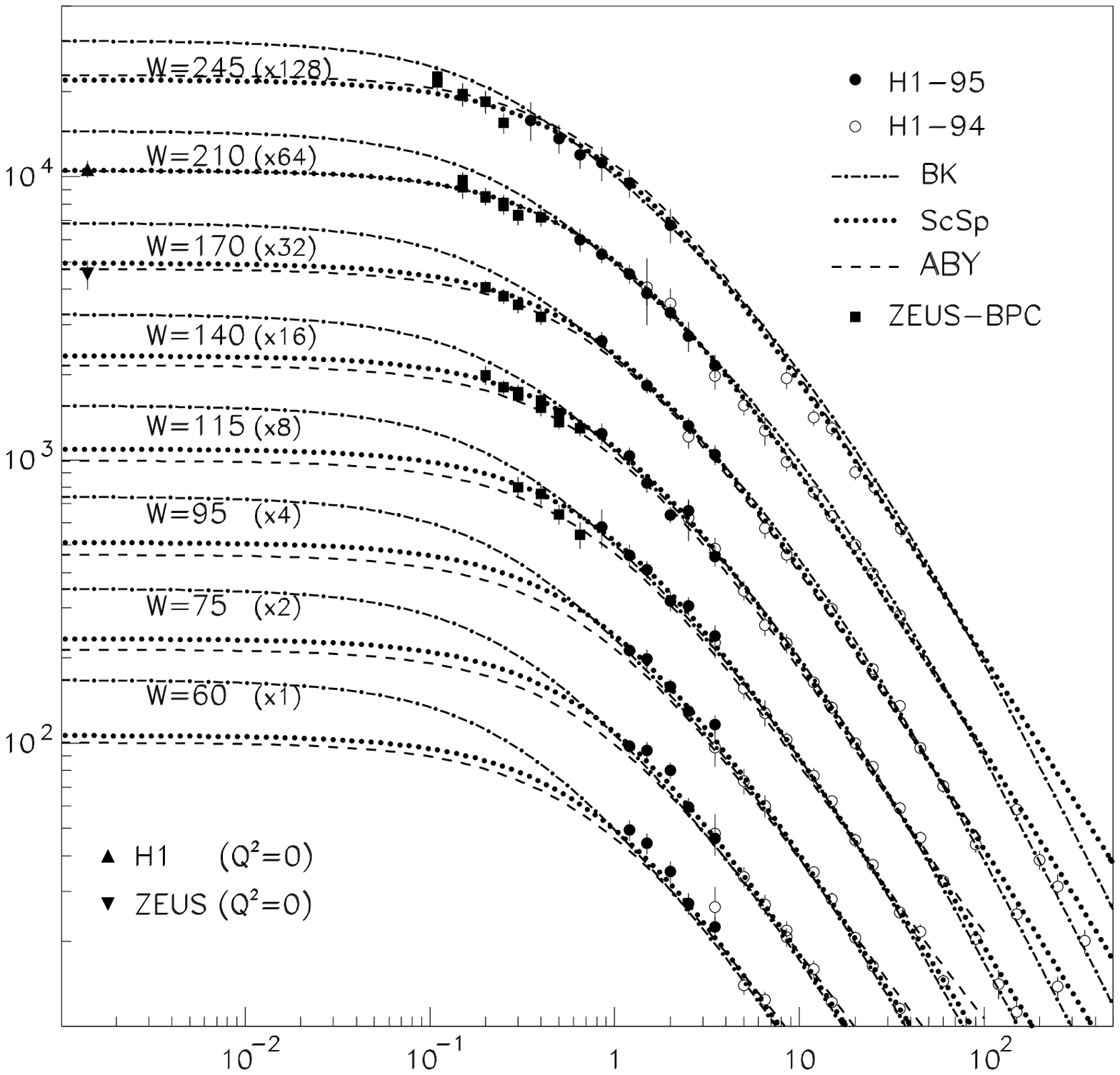} 
{120mm}{ Values of the total virtual photon-proton  
cross-section from ZEUS and H1 for various $W$ (in GeV), plotted as a function
of $\ln Q^2$. The curves show the GVD fits of Schildkneckt and Spiesberger~\protect\cite{Sch-Spi}.} 
{fig-SchSpi-sigma-Qtwo} 
 
Returning now to the region which is well described by 
perturbative QCD, it has been known for many years that 
conventional DGLAP  
evolution leads to a strong growth of \Ftwo\ as $x$ 
decreases~\cite{DLL-Rujula}. Ball and Forte~\cite{Ball-Forte} have shown that  
DGLAP also predicts a ``double asymptotic scaling"  
in the two variables $\rho$  
and $\sigma$ where 
\beqn 
\sigma & = & \sqrt{\ln \frac{x_0}{x} \cdot \ln \frac{t}{t_0}} \\ 
\rho & = &\sqrt{\ln  \frac{x_0}{x} \cdot \left(\ln \frac{t}{t_0}\right) 
^{-1}}\\
t & = & \ln \frac{Q^2}{\Lambda^2} 
\eeqn 
where $x_0 = 0.1$ and $t_0 = \ln (Q^2_0/\Lambda^2)$ with $Q^2_0 = 0.5$  
GeV$^2$, are the initial values from which the DGLAP evolution begins  
and $\Lambda = 185$ MeV is the leading order QCD $\Lambda$ value 
appropriate to four  
active flavours of quark. 
The observation of this scaling implies that DGLAP evolution is sufficient  
to explain qualitatively the behaviour of the HERA data; deviations from scaling at  
large $\rho$ (which is equivalent to small $x$ at fixed small \Qtwo)  
might indicate the importance of logarithmic terms not summed in 
conventional DGLAP evolution e.g.\ $\ln 1/x$ terms not accompanied 
by leading $\ln Q^2$ terms, often referred to as ``BFKL behaviour",  
although other  
explanations are also possible. \Fref{fig-H1-DAscaling} shows the H1  
\Ftwo\ values re-scaled by 
\beqn 
R'_F(\sigma, \rho) &=& 8.1 \exp \left( \delta \frac{\sigma}{\rho} +  
\frac{1}{2} \ln \sigma + \ln \frac{\rho}{\gamma}\right) 
\label{eq-rfp} 
\eeqn 
which removes the calculable model independent terms and leads to a  
linear rise with $\sigma$. The upper picture shows $F_2$ scaled by 
\beqn 
R_F & = & R'_F \exp{\{-2\gamma \sigma \}} 
\label{eq-rf} 
\eeqn 
In these equations $\gamma$ and $\delta$ are factors unambiguously predicted  
from NLO QCD.  
With sufficiently accurate data these can be fit and compared to the QCD  
expectation. The data do indeed exhibit  
approximate double asymptotic scaling; any possible deviations at high  
$\rho$ will require higher accuracy data to be established.  
\ffigp{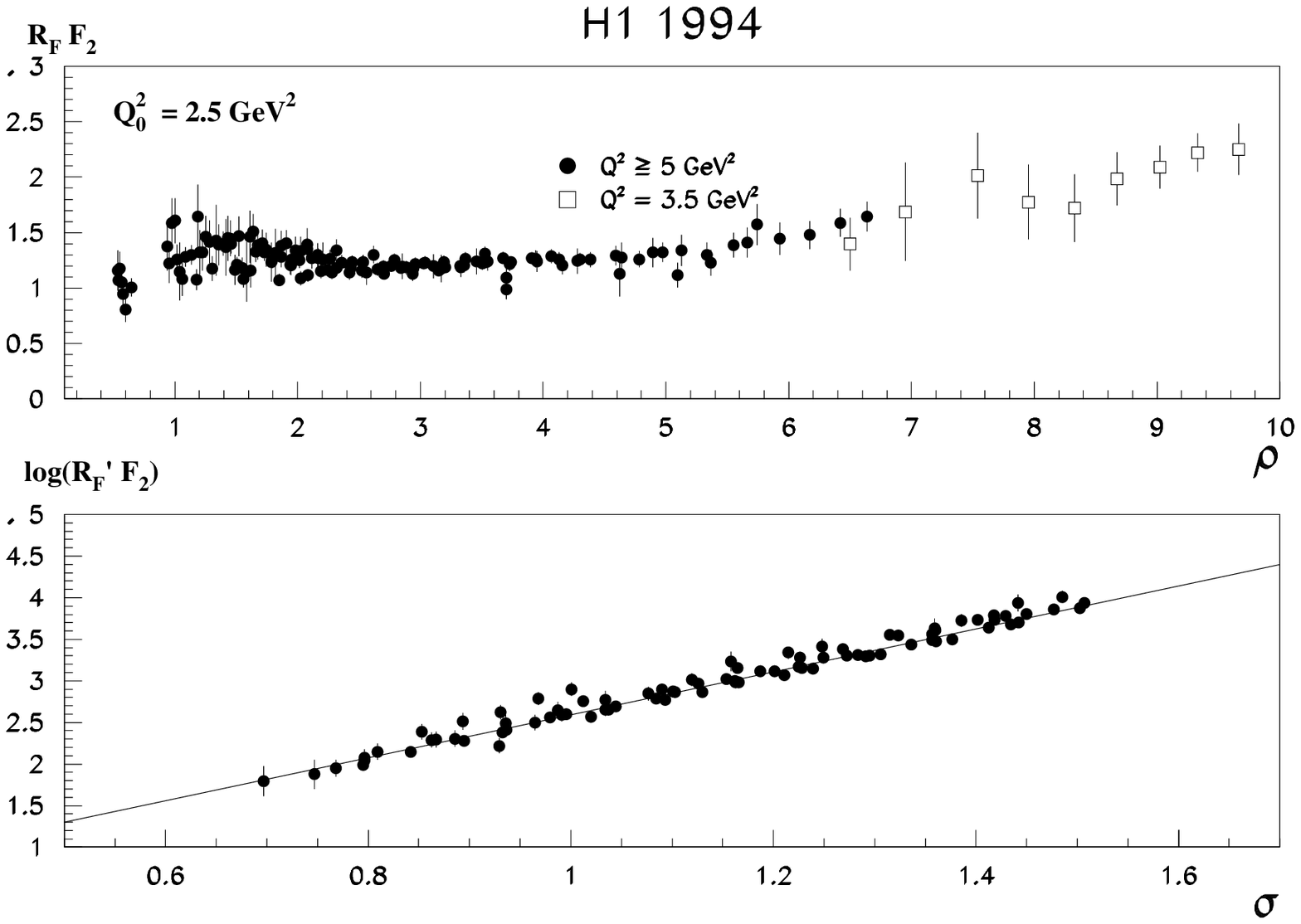} 
{90mm}{110mm}{The \Ftwo\ structure function from H1 scaled by $R_F$ and $R_F'$ . The data are binned in 
two \Qtwo\ bins  
as a function of $\rho$. In the plot against $\sigma$ only those points with  
$\rho > 1.5$ are  
plotted.} 
{fig-H1-DAscaling} 
 
The approach taken by Ball and Forte is one of several possible ways 
in which a perturbative QCD ansatz can be approximated to 
produce tractable formulae in particular kinematic conditions
(see also for example~\cite{Lopez,Ellis-KL,Ellis-HW,Forshaw-RT}). Their 
approach essentially corresponds to following the Altarelli-Parisi 
evolution and summing terms leading in  
$\ln Q^2$; another approach (see for instance~\cite{Askew})
 is to follow a BFKL evolution by 
summing the leading $\ln 1/x$ terms, leading to a power-type behaviour in 
$x$ of the form $x^{-\lambda_S}$. There have been attempts 
to unite these two approaches, using for instance the 
CCFM equation~\cite{CCFM} (see for instance~\cite{Kwiecinski-MS}). 
The work of  
Thorne~\cite{Thorne}, building on the work of Catani~\cite{Catani}, 
has demonstrated that a factorisation and renormalisation 
scheme-independent approach implies the necessity of  
summing {\it both} the leading  
$\ln Q^2$ {\it and} leading $\ln 1/x$ terms. This scheme gives a good fit 
to the HERA and fixed target data and also has some 
limited predictive power on the form of the structure functions 
at small $x$.   
 
\ffig{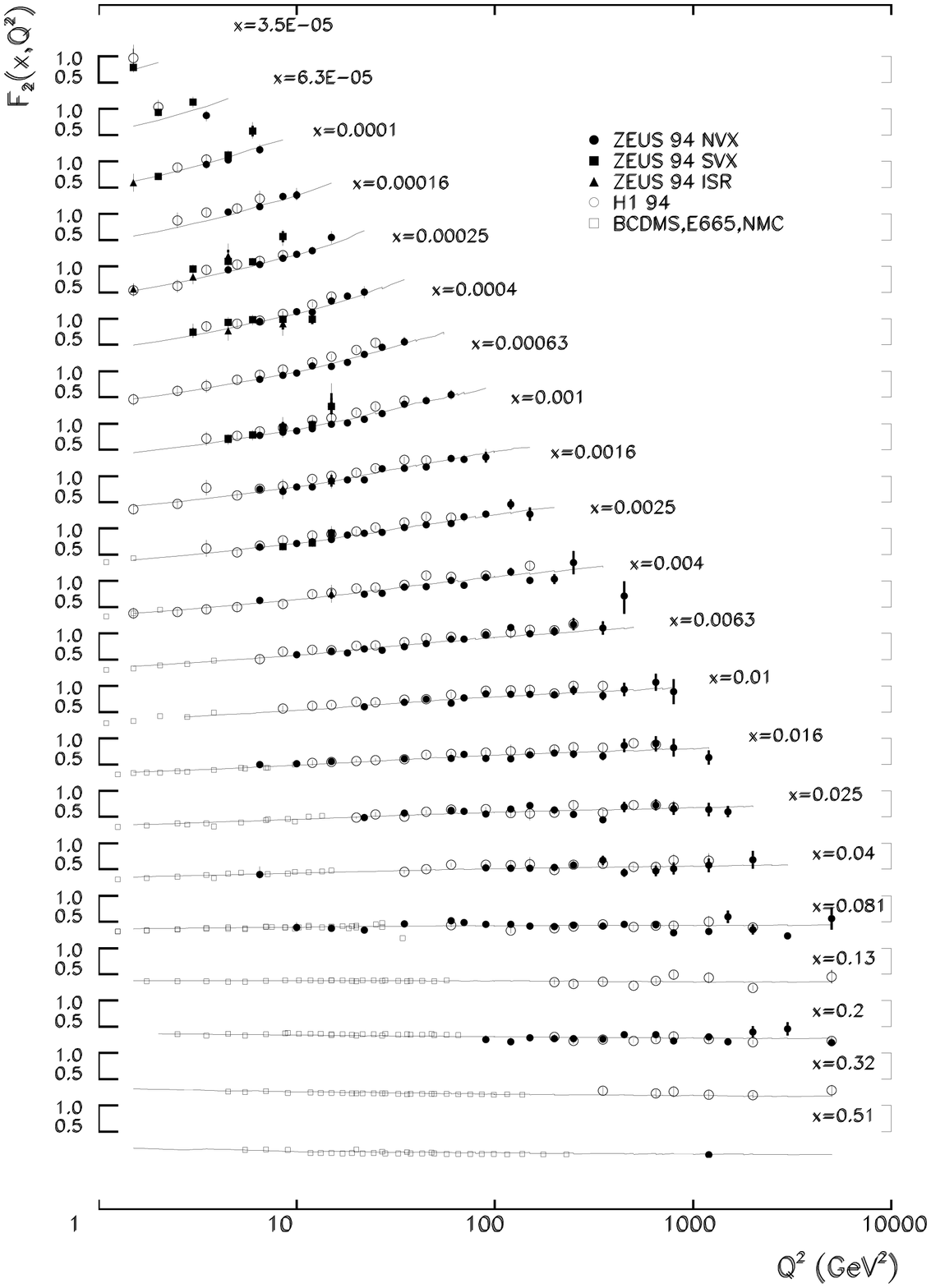} 
{120mm}{\Ftwo\ from ZEUS, H1,BCDMS~\protect\cite{BCDMS}, 
NMC95~\protect\cite{NMC-F2} and  
E665~\protect\cite{E665-F2} in $x$ bins plotted as a function of \Qtwo. The  
curve shows the prediction from a NLO QCD fit to the ZEUS data.} 
{fig-F2H1Z-xbins} 

Finally in this section we show in \fref{fig-F2H1Z-xbins} the \Ftwo\ data  
from both experiments  plotted as a function of \Qtwo\ 
for different $x$ bins. 
The data now extend over four orders of magnitude both in $x$ and  
\Qtwo. The pronounced change in slope as $x$ rises is clear evidence for  
scaling violations and the excellent agreement of the QCD fit with the  
data shows that these violations are consistent with the expectations of  
QCD. Since these scaling violations are generated by gluon emission and  
splitting, fits using a full QCD formalism would be expected to yield  
information on the gluon density in the proton. This leads us into   
section~\ref{sec-gluon}, in which the determination of the gluon  
structure function of the proton is discussed. Section~\ref{sec-F2c}  
discusses the first data on the 
determination of  
\Ftwo\ for particular quark flavours (i.e.\ $F^{charm}_2$), 
which is not only 
a major part of the inclusive $F_2$, but also will give us important  
new information on the gluon density. 
 
In conclusion, it is clear that the results on \Ftwo\ from HERA have had  
a profound effect on our understanding of QCD and have stimulated  
much theoretical activity.  The quality of the data is now such that QCD  
can be quantitatively tested. However, the difference in \Ftwo\ predicted  
at low $x$ if BFKL behaviour dominates compared to that from DGLAP is  
rather small. Moreover, there is a great deal of freedom in fitting the steep  
rise in $F_2$ in the DGLAP analyses, either in the MRS and CTEQ  
approaches by varying the steepness of the starting gluon distributions or  
in the GRV approach by changing slightly the starting \Qtwo\  
distribution, $Q_0^2$. The H1 collaboration~\cite{H1-gluon} 
have performed QCD fits to their  
\Ftwo\ data in which they have compared pure DGLAP evolution to a  
mixed DGLAP-BFKL formalism; both ansatzes give equally acceptable fits  
to the data. Currently the precision of the data is insufficient to  
distinguish these options.  
 
Our lack of knowledge of \Fl\  
will probably dominate the systematic error on \Ftwo\ at
the higher values of $y$ in the near future.  
Since \Fl\ is in principle sensitive to properties of QCD which cannot be  
explored from measurements of \Ftwo, it is also of intrinsic importance  
to measure \Fl. This measurement is however notoriously difficult, and 
forms the subject of the next section.  
\subsection{Measurement of \Fl} 
\label{sec-quarks-fl} 
The DIS differential cross-section given in equation~\ref{eq-Fl-sigma}  
shows that \Fl\ and $F_2$ have differing contributions depending on 
the value of $y$. One method for measuring \Fl\ would therefore be to 
measure the differential cross-section at fixed $Q^2, x$ but varying $y$. 
Since these quantities are related by \eref{eq-yq2sx}, this can only 
be achieved by changing $s$.  
 
One obvious way in which $s$ could be changed would be to change either the  
positron or proton beam energy by an amount large enough to produce 
a sizeable change in $y$ and therefore to separate out the contributions 
of \Fl\ and $F_2$ to the cross-section. Increasing the HERA
beam energies is severely limited by technology and cost;
unfortunately there are also severe 
penalties associated with reducing beam energies to the sort of level 
required to make a reasonable measurement of \Fl. These penalties 
include both a loss of luminosity and increase in systematic errors 
associated with for example the identification of the scattered electron. 
For these reasons, although there have been 
feasibility studies~\cite{Bauerdick} and the 
possibility continues to be discussed, the option of reducing the beam 
energies has not been used at HERA to date.  
 
An alternative possibility is to utilise the unavoidable tendency of the 
incoming positron to radiate a hard photon. Such hard radiation results in 
an effective reduction of $s$. Several studies of utilising such events to  
measure \Fl\ have indeed been carried out~\cite{Fl-gamma}.  
However, the radiation of 
hard photons is clearly less likely by a factor $\alpha$ and 
together with
the limitations of phase space on the emitted photon energy 
the fraction of such radiative events is very  small. This 
fact together with additional systematic effects which also 
plague such determinations mean that a useful measurement of 
\Fl\ using radiative events has yet to be achieved. 
 
The H1 collaboration has recently carried out an extraction of 
\Fl\ which utilises neither of these techniques~\cite{H1-Fl-sub}. 
Instead they choose the kinematic region which maximises the 
size of the contribution of \Fl\ to the differential cross-section 
($y \rightarrow 1$) and subtract the value of $F_2$. The disentanglement 
of the \Fl\ and $F_2$ contributions is achieved by using data at low $y$ 
where \Fl\ has a negligible contribution to the cross-section to determine  
$F_2$, and then evolving this 
value of $F_2$ to high $y$ using the NLO Altarelli-Parisi equations.  
Given that this procedure involves an extrapolation which makes 
strong assumptions both on the form of the extrapolation and on the value 
of \Fl\ in a different region of phase space, clearly this method is 
not as satisfactory as the two alternative methods mentioned above. 
It does however have the inestimable advantage of producing a significant 
determination of \Fl.  
 
The region of $y$ selected by H1 for the \Ftwo\ measurement was
 $0.6 < y < 0.78$,  
divided into six bins of \Qtwo\ between 8.5 and 35 GeV$^2$ and 
$x$ between 1.4 and $5.5 \cdot 10^{-4}$. 
In order to achieve such large values of $y$ it is necessary to 
identify very low energy scattered electrons,  
in this case between 6.5 and 11 GeV. Since high $y$ events also 
tend to produce current jets in the rear direction, there is significant 
background from hadronic and photonic energy deposits which mimic 
a scattered electron. Such factors were evaluated by modifying 
the normal electron finding strategy, which considered only the highest 
energy electromagnetic energy deposits in an event, to include also the  
second highest deposits. A careful estimation of the photoproduction 
background was also made using ``tagged" photoproduction events, which 
have the scattered electron detected in the luminosity detector, but which 
otherwise satisfied the selection criteria. The photoproduction background 
was never larger than 20\% and fell to $\sim 5$\% at the highest values 
of \Qtwo. The systematic errors on the cross-sections in the chosen  
high $y$ range were between 6 and 8\%.  
 
A QCD fit similar to the standard H1 analysis of $F_2$ was carried out 
but using only data with $y < 0.35$ and omitting fixed target data from 
NMC. BCDMS data~\cite{BCDMS} was used to constrain the high $x$
data. The 
Sigma method (see section~\ref{sec-kin-var-recons}) was used for the low $y$ 
data, while the electron method was used for the reconstruction of 
the kinematic variables in the high $y$ region used in the \Fl\  
determination. The starting point for the Altarelli-Parisi 
evolution of the $F_2$ determined from the low-$y$ data was 
\Qtwo\ = 5 GeV$^2$. The values for $F_2$ thus determined were 
subtracted from the measured cross-sections to produce the values 
for \Fl\ shown in \fref{fig-H1-Fl}. It can be seen that the extreme  
values of \Fl\ = $F_2$ and \Fl\ = 0 are excluded 
(at more than two standard deviations). The \Fl\ values are clearly  
dominated by the systematic error. They agree well with \Fl\ constructed  
from the parton distributions produced from the QCD fit to \Ftwo\ 
assuming the QCD prediction for \Fl\ (see \eref{eq-Fl}) .  
\ffigp{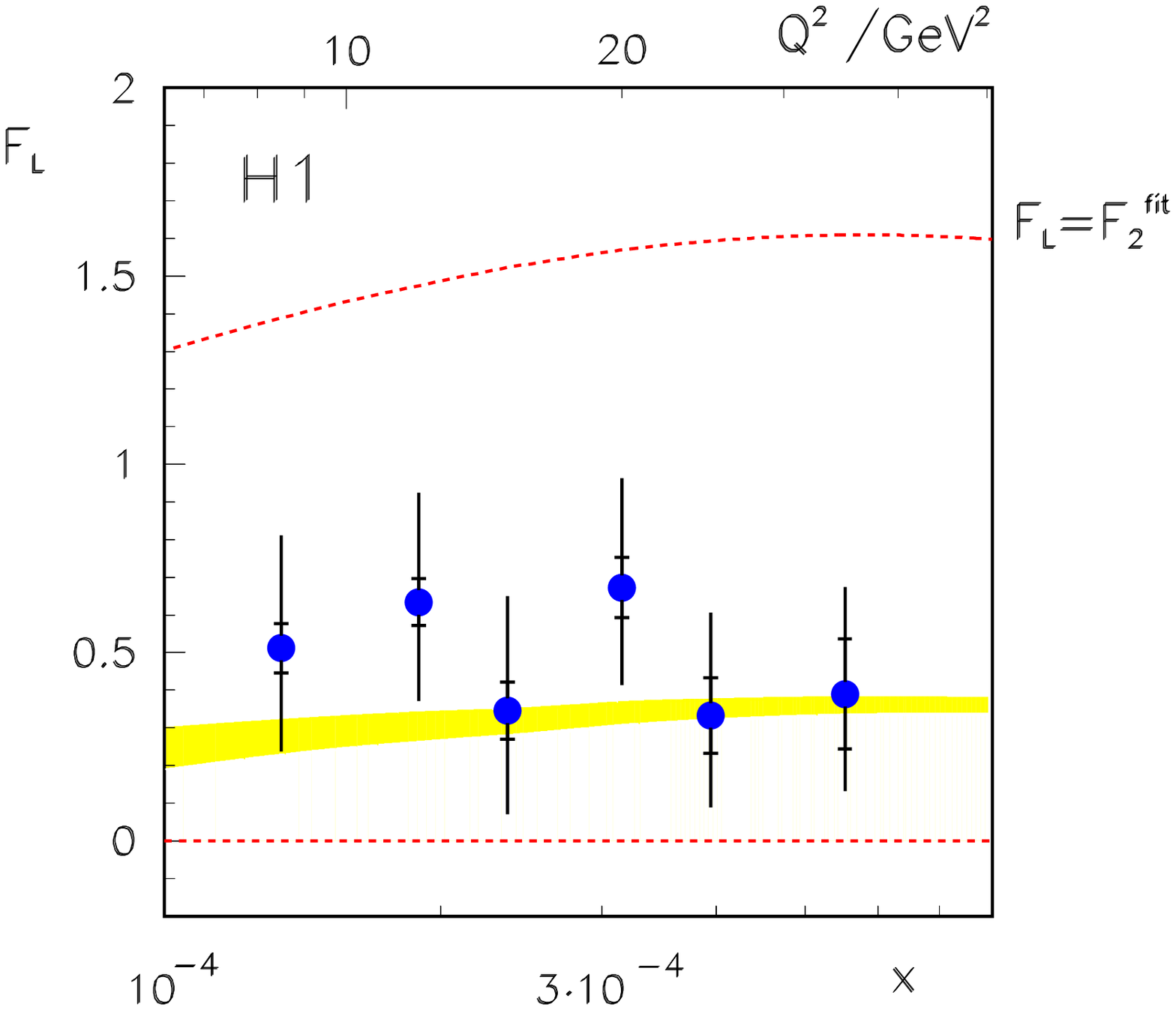} 
{70mm}{90mm}{\Fl\ as a function of $x$ for $<y> = 0.7$. The inner 
error bars correspond to statistical errors, the outer error 
bars to statistical and systematic errors in quadrature. The shaded 
band corresponds to the allowed range in \Fl\  from the 
parton distributions determined from the QCD fit for the low 
$y$ data evolved according to NLO QCD. The dashed lines 
correspond to the limits of \Fl\ = 0 and \Fl\ = $F_2$.} 
{fig-H1-Fl} 
 
It is clear that this is an interesting determination of \Fl\ 
at unprecedentedly low $x$; equally clearly, it is dominated by 
the systematic error and the theoretical assumptions employed. 
Further progress in the measurement of 
\Fl\ must come from either greatly increased luminosity 
(via radiative events) or from a reduction in one or both of the beam 
energies at HERA. 
\section{The gluon density inside the proton} 
\label{sec-gluon} 
Several different methods can be used at HERA to estimate the gluon  
structure function of the proton. The ``classic'' 
way of measuring $G(x)(\equiv xg(x))$ 
is to measure the scaling violations in $F_2$ 
as a function of \Qtwo. Other processes can give information, although 
usually in a rather more model dependent way. The crucial element is that  
something in the process should provide a hard scale about which a QCD  
evolution can be anchored. This can either be the \Qtwo\ of the process, a  
large \pt\ jet, or a large mass from heavy quark production.   The rate of  
multi-jet events from boson-gluon fusion processes and their kinematic  
properties allow a reconstruction of the momentum density of the initial  
gluon. Another source of information is the measurement of vector  
meson production in deep inelastic scattering. We  
also mention for completeness two photoproduction processes, \Jpsi\  
production and inclusive charm production, which, because of the hard  
scale given by the charm quark mass, can also be used to determine the  
gluon density within the proton.  

The importance of these methods in determining $g(x)$ is not only the 
different experimental conditions involved. They also have different
sensitivities to the gluon density. In scaling violations $g(x)$ is 
proportional to the derivative of the measured cross-sections; for jet
production and open charm production $g(x)$ is directly proportional to
the measured cross-sections; whereas in vector meson production the 
cross-section is proportional to $\|g(x)\|^2$. Determinations by all
of these methods therefore, while subject to differing degrees of model
dependency, give important constraints and cross-checks on the value of
the gluon density over a wide kinematic range.
 
It is important to note that the gluon distribution is not uniquely defined  
in NLO QCD (see section~\ref{sec-dsigdo-dis}). Both ZEUS and H1 use 
the $\overline{MS}$  
scheme~\cite{MSbar-scheme} in their determinations.  
\subsection{$G(x)$ from scaling violations in $F_2$} 
\label{sec-gluon-scale-viol} 
The ZEUS and H1 data presented in \fref{fig-F2H1Z-xbins} show  
qualitatively the behaviour to be expected from QCD, i.e.\ a growth in the  
sea quark density at lower $x$ which increases as \Qtwo\ increases. It is  
therefore instructive to make a quantitative comparison with QCD 
via a full QCD fit to next-to-leading order. As we have already  
discussed, this  
has also been done by a number of groups, who make global fits to  
all available data in order to produce the ``best-fit'' parton  
distributions. The  
approach of the HERA experiments is somewhat different, in that they use  
the minimum amount of extra data necessary to produce a reasonably  
constrained fit, particularly at high $x$, which is 
kinematically difficult to  
access at HERA, and are able to make full use of their knowledge of the  
correlations between the experimental  
errors~\cite{H1-1996-F2,ZEUS-gluon}.  
 
\Eref{eq-H1QCDfit-form} shows the functional forms at an initial \Qtwo\  
value of $Q^2_0 = 4$ GeV$^2$ used by the H1 collaboration in their QCD 
fit on their 1993 data.  
\beqn 
xg(x) &=& A_gx^{B_g}(1-x)^{C_g}\nonumber \\ 
xq_{NS}(x) &=& A_{NS}x^{B_{NS}}(1-x)^{C_{NS}}(1+D_{NS}x) \\ 
xq_{SI}(x)&=&A_{SI} x^{B_{SI} }(1-x)^{C_{SI}}(1+D_{SI}x)\nonumber 
\label{eq-H1QCDfit-form} 
\eeqn 
The singlet quark density is defined as  
$q_{SI}=u+\bar{u}+d+\bar{d}+s+\bar{s}$. The non-singlet quark density is  
given by $q_{NS}=u+\bar{u}-q_{SI}/3$. ZEUS used a rather different  form, as  
shown in \eref{eq-ZEUSQCDfit-form}. 
\beqn
xg(x)&=&A_gx^{B_g}(1-x)^{C_g}\nonumber \\ 
xq_{NS}(x)&=&A_{NS}x^{B_{NS}}(1-x)^{C_{NS}} \\ 
xq_{SI}(x)&=&A_{SI} x^{B_{SI} }(1-x)^{C_{SI}}(1+D_{SI}x + E_{SI}  
\sqrt{x})\nonumber 
\eeqn 
\label{eq-ZEUSQCDfit-form} 
 
It is important to  
impose a constraint on the momentum fraction carried by the gluon at  
low \Qtwo, which in the H1 case was set at 0.44~\cite{NMC} at $Q_0^2 = 4$  
GeV$^2$, the point from which their evolution begins. The momentum  
sum rule then gives the normalisation of the singlet quark density.  
ZEUS preferred to input the valence quark distribution directly from the  
MRSD-$'$ parameterisation, assuming that the strange quark distribution is  
20\% of the quark sea and using $Q_0^2 = 7$ GeV$^2$. Only the three  
light quarks are involved in the DGLAP evolution. The effect of the charm  
quark was included via the photon-gluon fusion process as discussed  
in~\cite{GRV-charm1,GRV-charm2}; ZEUS used the leading-order  
calculation for this effect for the analysis of the 1993 data, while H1 used the next-to-leading order according to \cite{Laenen-charm}. At the \Qtwo\ of  
interest here, the effects of the $b$ quark can be neglected.  
However, it is important~\cite{Marciano-charm} to ensure that  
$\alpha_s(Q^2)$ remains continuous across the $c$ and $b$ thresholds.  
Since diffractive events (see  section~\ref{sec-diffraction}) presumably 
occurred in fixed target experiments but were not separated out, diffractive 
events were also retained in these analyses.  
 
Since the whole basis of the DGLAP equations is a strong link between  
gluon and sea, with the sea quark distributions being driven by the gluon  
distribution at small $x$, it has been conventional until recently to assume  
that these distributions have identical $x$ behaviour. However, the quality  
of the HERA data, particularly when taken in conjunction with other  
experimental data, is now sufficiently good to indicate a preference that  
these distributions should be decoupled~\cite{MRS-pin-glue}. ZEUS and  
H1 have therefore performed fits in which the $x$ dependence of the sea  
and gluon is allowed to be different. 
 
\Fref{fig-ZEUS1993-gluon} shows the results of the full  
QCD analysis from the 1993 ZEUS data~\cite{ZEUS-gluon}.   
\ffigp{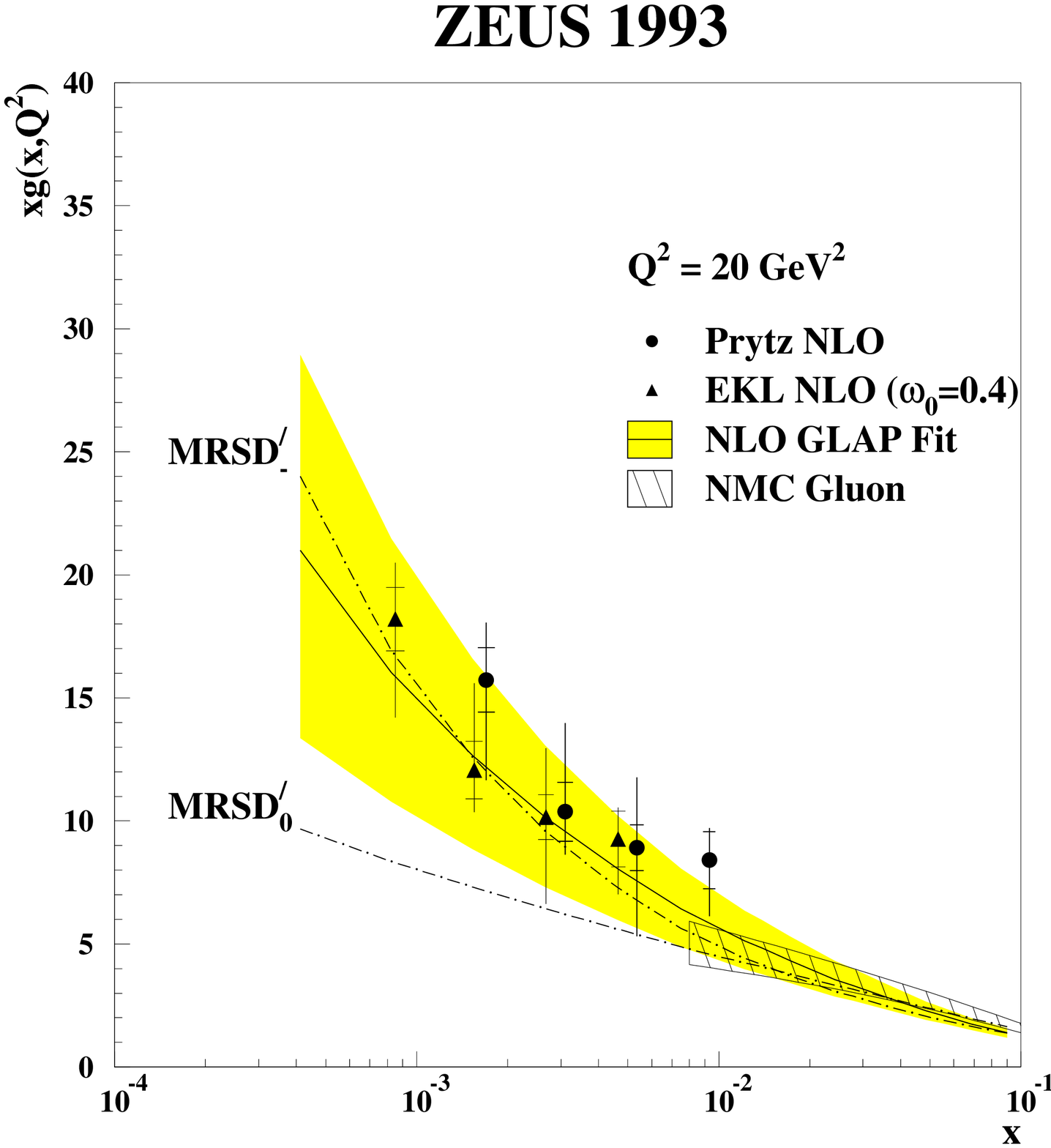} 
{70mm}{100mm}{$xg(x)$ as a function of $x$ as determined from scaling  
violations in the 1993 ZEUS $F_2$ data. Two approximate  
determinations from Prytz and Ellis, Kunszt and Levin are shown. The 
angled shaded  region shows a determination at higher $x$ from the NMC 
experiment. The shaded region shows the full QCD fit to the ZEUS 1993 
data. Also shown as dashed-dotted lines are the $MRSD_0'$ and  
$MRSD-'$ parameterisations of 
$G(x)$~\protect\cite{MRS}. 
Also shown are two methods using approximations to the QCD evolution 
equations to extract $G(x)$ in bins of $x$, due to 
Prytz~\protect\cite{Prytz}, and Ellis, Kunszt and  
Levin~\protect\cite{Ellis-KL}.}{fig-ZEUS1993-gluon} 
The Prytz method~\cite{Prytz} uses a Taylor expansion about $x/2$, ignoring the  
contribution from the quarks in the DGLAP evolution equation, to give  
the approximate result at leading order 
\beqn 
xg(x,Q^2) &=& \frac{\partial {F_2(x/2,Q^2)}}{\partial{\log (Q^2)}} 
\left[\frac{40}{27}\frac{\alpha_s(Q^2)}{4\pi}\right]^{-1}  
\label{eq-PrytzLO} 
\eeqn 
which at next-to-leading order becomes 
\beqn 
xg(x,Q^2) &=&\frac{\partial {F_2(x/2,Q^2)}}{\partial{\log (Q^2)}} 
\left[\left(\frac{40}{27} +  
7.96\frac{\alpha_s}{4\pi}\right)\frac{\alpha_s(Q^2)}{4\pi}\right]^{-1} 
\\ \nonumber 
&& - \frac{5\alpha_s}{9\pi} N(x/2, Q^2) \left[\frac{40}{27} +  
7.96\frac{\alpha_s}{4\pi}\right]^{-1} 
\label{eq-PrytzNLO} 
\eeqn 
where $N(x/2,Q^2)$ is a correction function which depends on the already  
measured gluon density at high $x$. The EKL~\cite{Ellis-KL} method 
assumes a particular  
functional form for $g(x)$ and includes both the gluon and singlet quark  
contribution to the DGLAP evolution.  
 
\Fref{fig-ZEUS1993-gluon} shows that these approximate methods are consistent with the full  
ZEUS QCD fit, which in turn matches well to a determination of $G(x)$ by  
NMC~\cite{NMC}. H1 have also compared the leading order Prytz  
approximation with their full QCD fit and find good agreement.  

\Fref{fig-H1ZEUS-1994-gluon} shows the results of the QCD analyses of  both  
experiments for the 1993 data. They are in very 
good agreement, and map on well to the  
NMC gluon determination at $x > 10^{-2}$.  
\ffig{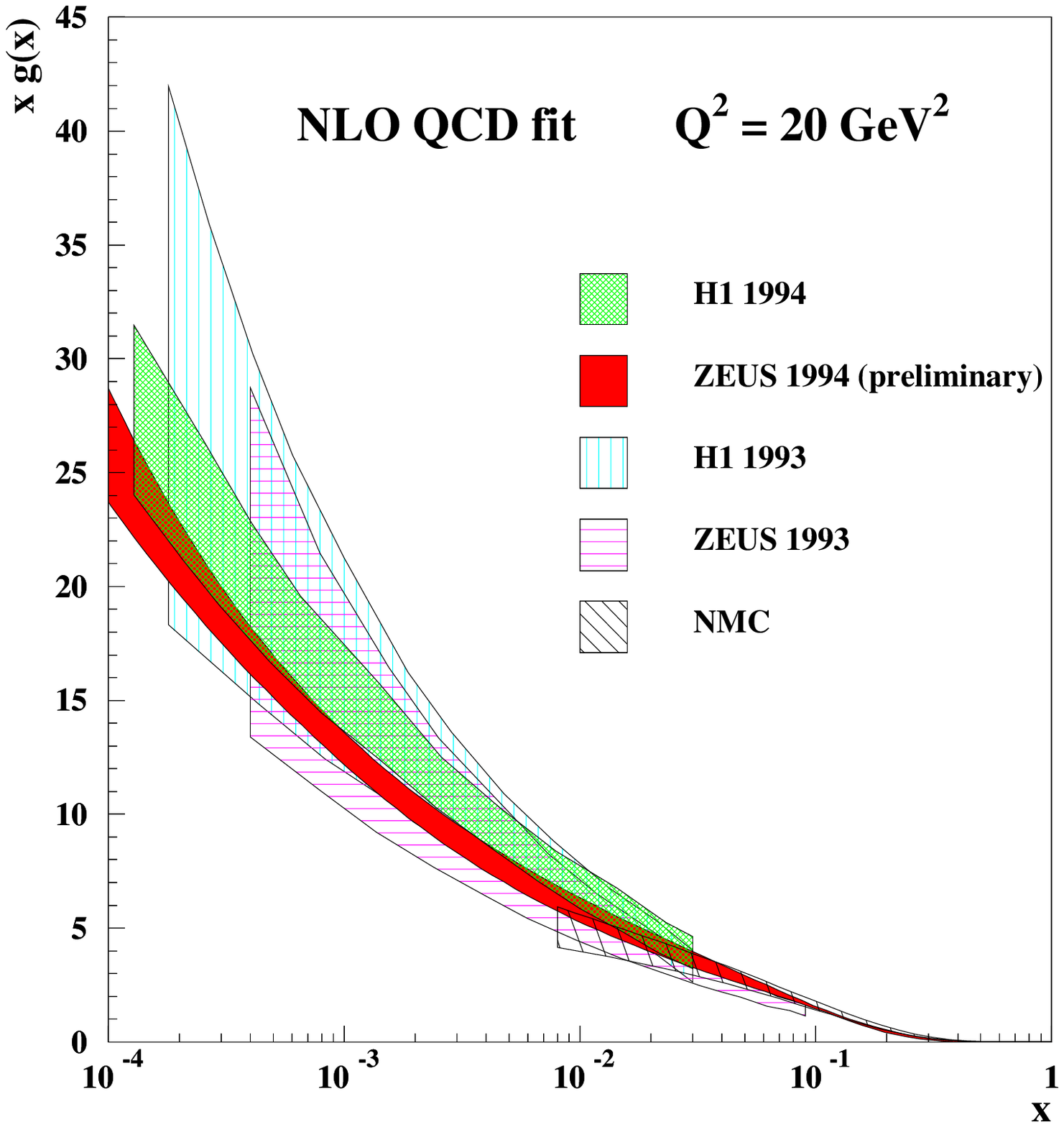} 
{100mm}{The gluon density $xg(x)$ as determined by the H1 and ZEUS 
collaborations using NLO QCD fits to both the 1993 $F_2$ data and 
the 1994 data, together with the 
determination at higher $x$ from NMC~\protect\cite{NMC}.} 
{fig-H1ZEUS-1994-gluon}  
 
The shaded areas in \Fref{fig-H1ZEUS-1994-gluon} show the allowed 
regions of the  
full NLO QCD fits in which the statistical and systematic errors, and the  
point-to-point correlations between them, are taken properly into account.  
A comprehensive list of the systematic effects studied can be found in the  
relevant experimental publications~\cite{H1-gluon,ZEUS-gluon}. Whereas  
in principle the value of $\alpha_s$ can be determined from the fit, in  
practice the scaling violation data is relatively insensitive to it. However,  
Ball and Forte have described a procedure in the double scaling regime  
which is sensitive to $\alpha_s$ using a fit of truncated parton  
distributions from global analyses to the HERA  
data at low $x$~\cite{Ball-Forte-alphas}. As the HERA data become more  
accurate, this may well become an important technique. Nevertheless, the  
attitude of the experiments until now has been to estimate the systematic  
error by  
redoing the fits using the limits of error values of $\alpha_s$. 
It transpires  
that the resultant additional systematic error is negligible at low $x$.  
 
The greatly increased statistics in the 1994 data allow a significant 
reduction in the error of the gluon determination. H1~\cite{H1-1996-F2}  
have modified their 1993 
fitting procedure to parameterise the parton distributions 
at \Qtwo\ = 5 GeV$^2$ and to allow the non-singlet distributions to be split 
into separate $u$ and $d$ quark distributions, viz. 
\begin{eqnarray} 
xg(x)&=& A_gx^{B_g}(1-x)^{C_g} \nonumber \\ 
xu_v(x)&=& A_{u}x^{B_{u}}(1-x)^{C_{u}}(1+D_{u}x+E_{u}\sqrt{x}) \nonumber \\
xd_v(x)&=& A_{d} x^{B_{d} }(1-x)^{C_{d}}(1+D_{d}x+E_{d}\sqrt{x}) \\ 
xS(x)&=& A_{S} x^{B_{S} }(1-x)^{C_{S}}(1+D_{S}x+E_{S}\sqrt{x}) \nonumber 
\label{eq-H1-1994-gluon} 
\end{eqnarray} 
In order to constrain the valence quark densities at high $x$, it is 
necessary to include some fixed target data from BCDMS~\cite{BCDMS} 
and NMC~\cite{NMC}. The H1 $F_2$ data were fit for 
\Qtwo\ $> 5$ GeV$^2$. The relative normalisation of these data sets and 
the H1 normal and shifted vertex data were allowed to vary 
within the quoted errors; no data set requires a normalisation 
change of more than $\pm$ 4\%. The overall $\chi^2$ for the fit was 
acceptable. The result of the fit in \Qtwo\ bins as a function of $x$ 
is shown in \fref{fig-H1-1994-QCD}. The quality of the fit and the  
change of slope as a function of \Qtwo\ can clearly be seen.  
\ffigp{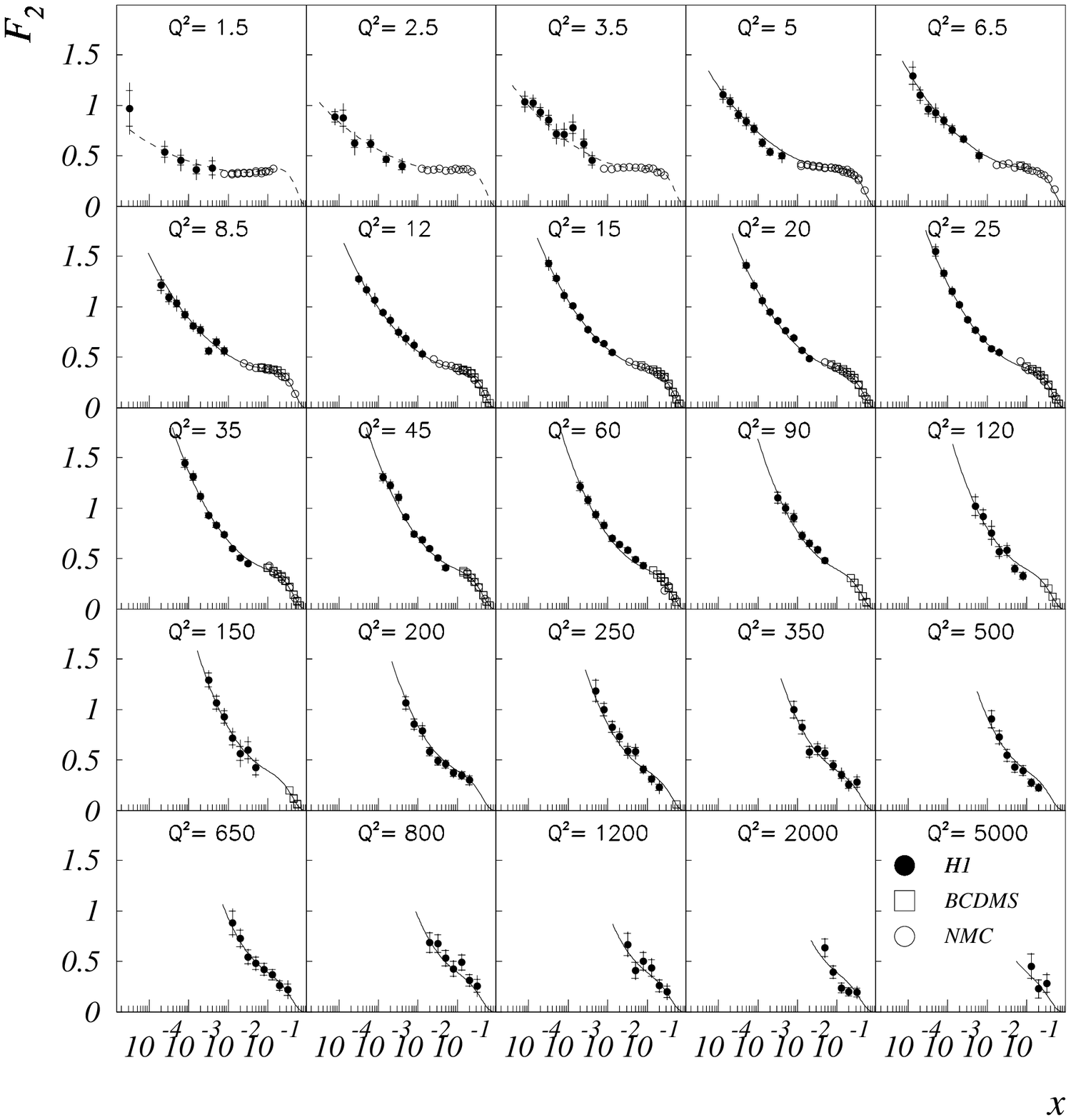} 
{160mm}{120mm}{The H1 $F_2$ data as a function of $x$ in bins of \Qtwo.  
The inner error bar is statistical, while the full error bar  
represents statistical and systematic errors added in quadrature. The 
overall scale uncertainty due to the luminosity is not included. 
The full curves represent the result of the QCD fit described in the 
text. The dotted lines represent the result of that fit evolved 
to lower \Qtwo\ according to the QCD prescription.} 
{fig-H1-1994-QCD}  
  
ZEUS have carried out a similar fit to their 1994 data, but this 
analysis is as yet still preliminary. \Fref{fig-H1ZEUS-1994-gluon} 
shows the result of the gluon determination from the 1994 data
from both experiments 
as a function of $x$, as well as the 1993
determinations. The 1994 data are in good agreement with the 1993
determinations and both experiments show a steep rise as 
$x$ falls. 
While the two experiments are compatible, the 1994 H1 
result does seem to be somewhat steeper than that of ZEUS.  
The GRV parameterisation has problems in fitting
the relatively shallow ZEUS data and maintaining a good
fit at higher $x$; the higher precision 1995 and 1996 data,
which are now starting to be analysed, may well be able to
rule out some of the {\it ans\"{a}tze} currently in use.  
 
In conclusion, the measurement of the gluon distribution via a NLO QCD  
analysis of scaling violations in $F_2$ has extended our knowledge of  
$G(x)$ by almost two orders of magnitude in $x$.  
 Scaling violations, while a relatively insensitive  
technique for measuring $G(x)$, remain the least model dependent  
method available.  
\subsection{$G(x)$ from jet rates  in DIS} 
\label{sec-gluon-jetrates} 
In the previous section we discussed the measurement of the gluon  
distribution by means of a fully inclusive DIS measurement, the scaling  
violations in $F_2$ as a function of \Qtwo. It is possible by studying more  
exclusive processes to obtain greater sensitivity to the gluon distribution,  
although usually at the price of greater model dependence. One such  
method is to study jet production in a region of phase space where  
gluon-induced reactions can be cleanly isolated.  
\Fref{fig-H1direct-feyn} shows the process of boson-gluon fusion, which  
is clearly sensitive at the Born level to the gluon density in the proton.  
Since each of the quarks at high enough \Qtwo\ will give rise to a jet,  
such a reaction can be isolated in a sample of ``2+1" jet events, where the  
proton remnant jet referred to as the ``+1" is predominantly lost in the  
beam pipe. Unfortunately, similar processes initiated by a quark from  
the proton rather than a gluon and where the second visible jet originates  
from hard gluon radiation, known as the ``QCD Compton" process, also  
populates this sample. To leading order, the total rate of ``2+1" jets is  
proportional to 
the sum of these two processes,  $\alpha_s (A \cdot g + B \cdot q)$, where  
$g$ and $q$ refer to the gluon and quark densities in the proton  
respectively and $A$ and $B$  can be calculated in perturbative QCD.  
Clearly it is possible in principle to measure $\alpha_s, g$ and $q$ from  
the ``2+1'' jet events. However in practice it has until now been necessary  
to either assume $\alpha_s$ from other measurements and measure  
parton distributions or {\it vice-versa}. The latter procedure 
is discussed in  
section~\ref{sec-alphas}. In this section we assume that $\alpha_s$ can be  
fixed from other processes and concentrate on the determination of $g(x)$.  
\ffigp{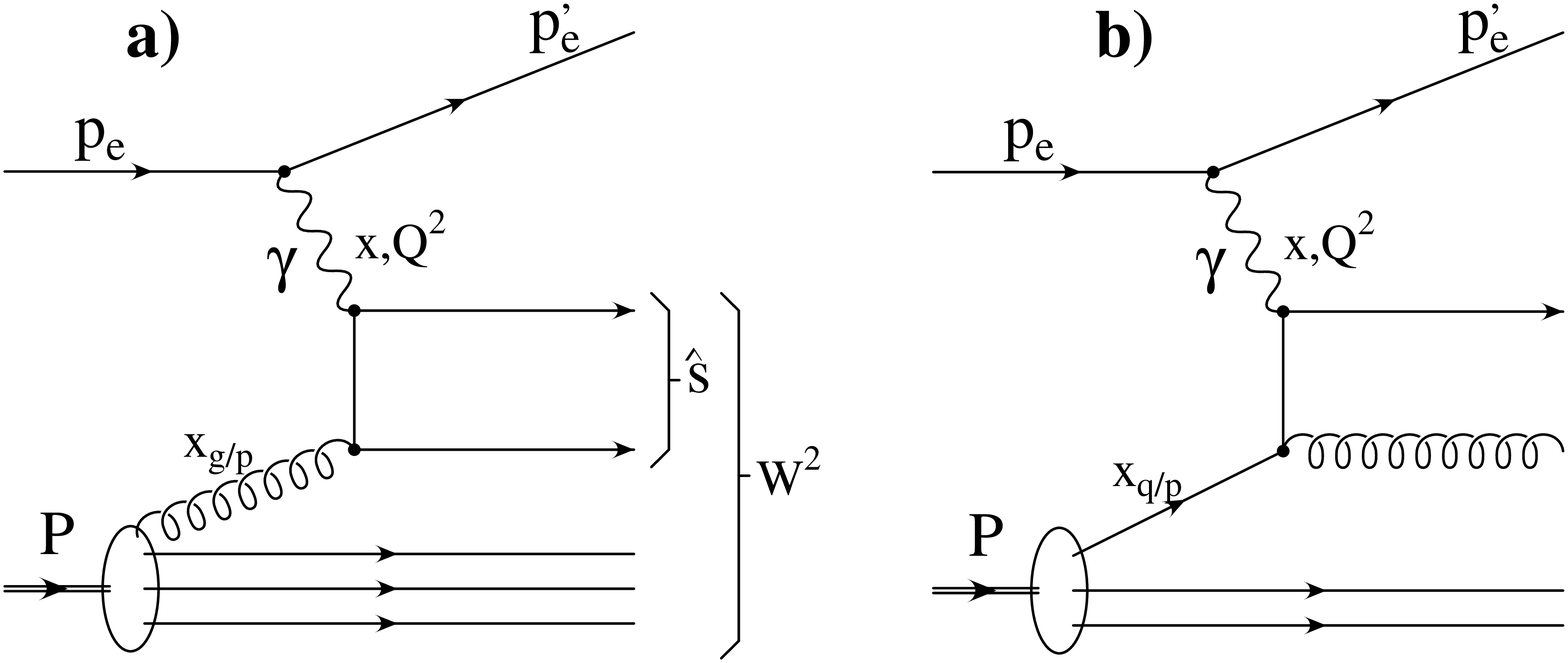} 
{50mm}{80mm}{Feynman diagrams appropriate  to ``2+1'' jet production in DIS.
The first, a), shows the boson-gluon fusion process; b) shows the QCD Compton 
process. Note that hard gluon radiation from the struck quark after it
has absorbed the virtual photon is also a QCD Compton 
process, but it not shown here.} 
{fig-H1direct-feyn} 
 
In the processes illustrated in \fref{fig-H1direct-feyn}, the $x$ of the  
parton from the proton ($x_{i/p}$) is not identical with the ``Bjorken-$x$"  
seen by the virtual photon. Indeed 
\begin{equation} 
x_{i/p} = \frac{\hat{s}+Q^2}{ys} = x\left(1+\frac{\hat{s}}{Q^2}\right) 
\label{eq-direct-xip} 
\end{equation} 
The hard sub-process energy squared, $\hat{s}$, can be measured 
either directly from 
the invariant mass 
of all particles $p_j$ of the two jets belonging to 
the hard sub-system 
\begin{equation} 
\hat{s} = \left( \sum_{j} p_j \right)^2 
\label{eq-direct-shat-sump} 
\end{equation} 
or from the jet directions in the 
hadronic centre-of-mass system 
\begin{equation} 
\hat{s} = 
W^2 e^{-(\eta_{1}^*+\eta_{2}^*)} 
\label{eq-direct-shat-rapidity} 
\end{equation} 
where $\eta_{1}^*$ and $\eta_{2}^*$ are the pseudo-rapidities 
($\eta = - \ln (\tan \theta /2$))
of the partons taking part in the hard sub-process.  
 
The struck partons in both the BGF and QCD Compton processes emerge  
into the detector as jets of particles whose reconstructed energies and  
directions approximate to those of the partons. The ability of the analysis  
to discriminate on kinematic grounds between the BGF and QCD  
Compton processes is therefore crucially dependent on the quality of the  
jet measurement. 

The event selection in the H1 analysis~\cite{H1-gluon-jetrates} is based on 
the identification of exactly two  
jets via the cone algorithm with 
$\Delta R = \sqrt{\Delta\eta^2  
+ \Delta\phi^2} = 1$ and \pt$^*  > 3.5$ GeV. A 
series of kinematic cuts is made  designed to enhance the proportion of  
``2+1'' jet events emanating from the BGF process and to improve the jet  
finding efficiency. In order to improve the resolution on $\hat{s}$ and  
hence on 
$x_{g/p}$, both equations~\ref{eq-direct-shat-sump} and 
\ref{eq-direct-shat-rapidity} 
 are evaluated and only those events with $|\Delta \sqrt{\hat{s}}| \leq 10$  
GeV, where $\Delta \sqrt{\hat{s}}$ is  the difference between  
$\sqrt{\hat{s}}$ found from the two reconstruction methods,  
are retained in the analysis. This  
requirement removes about 20\% of candidates, and not only suppresses  
tails in the resolution but also discriminates against contamination from  
``1+1'' jet events, where the two methods in general produce different  
answers.  The sample produced in this way consists of 328 ``2+1'' jet events.  
 
Good agreement is observed between the data and Monte Carlo predictions  
using the LEPTO 6.1 generator, which is used both to estimate the  
acceptance for the BGF process and also to calculate the background from  
quark initiated contributions such as the QCD Compton process. The 
cross-section can be parameterised as 
\beqn 
\sigma^{2+1}_{obs.}(x_{i/p}^{rec}) & = & 
\int M(x_{g/p},x_{g/p}^{rec}) 
x_{g/p} g(x_{g/p}) \mbox{\rm d}x_{g/p} \\ \nonumber  
& + & \sigma^{QCD-C}_{MC}(x_{q/p}^{rec}) + 
\sigma^{QPM}_{MC}(x_{q/p}^{rec}) 
\label{eq-direct-sigma21} 
\eeqn 
After subtraction of the backgrounds from QCD Compton processes 
($\sigma^{QCD-C}_{MC}$), and from ``1+1" jet events faking 
``2+1" jet events ($\sigma^{QPM}_{MC}$), the gluon  
distribution may be deduced by de-convoluting with the function $M$,  
which contains the QCD matrix element and the effects of hadronisation. 
 After unfolding in five bins the results shown  
in \fref{fig-H1direct-results} are obtained.  
%
\ffig{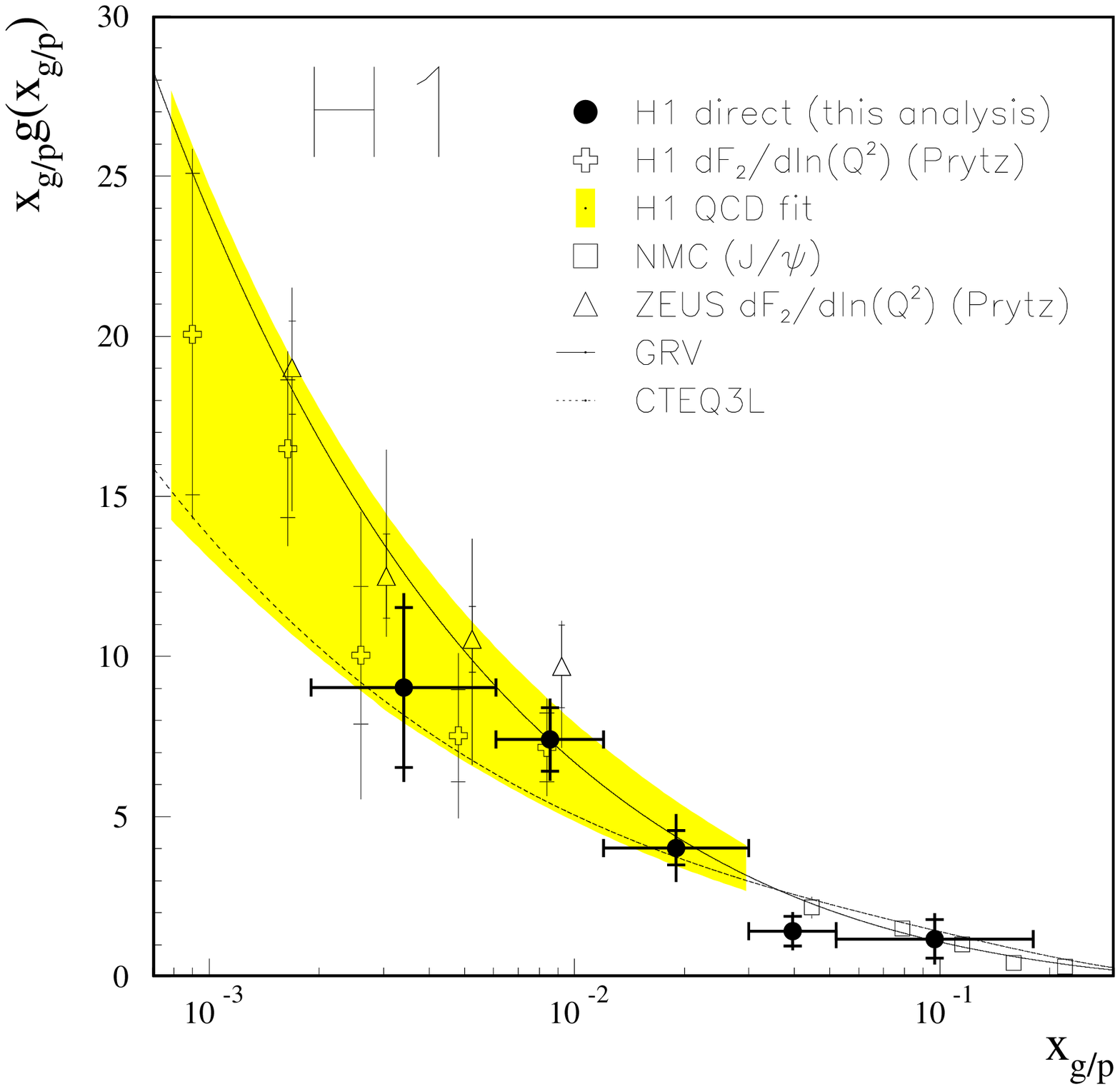} 
{80mm}{The measured gluon density at $<Q^2> \sim$ 30 GeV$^2$ 
as a function of the fractional gluon momentum as determined in
the H1 ``2+1" jet analysis. Also shown are the  
determinations by H1 and ZEUS referred to in  
section~\ref{sec-gluon-scale-viol} at $Q^2$ = 20 GeV$^2$ as well as a 
determination from $J/\Psi$ production 
by NMC ~\protect\cite{NMC-g-Jpsi} evolved to $Q^2$ = 30 GeV$^2$} 
{fig-H1direct-results} 

The unfolding procedure produces a full covariance matrix of the  
statistical errors. The systematic error in each bin is obtained from  
reasonable variation in the assumptions used in the analysis. The largest  
systematic effects are the uncertainty in the absolute hadronic energy scale  
in the H1 Liquid Argon calorimetry at low $x$ and the variation in the cut  
on $|\Delta \sqrt{\hat{s}}|$ at high $x$. The error on the BGF and QCD  
Compton acceptance was estimated by using different models in the Monte  
Carlo, e.g.\ HERWIG. \Fref{fig-H1direct-results} shows good  
agreement between this analysis and other methods from H1, ZEUS and  
NMC. The data from this analysis lies in an $x$ region 
between those from other methods at  
HERA and  NMC~\cite{NMC-g-Jpsi}. It demonstrates the consistency  over  
the full $x_g$ range and also exhibits a strong rise in the gluon density as  
$x_g$ falls.  
 
Subsequent to this analysis, which uses leading order matrix elements,  
Mirkes and Zeppenfeld~\cite{Mirkes-Zepp}  derived the next to leading  
order corrections. The main effect of these corrections is to reduce the  
dependence on the renormalisation and factorisation scale. Although the  
effect on the $\hat{s}$  and $x$ distributions is quite large, the effect on  
the final parton distribution functions is relatively small due to  
cancellations. However, the unfolding of the gluon density becomes  
significantly more complicated.  
\subsection{Light vector meson production}  
\label{sec-gluon-vectorm} 
In DIS, the hard scale in the production of light vector mesons  such as the  
$\rho$ and $\phi$ is fixed by the \Qtwo\ of the deep inelastic process.  
The link between this phenomenon and the gluon density in the proton is  
found in many perturbative QCD models of the production 
mechanism~\cite{Ryskin,Brodsky}. Such models consider the reaction to 
proceed in leading order via the colour singlet exchange of two gluons. For  
example, 
in the model of Brodsky {\it et al.}~\cite{Brodsky} the cross-sections of  
vector mesons produced 
by longitudinally polarised photons  
are proportional to the square of the gluon density in the proton:  
\beqn 
\hspace{-0.4cm} \left. \frac{d\sigma _L}{dt}\right|_{t=0}(\gamma^*N  
\rightarrow V^0N)  & =  &   
\frac{A}{Q^6} \alpha_s^2(Q^2) \nonumber\\
& \cdot & \left| \left[ 1  + i\frac{\pi}{2}  
(\frac{d}{d \ln x})   
\right] xg(x,Q^2) \right|^2  
\label{brodsky} 
\eeqn 
where $A$ is a constant predicted by the model.  In principle this quadratic  
dependence increases the sensitivity to the  
gluon density compared to methods such as those described earlier, which  
are related to single gluon processes. However, the extraction of the gluon  
distribution from vector meson data 
using these methods is only correct to leading order and 
furthermore relies on explicit theoretical assumptions which currently  
cannot be unambiguously established by the data.  
 
Deep inelastic vector meson production has a very distinctive and clean  
signature at HERA. This means that the data selection is straight-forward.  
Two charged tracks consistent with being hadrons 
and a scattered positron indicative 
of deep inelastic scattering are required.  
Kinematic cuts are used to reduce the photoproduction background 
caused by the mis-identification of energy deposits in the calorimeter 
as a scattered positron. 
Both H1 and ZEUS see clear signals for $\rho^0$  production. The 
signal obtained by H1 is shown in~\fref{fig-rho-mpipi}, 
where a clear $\rho$ signal with the expected mass and width is 
observed with very little background.  
\ffig{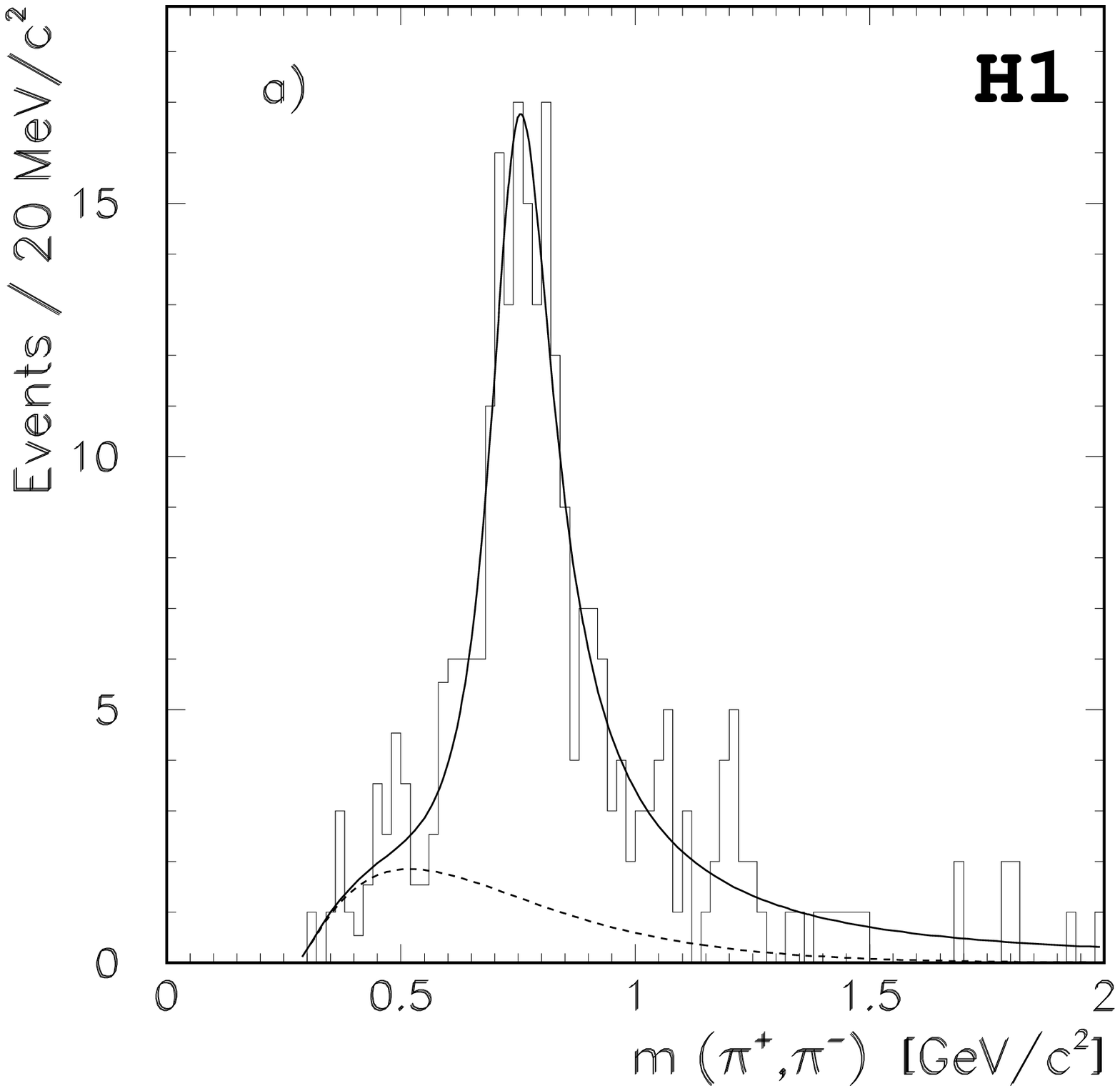} 
{70mm}{The $\pi^+ \pi^-$ invariant mass distribution for $\rho$ 
candidates in deep inelastic scattering for $7 < Q^2 < 25$ GeV$^2$ and 
$40 < W < 130$ GeV.} 
{fig-rho-mpipi} 
\Fref{fig-H1-rhoJpsi-sigma}a 
shows the total elastic $\rho$ cross-section as a  
function of $W$ as measured by ZEUS~\cite{ZEUS-DIS-rho} and  
H1~\cite{H1-DIS-rho} in two bins of \Qtwo. 
NMC~\cite{NMC-rho} points at lower $W$ are also shown. 
The ZEUS data seem substantially  
higher than those of H1, although the quoted overall normalisation  
uncertainties of 31\% for ZEUS and 7\% for H1 mean that the discrepancy is  
not significant. The slope of the H1 cross-section as a function 
of $W$ seems to lie between the ``soft''  
value which fits the $\rho$ photoproduction  
data~\cite{DL-rho}, and the ``hard'' value appropriate to QCD-type  
models~\cite{Brodsky}, whereas that of ZEUS is compatible with the  
``hard'' expectation. 
Only ZEUS therefore has used the $\rho$ data to extract a  
gluon distribution.   
\ffigp{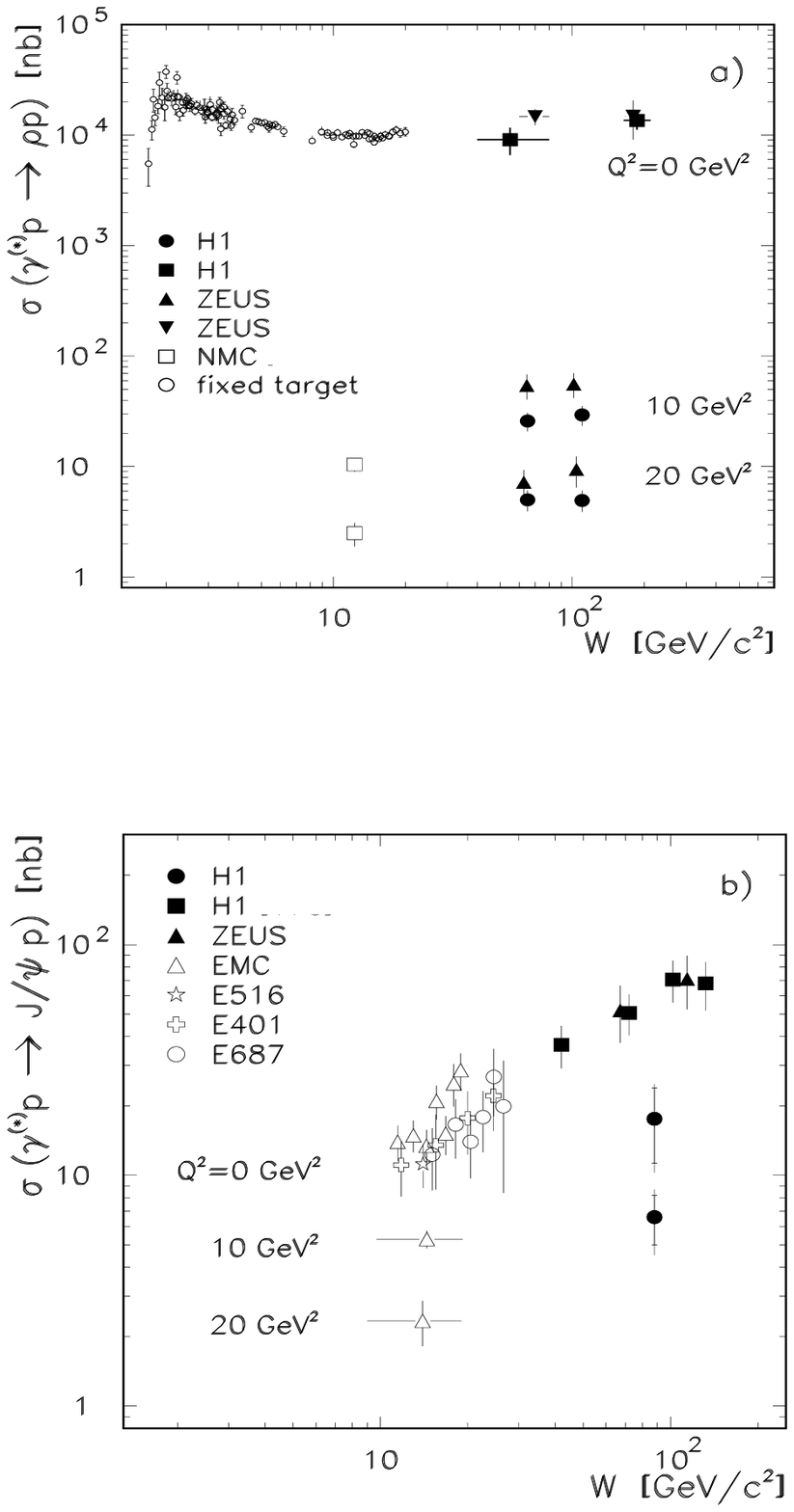} 
{120mm}{80mm}{$W$ dependence of the elastic DIS vector meson cross-section
in fixed target and HERA experiments a) for $\rho$ production
(computed
for $m_{\pi^{+}\pi^{-}} < 1.5$ GeV) and b) for \Jpsi\
production (see section~\protect\ref{sec-gluon-hq}).
For the $\rho$ data, an
overall normalisation uncertainty of 31\% for ZEUS and
20\% for NMC is not included in the plot.}
{fig-H1-rhoJpsi-sigma}        
 
The $\gamma^{\star} p$ cross-section for $\rho$ production from ZEUS 
is shown in \fref{fig-sigmarho-x} 
as a function of $x$ in the two measured  
\Qtwo\ bins. The ZEUS measurement for the gluon distribution from the  
$F_2$ scaling violations can be used within the framework of the model of 
Brodsky {\it et al.} to obtain a prediction for $\rho$ production which is 
also shown in \fref{fig-sigmarho-x}. The $\rho$ data is consistent with 
this prediction. 
\ffigp{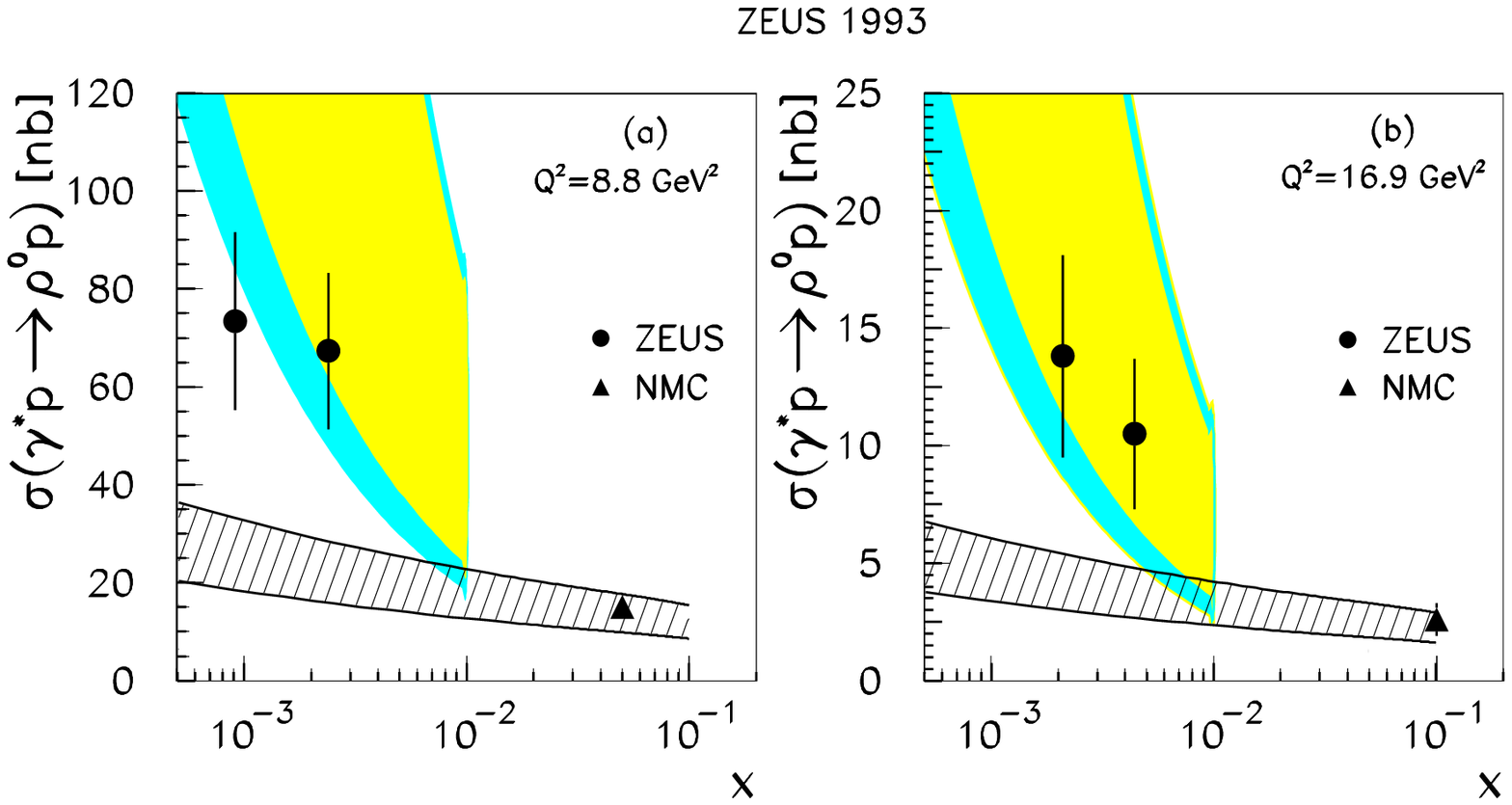} 
{70mm}{130mm}{a) $\sigma (\gamma^* p \rightarrow \rho^0 p)$,  
as a function of $x$ at \Qtwo\ = 8.8 GeV$^2$.   
The errors shown include both statistical  
and  systematic uncertainties added in quadrature. 
Also shown is the NMC result~\protect\cite{NMC-rho}.  
The dashed curve is a prediction from~\protect\cite{DL-rho}. 
The hatched area corresponds to the predictions of Brodsky {\it et al.} 
\protect\cite{Brodsky}  
for $x<0.01$, where the upper and lower limits 
are obtained from the limits on $xg(x)$ from \protect\cite{ZEUS-gluon}.  
b) a similar plot for data at \Qtwo\ = 16.9 GeV$^2$ .} 
{fig-sigmarho-x} 
\subsection{Heavy quark processes in DIS and photoproduction} 
\label{sec-gluon-hq}
 
There are two particular processes of interest in heavy quark production.  
The first is the production of  
vector mesons containing heavy quarks, in particular J/$\psi$ production.  
In some models the elastic process proceeds via a two gluon exchange mechanism  
very  
similar to that described above in deep inelastic $\rho$  
production~\cite{Ryskin}.  
Inelastic J/$\psi$ production can result from photon-gluon fusion  
followed by the emission of a second gluon from one of the heavy quark  
lines. Although produced dominantly at $Q^2 \sim 0$, deep inelastic \Jpsi\  
production has also been observed. The second class of processes consists  
of inclusive production of charmed particles, typically observed either via  
detecting $D^*$ mesons or high-\pt\ leptons and also seen in  
both photoproduction and deep inelastic regimes. Here the dominant  
production mechanism is photon-gluon fusion. In the case of  
photoproduction, the hard scale provided by the charm quark mass  
allows QCD to be used to interpret the measured cross-section in terms of a 
gluon density.  
\subsubsection{Elastic J/$\psi$ photoproduction} 
\label{sec-gluon-hq-jpsi-el-photo} 
One particularly clean channel is elastic \Jpsi\   
production. Pairs of oppositely charged leptons, either with well  
reconstructed charged tracks matching energy deposits in the 
calorimeter consistent with \eplus\ or \emin, or with charged tracks  
pointing to hits in the muon chambers consistent with $\mu^+$  or 
$\mu^-$, are selected from events with no other activity in the   
detectors. The effective mass of the leptons shows a clear peak at the  
\Jpsi\ mass with very little background. 
 
The total $\gamma^*p$ 
cross-section obtained from these signals can be 
plotted as a function of $W$. ZEUS data from 1993 have been 
published~\cite{ZEUS-Jpsi-photo}
which, in conjunction with data from fixed target experiments, showed 
that the \Jpsi\ cross-section was rising quickly with $W$, much faster than 
would be predicted from models based on vector meson dominance or 
Regge theory. These data, together with results from H1 based on the 
1994 running, confirm this from HERA data alone. \Fref{fig-Jpsi-sigma} 
shows the H1 and ZEUS $\gamma^*p$ cross-sections plotted as a  
function 
of $W$ together with fixed target data from Fermilab. The HERA data lie 
substantially above the prediction from the ``soft" Donnachie-Landshoff 
model~\cite{Jpsi-dl}. A fit to all the data gives a slope with 
$W$ of $0.90 \pm 0.06$, to be compared with the Donnachie-Landshoff  
prediction of 0.32 for $t = 0$. A similar conclusion is drawn from the  
analysis of the ZEUS 1994 data~\cite{ZEUS-Jpsi-photo-1997}.  
\Fref{fig-Jpsi-sigma} shows the results of calculations using a
two gluon exchange model, the  
Ryskin model~\cite{Ryskin}, 
with two different forms for the gluon. There is clearly some discrimination  
between different gluon distributions, with the steeper gluon from  
GRV being disfavoured.  
\ffig{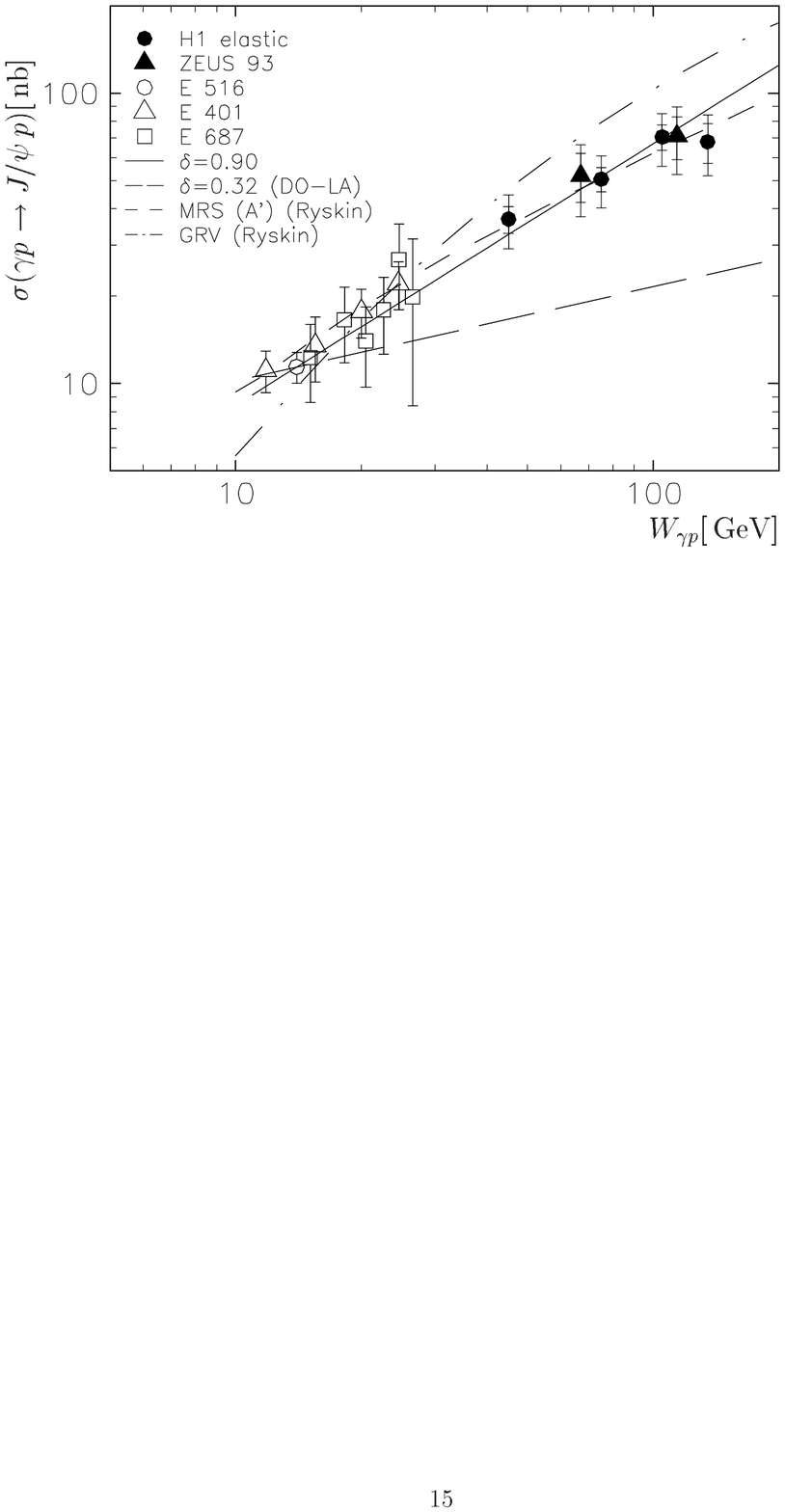} 
{100mm}{The \Jpsi\ cross-section as a function of $W$.  
The inner error bars on the HERA data points are statistical, the outer error 
bars show the systematic and statistical errors added in quadrature. 
The fixed  
target data points~\protect\cite{Jpsi-E516,Jpsi-E401,Jpsi-E687} have been  
where necessary scaled 
with the most recent \Jpsi\ branching ratio; the error bars
correspond to both statistical 
and systematic errors added in quadrature. The full and long-dashed curves   
obey $W^\delta$ 
with $\delta = 0.90$ and 0.32 respectively. The other curves show  
calculations from the Ryskin 
model~\protect\cite{Ryskin} with differing parameterisations of the  
gluon.} 
{fig-Jpsi-sigma} 
\subsubsection{Inelastic J/$\psi$ photoproduction} 
\label{sec-gluon-hq-jpsi-inel-photo} 
Inelastic \Jpsi\ production is thought to proceed by photon-gluon fusion 
followed by the emission of either a hard gluon or several soft gluons. 
The former is a colour-singlet process whereas the latter is a colour-octet. 
An inelastic \Jpsi\ sample can be obtained by a similar selection on 
lepton pairs as in the elastic case but requiring additional tracks to  
emanate from the interaction point. In practice the greater activity present 
in inelastic events makes electron identification more difficult so that 
muons pairs are normally used. Both H1~\cite{H1-Jpsi-photo} and  
ZEUS~\cite{ZEUS-Jpsi-photo-inelastic} have presented clear inelastic signals 
with small backgrounds.  
 
The inelastic sample can be clearly distinguished from the elastic sample by  
cutting on the elasticity variable $z$ as defined by  
\begin{equation} 
z=\frac{y_{\psi}}{y} \qquad \mbox{with} \qquad 
                      y_{\psi}= \frac{(E-p_{z})_{J/\psi}}{2\,E_{e}}. 
\label{eq-Jpsi-zdef} 
\end{equation} 
For elastic events $y_{\psi}=y$ and thus $z=1$.  
The inelastic sample is defined between $0.45 \leq z \leq 0.90$. 
The shape of the 
cross-section variation with $W$ is relatively insensitive to different 
parton distributions. However, the absolute normalisation
of the cross-section does show a significant dependence on
parton distribution sets, and the variation of the cross-section as a function 
of $z$ is sensitive to the importance of colour singlet or octet production 
mechanisms. 

\Fref{fig-ZEUS-Jpsi-inel-fig4} shows the inelastic cross-section as a
function of $W$. In addition to the ZEUS and H1 points, fixed target
data from the FTPS~\cite{Jpsi-E516}, NA14~\cite{Jpsi-NA14}, and 
EMC~\cite{Jpsi-EMC} collaborations is shown. There
is clear discrimination between models, depending on the assumed
values of $m_c$ and $\Lambda_{QCD}$, which implies that future
data may well be able to give information on the gluon distribution
in the proton. 
\ffig{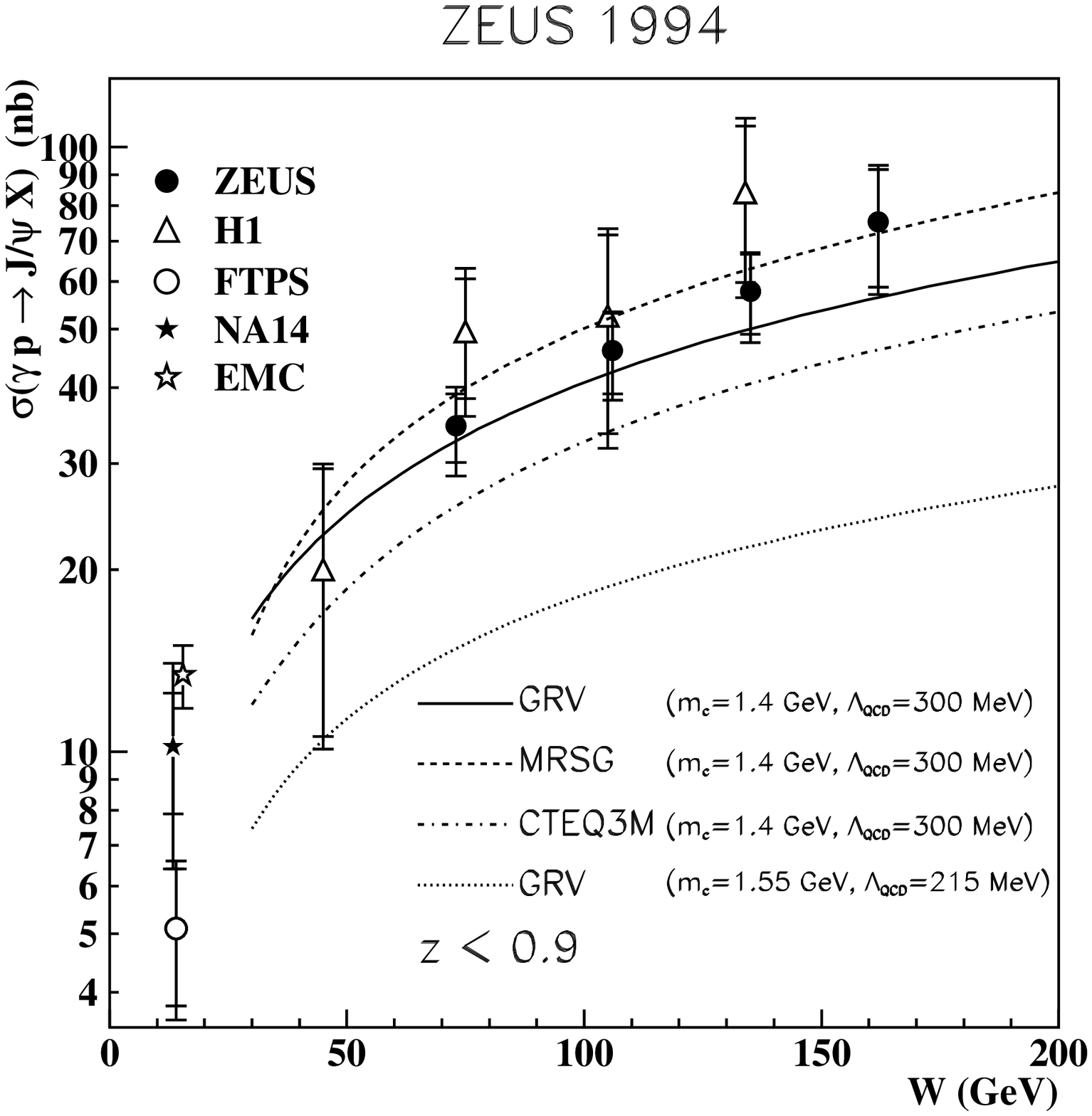} 
{70mm}{The direct inelastic \Jpsi\ photoproduction cross section as a 
function of $W$ for $z<0.9$. 
Data from ZEUS, H1, 
FTPS~\protect\cite{Jpsi-E516}, NA14~\protect\cite{Jpsi-NA14} and 
EMC~\protect\cite{Jpsi-EMC} 
are shown. The ZEUS result at the lowest $W$ value is obtained with the 
muon channel only, whereas
the other ZEUS measurements come from the combination of the electron 
and muon results. 
Except for the latter three points, the 
inner error bars indicate the statistical 
uncertainty, the outer error 
bars the quadratic sum of the statistical and systematic
uncertainties.  For the high $W$ ZEUS points, the inner error bars represent 
the statistical and decay 
channel specific errors added in quadrature, the outer ones the statistical, 
decay specific and common systematic errors added in quadrature. 
The lines correspond to the NLO prediction from
\protect\cite{Jpsi-Kramer} assuming the GRV \protect\cite{GRV} (continuous), 
MRSG \protect\cite{MRS-pin-glue} (dashed) and CTEQ3M \protect\cite{CTEQ} 
(dotted-dashed) 
gluon distributions with $m_c=1.4$ GeV and $\Lambda_{QCD}=300$ MeV, 
the dotted curve was obtained with GRV, $m_c=1.55$ GeV and 
$\Lambda_{QCD}=215$ MeV . The curves are scaled up by a factor 
of 1.15 to take into 
account the contribution from $\psi' \rightarrow \Jpsi X$.}
{fig-ZEUS-Jpsi-inel-fig4} 

\Fref{fig-ZEUS-Jpsi-inel-dsdz} shows the cross-section 
as a function of $z$ as measured by ZEUS and H1. 
The solid line shows the expectation 
from a colour singlet exchange model 
and agrees well with the data. The colour octet  
model is strongly peaked towards $z=1$ and thus the octet contribution  
allowed by this data is very small. The dashed line shows the 
expectation~\cite{Jpsi-Cacciari} if  
the octet normalisation were fixed to explain the excess production seen in  
the CDF data~\cite{Jpsi-CDF}. It is clearly ruled out by the HERA data. 
\ffig{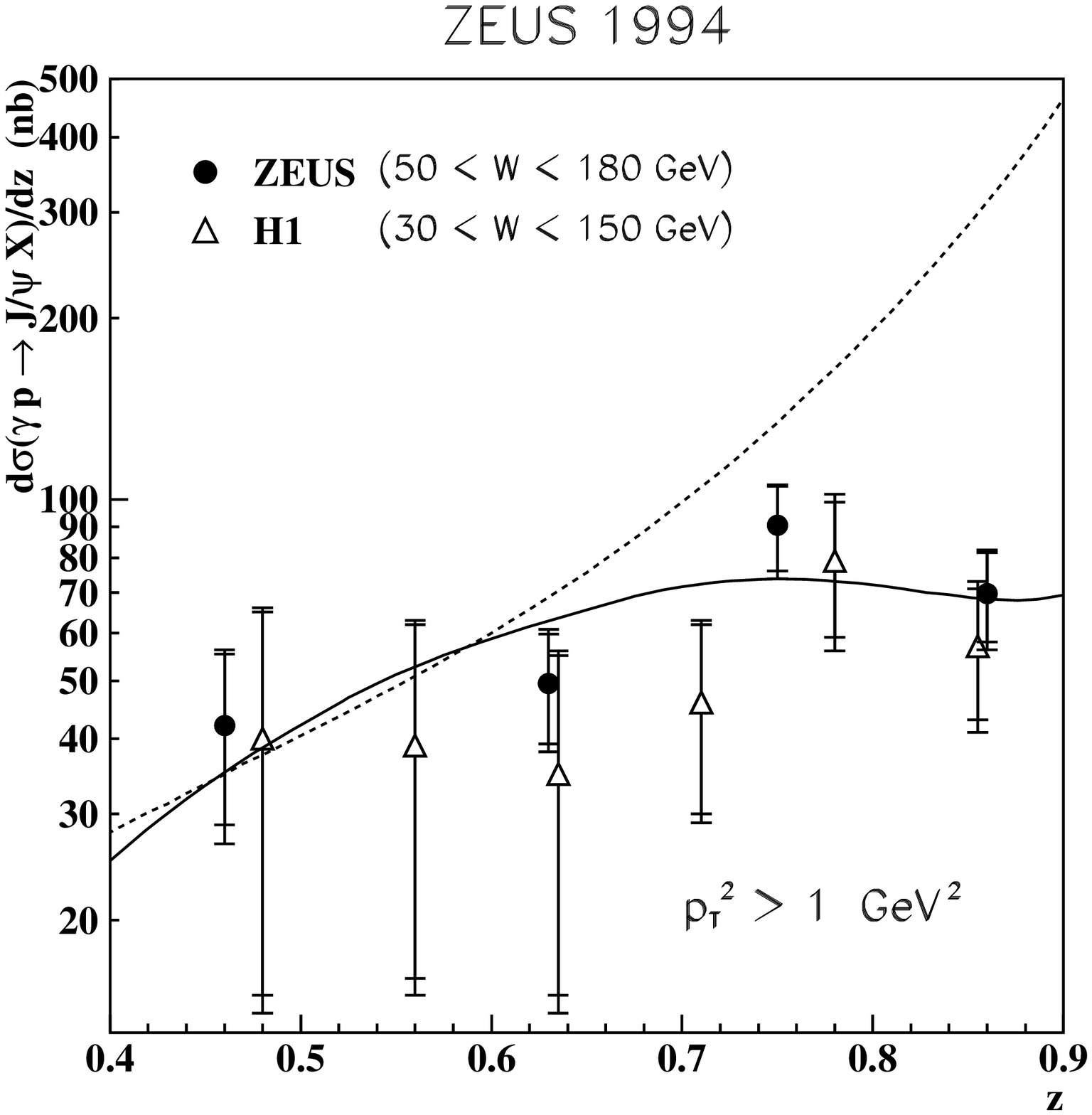}
{70mm}{Differential cross section $d\sigma/dz$ for the inelastic 
J/$\psi \rightarrow \mu^+ \mu^-$ sample with $50 < W < 180$~GeV 
and \pt$^2 > 1$ GeV$^2$. 
Data from ZEUS and H1 are shown.
The inner error bars indicate the statistical uncertainties, the outer error 
bars the quadratic sum of the statistical and systematic uncertainties.
The NLO computation \protect\cite{Jpsi-Kramer} with the 
GRV \protect \cite{GRV} 
structure function, $m_c=1.4$ GeV and $\Lambda_{QCD}=300$ MeV   
is shown as a solid line. The dashed line is given by the 
sum of the colour-singlet and the colour-octet leading order 
calculations \protect\cite{Jpsi-Cacciari}. In the theoretical curves 
the 15\% contribution of the $\psi'$ has not been included.}
{fig-ZEUS-Jpsi-inel-dsdz} 
\subsubsection{Elastic \Jpsi\ production in Deep Inelastic Scattering} 
\label{sec-gluon-hq-jpsi-el-DIS} 
 
Elastic \Jpsi\ production has been observed at both  
H1~\cite{H1-DIS-rho} 
and ZEUS~\cite{ZEUS-Jpsi-photo} at \Qtwo\ up to 40 GeV$^2$. The  
selection 
criteria are similar to those for the elastic \Jpsi\ in photoproduction 
discussed in section~\ref{sec-gluon-hq-jpsi-el-photo}, with the additional 
requirement of a well-identified scattered positron.  
 
In contrast to the behaviour observed in $\rho$ production, the 
\Jpsi\ cross-section seems to show a steep rise as a function of $W$ for  
all values of \Qtwo\, including \Qtwo\ = 0. This implies that the hard 
scale provided by the charm quark mass is already sufficient to allow 
the QCD ``hard pomeron"-like behaviour to be observed in photoproduction.  
\Fref{fig-H1-rhoJpsi-sigma} shows the value of the total virtual  
photon-proton cross-section for $\rho$ and \Jpsi\ as a function of $W$ for 
different \Qtwo\ bins. Despite the fact that at \Qtwo\ = 0
and low $W$, \Jpsi\ production  
is suppressed by a factor of almost 1000, at $Q^2 \sim 20$  
GeV$^2$ 
the \Jpsi\ and $\rho$ cross-sections are almost equal. In fact, H1  
obtain~\cite{H1-DIS-rho} 
$\sigma(\Jpsi)/ \sigma(\rho) = 0.64 \pm 0.13$ for \Qtwo\ = 10 GeV$^2$ 
and $1.3 \pm 0.5$ for \Qtwo\ = 20 GeV$^2$. This implies that, as 
expected in both soft and hard models of vector meson production,  
approximate 
$SU(4)$ symmetry is restored for $Q^2$ $ \gg m_{\tiny {V}}^2$. The  
fact that 
the \Jpsi\ : $\rho$ ratio is so high at  
\Qtwo\ as low as  $\sim 10$ GeV$^2$ is perhaps 
surprising and is not expected in  
some QCD models which take into account the effects of quark Fermi motion  
within the light-cone wave function of the
\Jpsi~\cite{Jpsi-DIS-Frankfurt}.    

\subsubsection{Production of open charm} 
\label{sec-gluon-hq-charm} 
Finally in this section we discuss the use of heavy quark resonances other 
than vector mesons to determine the gluon distribution.  
The dominant mechanism for production of charm-anticharm quark pairs 
at HERA is photon-gluon fusion. In deep inelastic scattering, a gluon from  
the proton can split into 
a $c\overline{c}$ pair which interacts directly with the 
virtual photon. This process is also important in photoproduction, as is  
the additional process in which a constituent of the photon interacts 
with the heavy quark or antiquark in a so-called 
``resolved photon'' interaction. In either case the cross-section for 
charm production depends on the probability of finding a gluon in the 
proton and therefore measurement 
of these cross-sections in principle allows an unfolding of $G(x)$. 
 
Two distinct methods which isolate charm production have been used. 
The first uses the kinematic characteristics of heavy quark 
decay to enhance the charm signal. A particularly useful 
method~\cite{SLAC-D*} uses the 
decay $D^* \rightarrow D \pi_{slow}$, where the 
slow pion, $\pi_{slow}$, is restricted to a very small area of phase space just above 
threshold. Several different $D$ decay modes have been used by ZEUS and 
H1. For the mode $D^0 \rightarrow K \pi$ a clear signal is produced by
a cut on the mass difference between 
the $K\pi \pi$ and $K \pi$ combinations. Particle  
identification, for example by energy loss in drift chamber gas, has been 
used to help in the identification of charmed particles.  
The hard fragmentation of  
heavy quarks also permits the enhancement of the charm signal. The  
momentum fraction, $x_D$, of the $D$ candidate compared to the  
incident proton is evaluated in the $\gamma p$ centre-of-mass system. 
True $D$ candidates tend to have large $x_D$, a fact which also implies 
that the leading particles in jets will preferentially come from $D$ mesons. 
By utilising these effects a clear charm signal can be obtained by suitable  
selections on both leading particles and $x_D$.  
 
The second method to enhance charm uses high \pt\ 
leptons, usually muons, as a 
signal for inclusive charm production. Since bottom production is 
negligible at current HERA luminosities, a cleanly identified high \pt\  
lepton is an excellent signature for charm production.  
\\ 
\\ 
{\bf \thesubsubsection.1 Photoproduction of charm} 
\label{sec-gluon-charm-photo} 
\\ 
\\ 
\Fref{fig-sigmac} shows the cross-sections obtained by 
ZEUS~\cite{ZEUS-photo-D*} and H1~\cite{H1-photo-D*}.
The signals were obtained using the methods  
discussed in the previous section, and are plotted as a function of  $W$, 
together with low energy data from fixed target  
experiments~\cite{lowW-charm}.  
The solid curve shows the predictions of a next to 
leading order QCD model by Frixione {\it et al.}~\cite{Frixione}, using as input 
the proton structure function parameterisation of 
$MRSD-\,'$~\cite{MRS}  
and the 
photon structure prediction due to Gl\"{u}ck, Reya and 
Vogt~\cite{GRV-photon}. The dotted lines on either side of the solid 
line show the variation in the cross-section obtained using 
alternative parameterisations which are broadly compatible with the 
data on photon 
and proton structure functions. The size of these variations and the 
fact that they depend  on the photon structure as well as the 
proton structure imply that much more data will be necessary before 
this method can lead to an accurate determination of the gluon density 
in the proton.\\ 
\ffigp{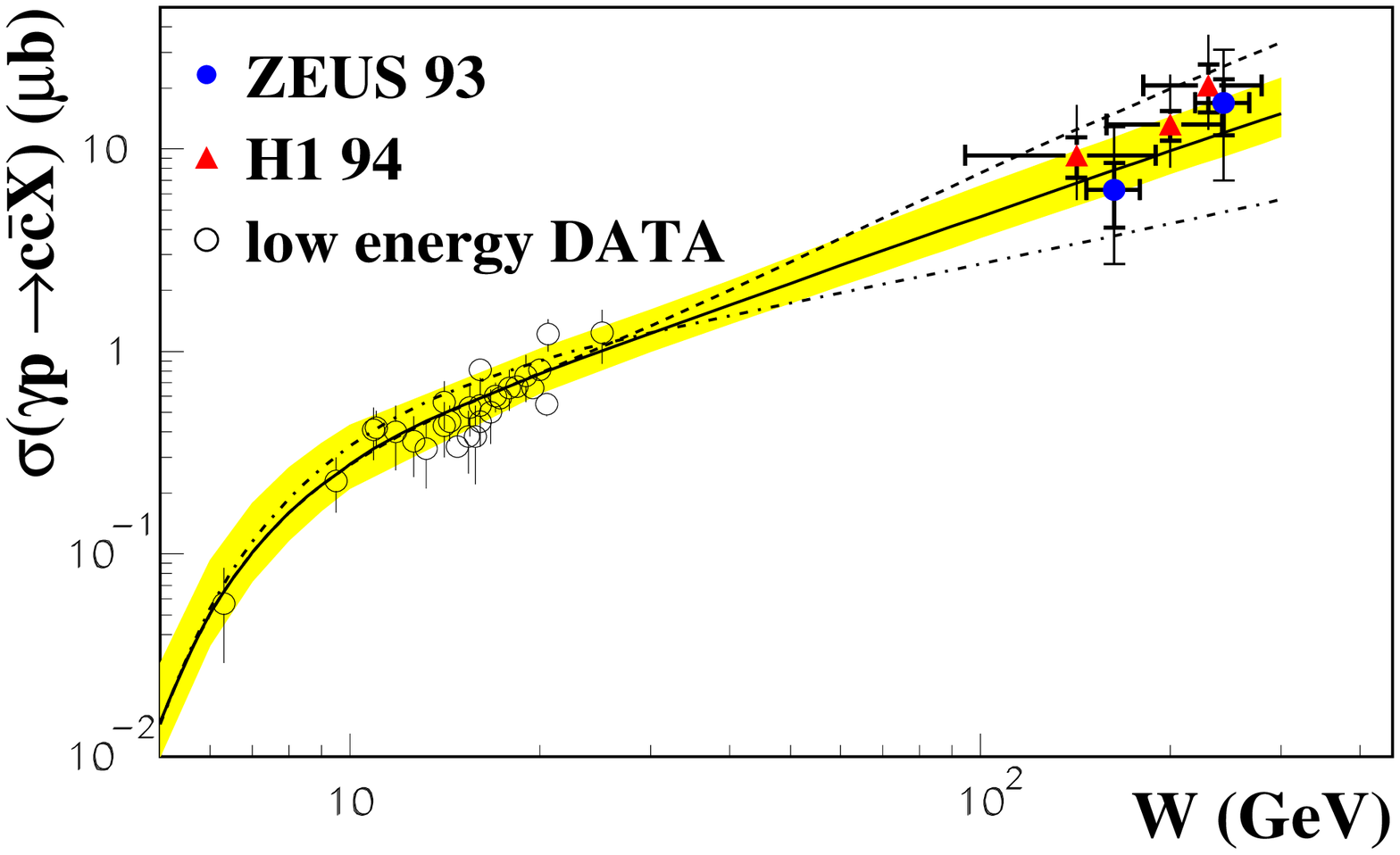} 
{70mm}{100mm}{The total cross-section for charm photoproduction as a  
function of $W$ derived 
from the observation of inclusive $D^*$ mesons. The solid circles  
represent the ZEUS results, while the triangles are data from H1.  
The open circles show data from fixed-target 
experiments~\protect\cite{lowW-charm}. The solid line shows the  
predictions of 
the model by Frixione {\it et al.}, based on the $MRSD-'$ and  
GRV(HO)~\protect\cite{GRV-photon} structure  
functions for the proton and photon respectively. The shaded area 
shows the uncertainty due to renormalisation scale. The higher dotted 
line shows similar predictions based on  
$MRSD-'$ and LAC1~\protect\cite{LAC}, while the lower dotted line  
is based on  
$MRSD_0$ and GRV(HO).} 
{fig-sigmac}  
\subsection{Summary} 
\label{sec-gluon-summary} 
The above data demonstrate that a consistent picture is emerging for 
the gluon density using a variety of different methods, including the  
``classic'' method of scaling violations in DIS. Now that next-to-leading  
order corrections are available, the measurement of $G(x)$ from jet rates  
should become more important. At the moment however, 
the other methods are really only consistency checks. A better 
understanding of the theoretical uncertainties but in particular more  
precise data are required before a measurement of $G(x)$ can be obtained 
from inclusive charm and vector meson production.  
\section{Determination of $F_2^{charm}$} 
\label{sec-F2c} 
Our discussion of the quark and gluon content of the proton  
concludes with the determination of the 
fraction of $F_2$ arising from the production of charm. This is 
not only a measure of part of the quark content of the proton, but, 
since it is dominated by boson-gluon fusion, it is very sensitive 
to the gluon density. Thus this topic unites the previous 
two sections.  
  
Both the H1 and ZEUS  
experiments have observed signals for the  
production of charmed mesons in deep inelastic scattering events.  
The experiments select DIS events as described in 
section~\ref{sec-quarks}. Both the ZEUS~\cite{ZEUS-F2charm} and
H1~\cite{H1-F2charm} collaborations have published data
using rather similar methods; 
we concentrate on a discussion of the H1 analysis in this section.  
 
Events containing $D$ mesons are selected via  
methods described in the previous section.  
The events are restricted by the requirement for high efficiency for 
the trigger and the reduction of photoproduction background to 
the kinematic region 
$y < 0.53$ and 10 GeV$^2 < Q^2 < 100$ GeV$^2$.  
H1 isolate signals  
from both $D$ and $D^*$ production, in the  
former into $K\pi$ and in the latter into 
$K\pi\pi_{slow}$. The 
$D^0$ events are isolated by employing a selection on $x_D$, whereas 
$D^*$ events are obtained by use of  
the kinematic properties of the slow pion. Clear signals are obtained 
for both channels. The two samples lead to differential 
cross-sections and other kinematic quantities which are in good agreement, 
justifying taking an average of both channels. This leads to a total 
charm cross-section of 
\beqn 
\sigma(ep \rightarrow c\overline{c}X) = (17.4 \pm 1.6 \pm 1.7 \pm 1.4)
\; {\rm nb}. 
\eeqn 
where the errors are statistical, systematic, and dependent on the 
Monte Carlo model used to extrapolate to the full phase-space,  
respectively.   
This cross-section is somewhat higher than that expected in 
theoretical predictions based on NLO  
calculations~\cite{Laenen,Riemersma}. 
 
The normalised $x_D$ distributions are able to discriminate between 
possible charm production mechanisms. The $x_D$ distribution is 
sensitive to both the fragmentation function of the charm quark 
and to the charm production spectrum. Assuming a universal 
fragmentation as determined dominantly by the LEP experiments, then 
the $x_D$ distribution strongly supports the boson-gluon 
fusion production mechanism for charm. 
 
The H1 data is binned into 9 bins in \Qtwo\ and $x$ and the cross-section 
determined per bin, corrected for QED radiation, is converted into 
$F_2$ using \eref{eq-F2} under the assumption that \Fl\ is zero. 
The result, together with the ZEUS data, which extends the data to
somewhat lower \Qtwo, is  
shown in \fref{fig-ZEUS-F2charm}.    
At the current level of accuracy, the data are in
reasonable agreement with the predictions from the GRV
parton distributions.   
Low energy data from EMC~\cite{EMC-charm} suggested a strong 
increase of the ratio $F_2^{charm}/F_2$ as a function of \Qtwo. Such 
an effect is not seen in the ZEUS and H1 data, which is compatible with no 
increase with \Qtwo. The average fraction which $F_2^{charm}$ 
contributes to $F_2$ is determined by H1 to be  
\beqn 
0.237 \pm 0.021 ^{+0.043}_{-0.039} \nonumber 
\eeqn 
a growth of about an order of magnitude compared to the EMC 
result, in agreement with the strong rise in the gluon 
density at small $x$. The ZEUS analysis leads to a similar result. 
Thus charm quark production forms a sizeable 
fraction of all the leading quarks produced in DIS, particularly 
at low $x$.   
\ffigp{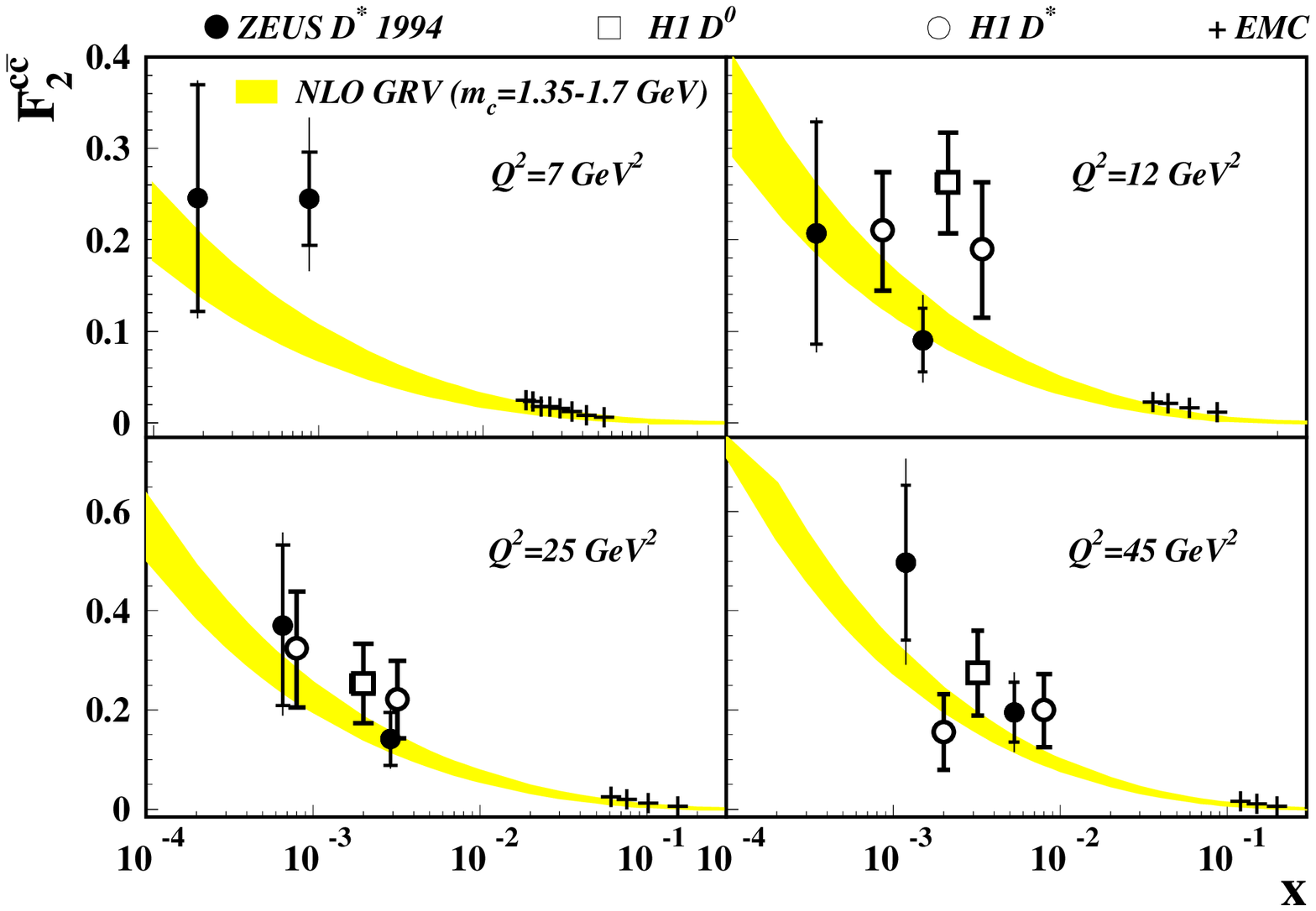} 
{80mm}{110mm}{$F_2^{charm}$ as measured from $D^0$ events by H1 (open squares)
 and 
$D^\star$ events by H1 (open circles) and ZEUS (closed circles).  
For the ZEUS data, the inner error bars  
on the data points are the statistical errors, the full bars 
correspond to statistical and systematic errors added in quadrature.  
The error bars on the H1 points correspond to the statistical
and systematic errors added in quadrature. 
The shaded band shows the result of the NLO QCD prediction
based on the GRV-HO parton distributions; the width of the
band corresponds to varying $m_c$ between 1.3 and 1.7  GeV.  
Data from 
EMC~\protect\cite{EMC-charm} are also shown as crosses.}  
{fig-ZEUS-F2charm}  
\section{Properties of the final state in Deep Inelastic Scattering} 
\label{sec-finalstate} 
The large increase in centre of mass energy at HERA greatly improves 
the sensitivity of studies in perturbative and non-perturbative 
aspects of QCD compared to fixed target experiments. Studies 
of inclusive distributions have been made in both the $\gamma^* p$ 
centre of mass  
system and in the Breit frame, and the general features compared with \ee\  
annihilation. In addition both H1 and ZEUS have studied details of  
fragmentation by the observation of inclusive strange neutral particle  
production. Finally, studies on perturbative aspects of QCD have been carried  
out by studying energy flows and inclusive jet production.  
\subsection{Inclusive charged particle distributions} 
\label{sec-inclusive-charged} 
\subsubsection{Distributions in the $\gamma^* p$ CM system} 
Inclusive particle distributions  in neutral current deep  
inelastic scattering are sensitive both to the 
hadronisation process and to the underlying QCD processes at the 
parton level. We first discuss charged particle distributions in 
variables aligned with the exchanged virtual photon; the 
fractional momentum along the virtual photon direction known as the 
Feynman variable, \xf $=2 p^*_l/W$, where $p^*_l$ is the track  
momentum along the virtual photon direction, and the momentum  
perpendicular to the 
virtual photon direction, \pts . In H1, charged tracks reconstructed in the  
central jet  
chamber were used in the analysis provided that they  
were of good quality and associated with the primary event vertex. In 
order to keep within the well-understood and high-acceptance region of 
the detector charged tracks were required to satisfy $0.15 <$\pt $~< 10$  
GeV and  22\degree $ < \theta <$ 158\degree. Charged tracks in ZEUS are  
reconstructed principally using the Central Tracking Detector. 
Only those charged particles with \pt $~> 0.2$ GeV and  
25\degree $ < \theta <$ 155\degree, where $\theta$ is the polar angle, 
are analysed. This restricts the range in \Qtwo\ and $W$ to  
$10 < \Qtwo < 160$ GeV$^2$ and $75 < W < 175$ GeV. 
 
\Fref{fig-xfpt-fig3} shows the differential charged hadron multiplicities for  
deep inelastic scattering events, excluding those with a large rapidity gap 
in the forward direction indicative of a diffractive type 
interaction~(see section~\ref{sec-diffraction}). The distributions from both  
H1~\cite{H1-xfpt} and ZEUS~\cite{ZEUS-xfpt} are shown as a function of  
\xf, while the ZEUS data is shown versus $p_{T}^*$. Also shown 
is the H1 and ZEUS $<p_{T}^{* 2}>$ as a function of \xf, the so-called 
``seagull'' plot. The H1 and ZEUS data are in good agreement, and show  
the typical features associated with hard gluon radiation. This is  
particularly visible in the seagull plot, where much greater \pt\ is  
developed than expected from the naive quark parton model.  
\ffig{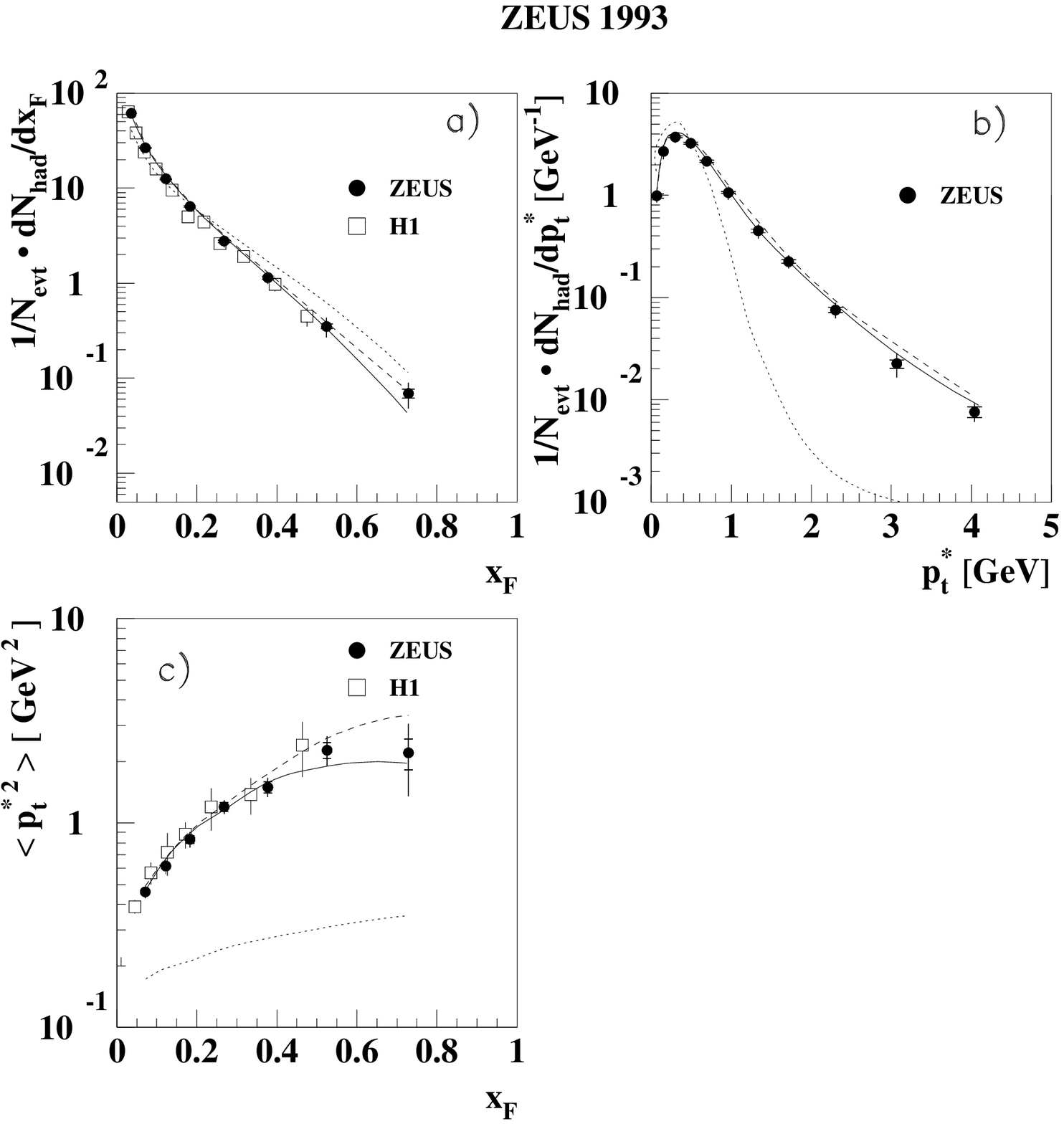} 
{140mm}{Differential normalised charged hadron multiplicities for  
DIS events as a function of a) \xf\ b) \pt$^*$ for \xf\ $> 0.05$ c)  
$<$\pt$^{*~2} >$ as a function of \xf. The solid line shows the result  
of the MEPS~\protect\cite{MEPS} Monte Carlo model, whereas the  
dashed curve is the result of  
CDMBGF~\protect\cite{CDMBGF}. The dotted curve shows the result of a  
phase-space model which  
does not include hard gluon radiation and therefore corresponds to the  
expectation from the naive quark-parton model. } 
{fig-xfpt-fig3}  
On the  
assumption that fragmentation is a long-distance phenomenon and does  
not depend on the details of the perturbative origin of the fragmenting  
quarks, it would be expected that the general features of DIS events would  
be similar to those from \ee, provided that one half of the multiplicity  
observed in \ee\ is compared with the DIS multiplicity (since the current  
jet corresponds to the fragmentation of one quark or antiquark, whereas in 
\ee\ both 
quark and antiquark fragment). This picture does indeed seem to be borne  
out by the HERA data. \Fref{fig-xfpt-ee} shows the ZEUS data in comparison  
to data from the DELPHI experiment~\cite{DELPHI-xf} at a similar mean  
$W$.  It can be seen 
that the ZEUS and \ee\ data are in good agreement with each other and 
with the solid line which is the prediction of the MEPS model. In contrast,  
the lower energy data from fixed target $\mu p$ scattering are consistent  
with the quark-parton model prediction and do not agree with the ZEUS  
data. Similar results have been obtained by H1~\cite{H1-nch}, who
also observe KNO scaling~\cite{KNO} with similar characteristics to
that observed in \ee\ annihilation, and discuss detailed
comparisons with Monte Carlo models and hadron-hadron scattering.  
 
\ffig{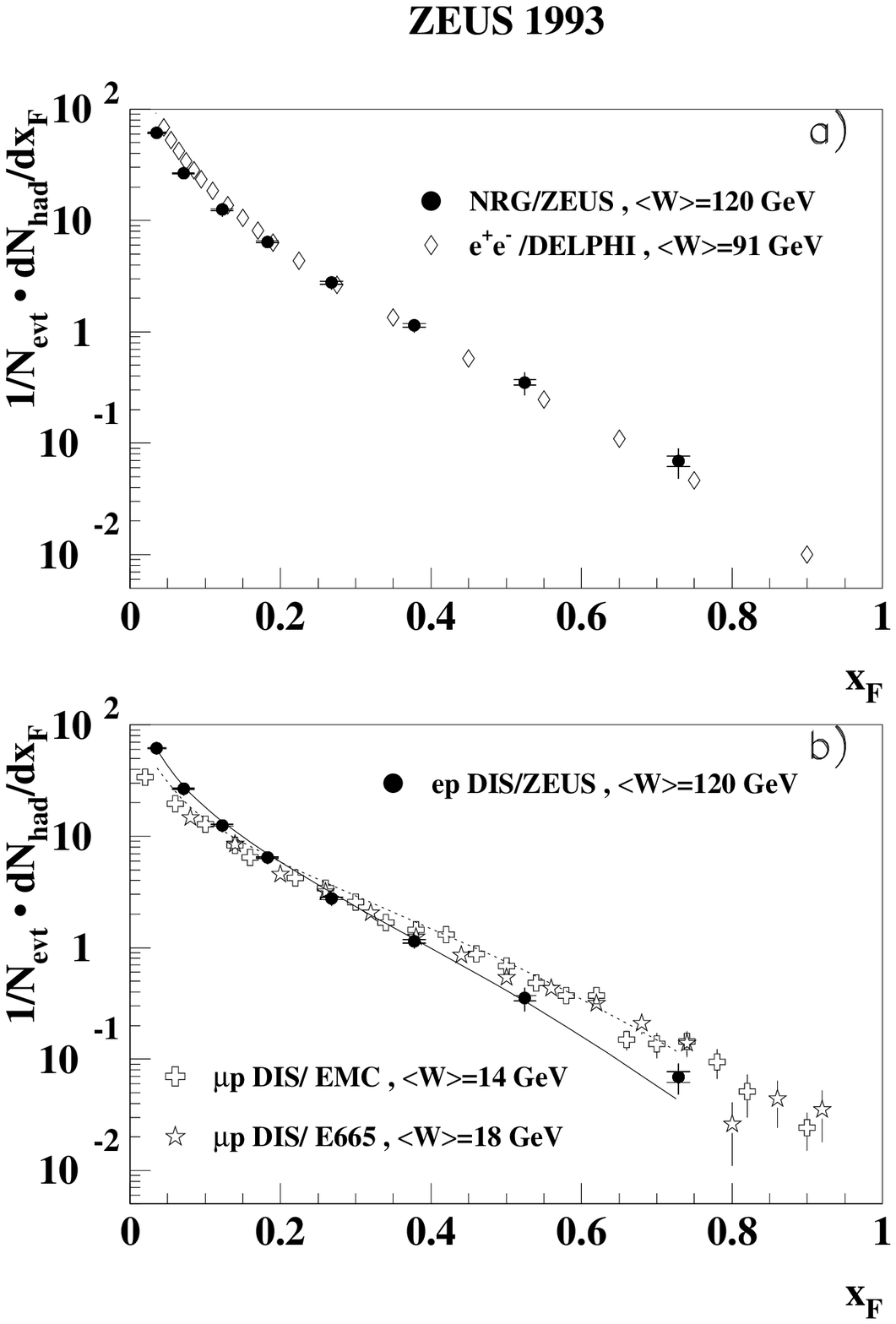} 
{80mm}{a)  Normalised \xf\ distributions for charged particles for 
the ZEUS  deep inelastic scattering data together with the 
distribution for \ee\ data from the DELPHI  
experiment~\protect\cite{DELPHI-xf} 
at similar  $<W>$. b) ZEUS data in comparison with fixed target data at  
much lower $<W>$ from EMC~\protect\cite{EMC-xf} and  
E665~\protect\cite{E665-xf}. The solid line is the  
prediction of the 
MEPS Monte Carlo model~\protect\cite{MEPS}. 
The dotted line shows the prediction from a 
phase space model excluding the effects of hard gluon radiation.} 
{fig-xfpt-ee}  
\Fref{fig-ptsq-W} shows $<\ptsq>$ as a function 
of $W$ for different ranges of \xf . The ZEUS data agree 
well with the predictions of the MEPS model, 
shown as a solid line, and show a much steeper rise in  $<\ptsq>$ than 
the fixed target data for a similar range in $W$. \Fref{fig-ptsq-Q}  
shows that $<\ptsq>$ for the ZEUS data increases strongly with \Qtwo, in  
contrast with the EMC data which show almost no rise over a similar 
range of \Qtwo . Thus the expected rise in $<\ptsq>$ 
due to hard gluon radiation can be more clearly seen in the HERA data  
than in the EMC data, where the phase space for gluon radiation 
is severely restricted. 
\ffig{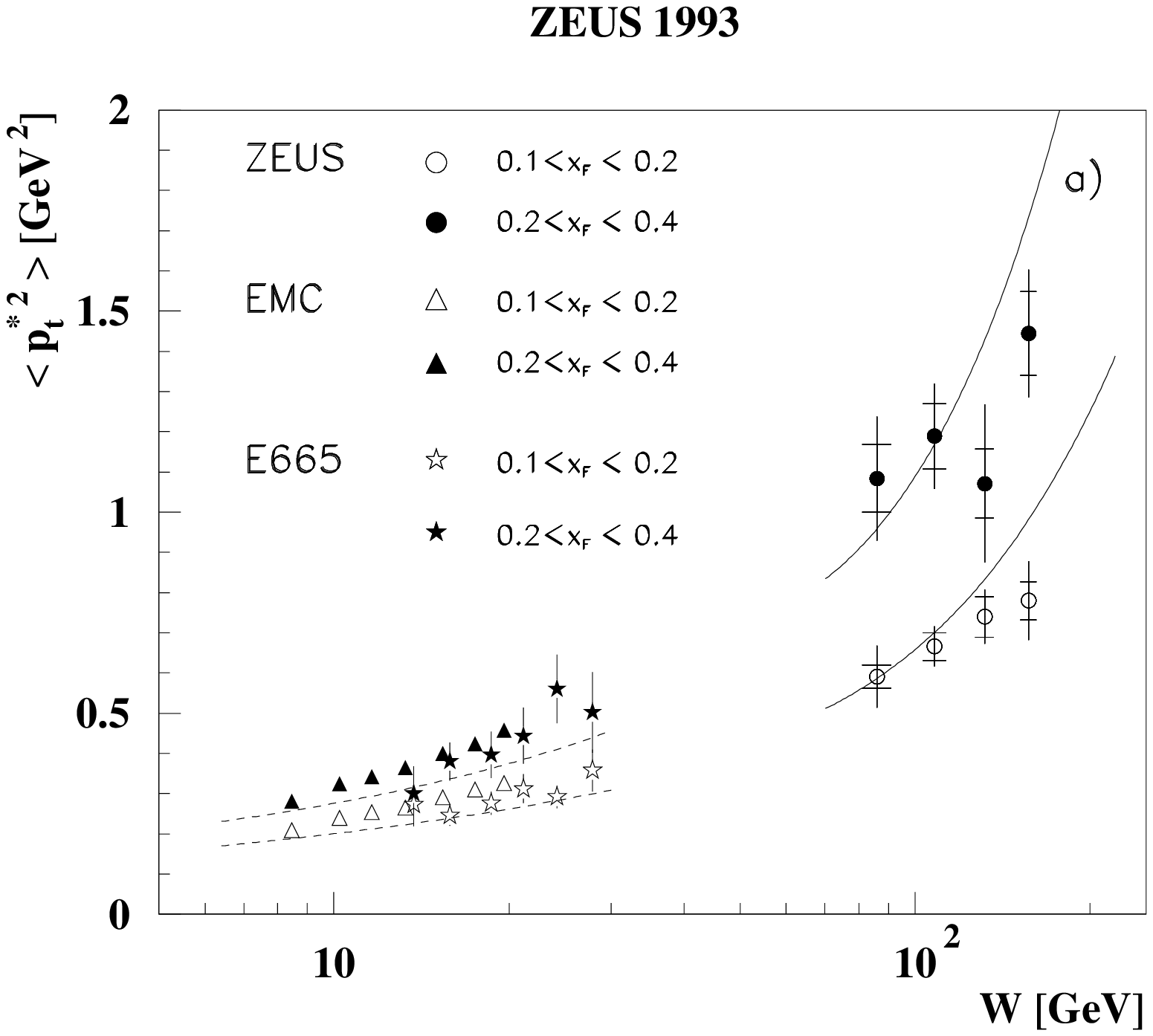} 
{70mm}{$<\ptsq>$ distributions in two ranges of \xf\ for charged  
particles for the ZEUS  deep inelastic scattering data together with 
the  
EMC~\protect \cite{EMC-ptsq} and E665~\protect \cite{E665-ptsq} 
distributions at lower $ <W> $. The solid line shows the prediction  
from the MEPS Monte Carlo model, the dotted line shows the prediction  
from a phase space model excluding the effects of hard gluon 
radiation.} 
{fig-ptsq-W} 
\ffigp{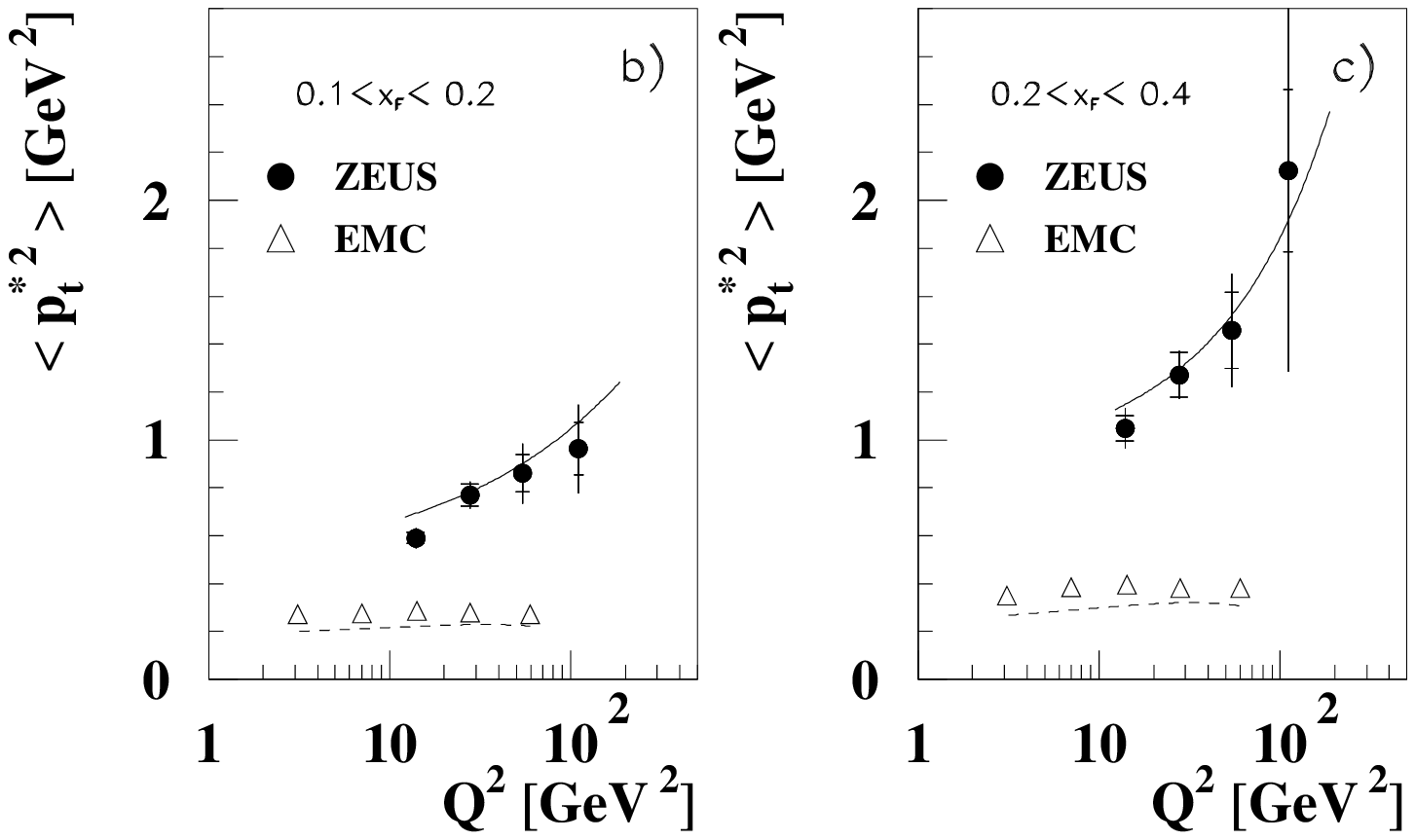} 
{70mm}{100mm}{$<\ptsq>$ distribution as a function of \Qtwo\ for the ZEUS 
and EMC experiments for two different ranges in \xf .  
The solid line shows the prediction from the MEPS Monte Carlo model, 
the dotted line shows the prediction from a phase space model 
excluding the effects of hard gluon radiation. } 
{fig-ptsq-Q}  

H1~\cite{H1-pt97} have also used the transverse momentum spectra
of charged particles as a function of \Qtwo\ and $x$ to study
various models of fragmentation and soft gluon 
radiation. For $x > 0.001$, all models give a good description of
the data. For smaller $x$, however, and in the central rapidity
region, the \pt\ distributions are significantly higher than the 
MEPS (LEPTO 6.4) and HERWIG 5.8~\cite{HERWIG} predictions, but in
good agreement with the ARIADNE 4.08 predictions. Such an excess of
high \pt\ in the central region would be expected in models in which
BFKL behaviour (see section~\ref{sec-eflow}) was important.
\subsubsection{Distributions in the Breit Frame} 
In $ep$ collisions, in contrast to the situation in \ee, there is no ``natural" 
frame in which to study fragmentation effects. In the  
$\gamma^* p$ centre of mass  system discussed in the previous section, 
the struck quark is back-to-back with the proton remnant in the naive  
quark-parton model. In the Breit frame, the struck quark is back-to-back 
with the exchanged virtual photon, so that in the naive quark-parton  
model the momentum of the quark is $Q/2$ before the interaction 
and $-Q/2$ after it, where \Qtwo\ is the virtuality of the exchanged photon.  
Since in this frame the quark scattering can be viewed  
as a $t$-channel process closely 
related to the $s$-channel \ee\ $\rightarrow q \overline{q}$, it is a  
particularly suitable frame in which to compare with \ee\ results.  
The direction of the outgoing struck quark defines a hemisphere to which 
particles may be assigned and considered as part of the current  
fragmentation region. However, there is ambiguity in this assignment  
which complicates the comparison with \ee\, particularly at low
$Q$.   

Both ZEUS~\cite{ZEUS-Breit} and H1~\cite{H1-Breit} have published  
results on charged particle distributions in the Breit frame. The variable 
used is normally $x_p$, where $x_p =  2p/\sqrt{s}$, together with the  
related variable $\xi = \ln{(1/x_p)}$. QCD in the modified leading log  
approximation predicts that at the parton level $\xi$ is approximately 
Gaussian and that the maximum of $\xi$ increases as $Q$.  
\Fref{fig-H1-Breit} shows the evolution in the
H1 data~\cite{H1-Breit-97} of both the maximum and the  
width of the $\xi$ distribution as a function of $Q$, together with  
predictions from a modified leading logarithm approximation (MLLA) 
QCD model~\cite{MLLA-model}. There is good  
agreement between the H1 and ZEUS~\cite{ZEUS-Breit}
data and between the DIS and \ee\  
data~\cite{Breit-ee} (note that the \ee\ data has been refitted to
make it directly comparable to the H1 results). 
Although the HERA data agree quite well with the MLLA  
predictions, other models with varying degrees of QCD coherence are 
also able to give good fits.  
H1 finds essentially no dependence of the peak of $\xi$ 
on the Bjorken $x$ of the DIS event, so that the 
evolution takes place only as  
a function of $Q$, as predicted by MLLA QCD.

In summary, the   
features of the HERA data do indeed generally
support QCD based models and the  
general process independence of fragmentation. 
\ffigp{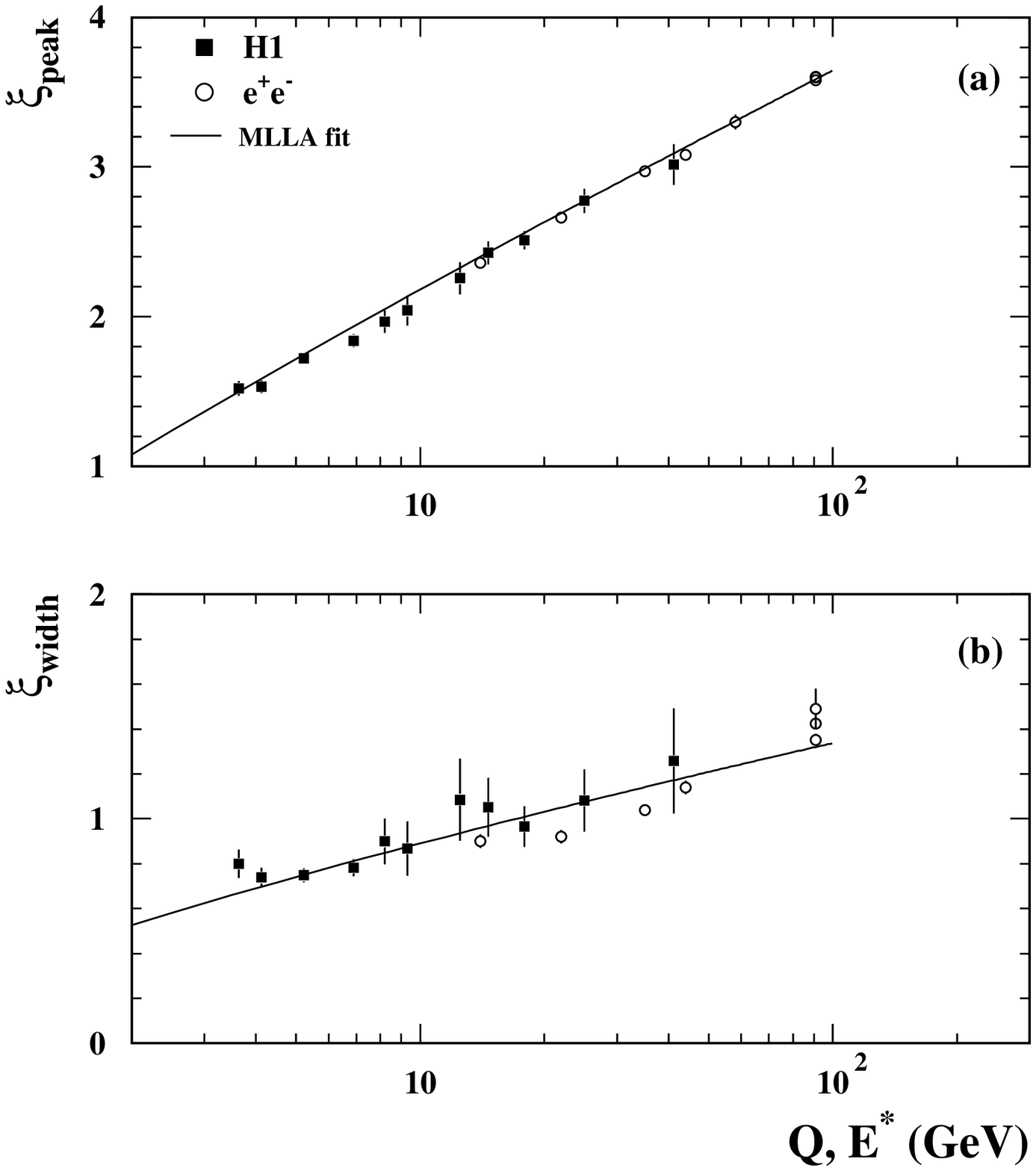} 
{100mm}{80mm}{a) The peak position and b) the width of the $\xi$ distribution  
according to a Gaussian fit. Data from H1 and 
\ee\ experiments are shown. The solid line is a 2 parameter MLLA fit to 
the H1 data.} 
{fig-H1-Breit}  
\subsection{Production of neutral strange particles in DIS} 
\label{sec-K0} 
In principle the study of strange particles can give information on both 
the strange quark content of the proton and the flavour properties of  
fragmentation. In practice however it is very difficult to disentangle these 
two aspects since the rate of strange particles is dominated by the 
production via fragmentation.  

The long lifetimes of the $K^0$ and  
$\Lambda$ mean that very clear signals with small backgrounds  
can be obtained by selecting tracks which are not associated with the  
primary production vertex. Reasonable agreement between the 
H1~\cite{H1-K0} and ZEUS~\cite{ZEUS-K0}  
results is observed. Both experiments observe that the rate of  
neutral strange particle production increases logarithmically with $W$, in  
line with expectations from QCD. By comparing the rates of 
$K^0$ and $\Lambda$ production with predictions of Monte Carlo 
models, both experiments find that the default value 
of 0.3 for the ratio of strange to non-strange quark production in 
fragmentation in the JETSET hadronisation scheme is too high, and prefer  
values nearer to 0.2, in agreement with 
observations at LEP~\cite{DELPHI-strangeness} and in previous DIS 
experiments~\cite{E665-strangeness}. However, there are some indications 
of a disagreement between ZEUS and H1 in the rate of $\Lambda$  
production; ZEUS observe that even with the  
smaller value for the strangeness suppression the data lie  
somewhat below the Monte Carlo, whereas H1 are broadly compatible.  
However, more data will be required to determine if this is a real  
disagreement and hence whether other fragmentation parameters of the models  
need to be tuned.  

H1 have also investigated the possible production of QCD instantons.  
Were such objects to be produced at HERA, large changes in the number 
of strange particles produced in the final state would be  
expected~\cite{QCD-instanton}. The production rate of $K^0$ is 
compared to standard fragmentation models plus an instanton model; 
there is no evidence for anomalous $K^0$ production, leading to 
an upper limit on the production cross-section for
instantons of 0.9 nb at 95\% confidence level. 
\subsection{Determination of $\alpha_s$ from jet multiplicity} 
\label{sec-alphas} 
At high \Qtwo\ the strongly accelerated struck quark from the proton 
can radiate hard gluons which may lead to resolvable jets in the 
final state. The rate of jet production is thus proportional to the value 
of the strong coupling constant, $\alpha_s$. Measurement of this jet rate  
in a kinematic region which minimises uncertainties from parton  
distribution functions and hadronisation can therefore lead to a  
measurement of $\alpha_s$. 

Both H1~\cite{H1-alphas} and ZEUS~\cite{ZEUS-alphas} 
have measured $\alpha_s$  using the above  
technique. Jets are found using the 
JADE jet algorithm~\cite{JADE-jet}. The rate of events containing two 
identifiable jets together with the forward jet resulting from the 
break-up of the proton (``2+1'' jet events) was compared with the rate 
expected in a Monte Carlo model including next to leading order QCD 
effects. The effect of hadronisation on this rate is large and varies  
depending on the exact criteria used in defining a jet. In order to 
reduce uncertainties in the hadronisation process and to avoid the 
very forward region which is not well modelled, the two experiments made  
the following cuts : \\ 
\\
H1: $10 < Q^2 < 100$ GeV$^2$; $y < 0.5$ 
and $Q^2 > 100$ GeV$^2$; $y < 0.7$\\ 
\\
ZEUS: $0.1 < y < 0.95$, $0.01 < x < 0.1$ and 
$120 < Q^2 < 3600$ GeV$^2$.\\ 
\\
For both sub-samples, the additional cut $W > 5000$ GeV$^2$ was also  
imposed. 
The variable $y_{ij} = m^2_{ij}/W^2$, where $i$ and $j$ are two massless  
objects (in this case jets) was used to define resolved jets.  
A cut in this variable, $y_{cut} > 0.02$,  
was applied by both experiments. In addition, H1 restrict the polar angle of  
the jets to $10^\circ < \theta_{jet} < 145^\circ$, which reduces the  
dependence on the influence of parton showers (10$^\circ$ cut) and  
improves the measurement of the jet energies (145$^\circ$ cut).  
 
With the above cuts good agreement was obtained between the data and NLO  
QCD models for the rate of ``2+1'' jet events. H1 
divided their data into 5 bins in  
\Qtwo\, while ZEUS used 3 bins. The values of $\alpha_s$ obtained by H1 are  
shown in \fref{fig-alphas} and indicate a variation 
with \Qtwo. 
\ffig{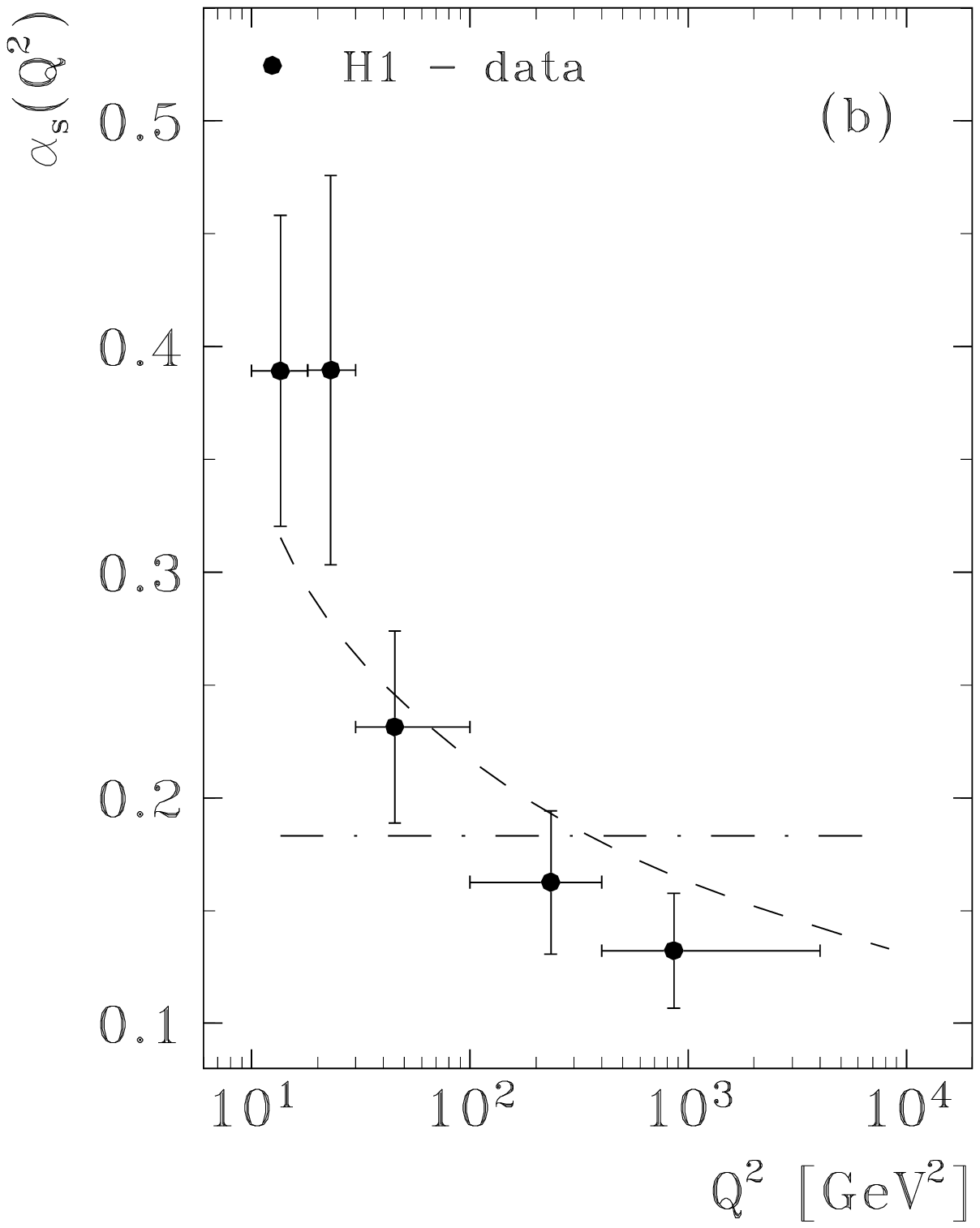} 
{70mm}{Measured values of $\alpha_s$ for five different $Q^2$ 
regions. Only statistical errors are shown. The dashed lines show the QCD 
prediction and the dashed-dotted line shows a fit assuming that $\alpha_s$ is  
constant.} 
{fig-alphas}  
Averaged over the full \Qtwo\ range and evolving to \Qtwo = \MZtwo ,  
the experiments obtain values of $\alpha_s$ at the $Z$ mass of  
\begin{eqnarray} 
\alpha_s(\MZ) & = &0.117 \pm 0.005 (stat.) ^{+0.008}_{-0.009} (syst.): 
{\rm (ZEUS)} \nonumber \\ 
\alpha_s(\MZ) & = &0.123 \pm 0.005 (stat.) \pm 0.013 (syst.): {\rm (H1)} 
\nonumber 
\end{eqnarray} 
which can be averaged to give $0.121 \pm 0.004 \pm 0.008$. 
This is in good agreement with the world average value from all 
determinations of $0.118 \pm 0.003$ ~\cite{PDG}. 
\subsection{Transverse energy flow and forward jet production} 
\label{sec-eflow} 
The distribution of transverse energy and/or production of forward jets is 
in principle a sensitive test of the underlying QCD dynamics. The large 
increase in the accessible kinematic domain provided by HERA leads to 
the possibility of exploring QCD in qualitatively different regimes. 
The DGLAP~\cite{DGLAP} approach to QCD evolution of partons re-sums  
the leading logarithmic terms of the form $\alpha_s \ln Q^2$ in the QCD  
expansion. This results in a ``strong ordering'' in the transverse 
momentum of the gluons emitted between the struck quark, or 
``current'' jet, and the proton remnant, such that gluons  
emitted near to the current jet direction have much larger \pt\  
than those emitted close to the proton direction. In contrast, the 
BFKL~\cite{BFKL} approach re-sums the leading logarithmic terms of the  
form $\alpha_s \ln 1/x$, resulting in a strong ordering in $x$ but no such  
ordering in transverse momenta. Thus 
one might qualitatively expect that, in a region in which the BFKL 
approach was appropriate (e.g.\ $1/x >> Q^2$), there would be an 
excess of transverse energy in the region between the proton remnant and  
the current jet compared to the  
prediction of a DGLAP based model.  In particular, BFKL dynamics would 
predict that the average level of transverse energy per unit pseudorapidity 
$\eta$ in the forward direction 
would grow as $x^{-1}$. Similarly, the production of jets with  
\pt$^2 \sim Q^2$ and $x_{jet} (= E_{jet}/E_p) >> x$ is strongly suppressed  
according to  DGLAP but not  BFKL. The kinematics of such jet  
production clearly imply that it will be dominantly in the forward  
direction close to the proton remnant.  
 
The energy observed as a function of $\eta$ has been shown by both 
H1~\cite{H1-eflow} and ZEUS~\cite{ZEUS-eflow}, and there does indeed 
seem to be an excess in the forward direction compared to the expectation  
from simple models based on standard DGLAP evolution. H1 prefer to  
work in the central rapidity region of the hadronic centre of mass system. At  
fixed \Qtwo\ the mean transverse energy rises with $1/x$ as predicted by 
the BFKL equation~\cite{H1-eflow-warsaw}. However, conventional DGLAP  
based models can also be tuned to reproduce the data by increasing 
the \Et\ produced in the hadronisation i.e.\ non-perturbative phase.  
 
H1 have also observed jets in the very forward  
region~\cite{H1-fjets,H1-fjets-warsaw}. The jets are well separated  
and show typical jet properties despite being in the 
difficult area very close  
to the proton remnant. The preliminary jet cross-section  
from~\cite{H1-fjets-warsaw} shows a steep rise as $x$ falls which is 
significantly above the DGLAP-based models and falls between a  
parton-level analytical BFKL calculation~\cite{Bartels-BFKL-jets} 
and the CDM1 Monte Carlo~\cite{CDMBGF}, based on the
colour dipole model of fragmentation. To the extent that some aspects 
of the CDM model mimic properties of BFKL, this is clearly a very interesting 
preliminary result, which needs to be compared to a full Monte Carlo model  
which explicitly contains the BFKL dynamics.  
 
In summary, although the behaviour of many aspects of the data at low  
$x$ is as expected from BFKL dynamics, conventional DGLAP-based  
models also seem able to describe the data after appropriate tuning. This 
is a similar situation to that discussed in section~\ref{sec-quarks-F2}.  
However, the ability to look at exclusive aspects of the DIS data, such as 
forward jet production, does seem to offer greater hope of finding a 
real discriminant between the DGLAP and BFKL regimes of QCD. This 
can only be convincing however when a comparison with a BFKL-based  
Monte Carlo model is possible.  
\section{The structure of diffraction} 
\label{sec-diffraction} 
By far the most likely processes to occur in the 
scattering of two hadrons are  
elastic scattering (in which the two hadrons emerge unscathed and with a  
relatively  
small deviation) or diffractive scattering (in which one of the two hadrons  
is excited into a relatively high mass state and the other remains unexcited  
and emerges with a large rapidity gap with respect to the excited
system.) The  
description of these phenomena has been a puzzle for many  
decades. In the Regge picture of strong interactions, these  
processes were 
considered to be mediated by a ``Pomeron'', which however did not  
correspond to any known real particle and had the quantum numbers of  
the vacuum. In the modern language of QCD, a self-consistent picture of 
diffraction is complicated since elastic and diffractive scattering 
are of their  
nature predominantly soft, low momentum transfer processes where the  
tools of perturbative QCD are not valid 
(for a review see Goulianos~\cite{Goulianos-review}).  
 
In the last few years   
a very striking class of events have been observed both in 
high energy proton-antiproton 
interactions~\cite{pp-rapgap-disc,D0-rapgap,CDF-rapgap} and in  
deep inelastic scattering~\cite{ZEUS-rapgap-disc,H1-rapgap-disc}, 
in which there are large gaps 
in rapidity space which contain no particles. 
A natural interpretation is that such events correspond to
diffractive processes~\cite{Feynman-diffraction,Bj-diffraction}. The  
very large 
phase space available at HERA allows a systematic  
study of the mechanism of diffraction at momentum transfers 
large enough for a perturbative QCD analysis to be valid. This in turn 
allows an investigation of possible partonic structure in the exchanged  
object. Without loss of generality, we shall refer to this object  
generically as ``the Pomeron''.  
 
In this section we will briefly discuss the kinematic characteristics 
of diffractive events at HERA, before proceeding to a more detailed study 
of their properties and nature. 
\subsection{Kinematic considerations in diffraction} 
\label{sec-diff-kin} 
As discussed in section~\ref{sec-dsigdo-dis}, fully inclusive deep inelastic 
scattering can be described in terms of two structure functions which in  
principle depend on two invariants. In the case of diffractive events, it is 
assumed that the process being observed is of the form 
\begin{equation} 
e + p \longrightarrow e + p + X 
\end{equation} 
where the proton is either observed using specialised silicon detectors far  
upstream of the interaction point or is inferred from the characteristics of  
the event\footnote{The excitation of the proton into an $N^*$ 
resonance may also satisfy the criteria for diffraction.}.  
An extra two invariants,  and if the
final-state proton is measured, in  
principle  
an extra two structure functions, are required to  
fully specify the process. Fortunately in the 
kinematic regime of interest the two extra structure functions are  
negligible~\cite{DelDuca-95}. The invariants are usually chosen to be the  
familiar $x$, \Qtwo\ 
together with two of the following set: 
\begin{eqnarray} 
\xpom &= &\frac{q\cdot (P-P')}{q\cdot P}\nonumber\\ 
\beta &=& \frac{-q^2}{2q\cdot (P-P')} \\ 
t &=& (P-P')^2 \nonumber 
\label{eq-xpom-beta} 
\end{eqnarray} 
where $P, P'$ are the initial and final momenta of the proton and $t$ 
is the four-momentum transfer at the incident proton vertex.  
The variables \xpom\ and  
$\beta$ can be expressed in terms of $W^2$ and $M_X$, 
(the invariant mass squared of the total hadronic system and the system  
excluding the scattered proton, respectively), as 
\begin{eqnarray} 
\xpom & \simeq & \frac{M_X^2 + Q^2} 
 {W^2 + Q^2} \simeq  x_{\pom/p} \\ 
\beta &=& \frac{x}{\xpom}  
   \simeq \frac{Q^2}{M_X^2 + Q^2} \simeq \, x_{q/\pom} 
\label{eq-LRG-variables} 
\end{eqnarray} 
In the kinematic regime of HERA, in which \Qtwo\ and $W^2$ can be  
very large and $t$ is small, \xpom\  
may be interpreted as the fraction of the proton's momentum carried by  
the Pomeron, while 
$\beta$ can be interpreted as the fraction of the momentum of the Pomeron 
carried by the charged constituent which interacts with the virtual photon.  
From these considerations the differential cross-section can be 
written, following \eref{eq-Fl-sigma}, as  
 \beqn  
\frac{d^4 \sigma} 
{dx dQ^2 d\xpom dt} & = & \frac{2\pi \alpha^2}{x Q^4}  
\left[   
\{ 1+(1-y)^2 \} 
 F_2^D(x, Q^2, \xpom,t) \right. \nonumber \\ 
 & - & \left. {y^2} \Fl^D(x,Q^2,\xpom,t)  
 \right] 
\label{eq-F2D4-sigma} 
\end{eqnarray} 
where for the moment we have ignored radiative corrections as well as 
the possibility of charged current 
interactions. It is also normal to assume, similar to the case for normal 
non-diffractive DIS, that \Fl\ is small and can be ignored, although, as will  
be seen below,  
the high gluon content of the Pomeron implies the possibility that this  
assumption may not necessarily be a good one.  
 
In soft hadronic physics and inclusive processes, a universal Pomeron 
which ``factorises", i.e.\ whose flux inside the proton is 
independent of the specific interaction, has been found to give a good  
representation of the data in a variety of 
processes~\cite{Pom-DL,IS}.
It is 
a reasonable, though unproven, hypothesis that similar factorisation occurs 
in the hard diffractive interactions discussed here. If this is indeed true, 
then equation~\ref{eq-F2D4-sigma} can be written as 
\begin{eqnarray}  
\frac{d^4 \sigma} 
{dx dQ^2 d\xpom dt} =  f_{P/\pom}(\xpom,t) \frac{d^2 \sigma}{d\beta dQ^2} 
\label{eq-factorised-sigma} 
\end{eqnarray} 
where $f_{P/\pom}(\xpom,t) $ is the universal Pomeron flux in the proton.  
Thus in analogy to the definition of the $F_2$ structure function we can  
identify a ``Pomeron structure function'' $F_2^{\pom}$ such that 
\begin{eqnarray} 
\frac{d^2F_2^D(x, Q^2, \xpom,t)}{d\xpom dt} &=& f_{P/\pom}(\xpom,t)  
F_2^{\pom}(\beta, Q^2) 
\label{eq-f2D-fact-sigma} 
\end{eqnarray} 
where $\beta$ plays the same role as Bjorken-$x$ in normal DIS. 
 
Finally, in results published to date the scattered proton has not been 
detected and hence $t$ is undetermined. Integrating over $t$ in  
\eref{eq-F2D4-sigma} leads to 
\begin{eqnarray}  
\frac{d^3\sigma_{diff}}{dQ^2 d\beta d\xpom} = \frac{2 \pi 
    \alpha^2}{\beta Q^4} \;  (1+(1-y)^2) \;   
F_2^{D(3)}(Q^2,\beta,\xpom) 
\label{eq-f2D3-sigma}  
\end{eqnarray} 
\subsection{Experimental results} 
\label{sec-diff-exp} 
In retrospect, it was entirely to be expected that diffractive 
processes would  
be important at HERA. Nevertheless, the discovery of the so-called ``rapidity  
gap events''~\cite{ZEUS-rapgap-disc,H1-rapgap-disc}, 
came as a great surprise to the ZEUS 
and H1 experimental teams. Not only did these events have a spectacular  
experimental signature, they also represented a significant fraction of the  
DIS cross-section, initial measurements suggesting about 10\%. The initial 
studies by the two experiments isolated diffractive events  
on the basis of a lack of activity between the current jet and the proton  
direction. By making cuts in each event on $\eta_{max}$, the largest  
pseudo-rapidity  in the detector 
having a significant energy deposit, or on the largest gap between  
significant energy deposits, a sample of diffractive events with high 
efficiency and little background could be obtained.  
 
Data taken by ZEUS and H1 in 1993 allowed a first measurement of the  
partonic structure of the Pomeron~\cite{H1-Pom-95,ZEUS-Pom-95}. The 
two collaborations found evidence that the factorisation hypothesis  
was approximately satisfied, 
i.e.\ that the flux was independent of the kinematics of the  
Pomeron-photon interaction, ($\beta$ 
and \Qtwo) but was a universal function of 
\xpom, namely \xpom$^{-\lambda}$. 
In addition they 
were able to determine the integrated structure function,  
\begin{equation} 
\tilde{F}_2^{D}(\beta,Q^2)= 
\int_{x_{I\!\!P_L}}^{x_{I\!\!P_H}} F_2^{D(3)}(\beta,Q^2,x_{I\!\!P})\ 
dx_{I\!\!P}, 
\label{eq-F2D2-sigma} 
\end{equation} 
in a number of bins of $\beta$ and \Qtwo. $\tilde{F}_2^D$ can be thought of  
as the ``structure function of the Pomeron", although it is important to 
bear in mind that, since the integration over \xpom\ in general covers a 
range kinematically inaccessible for certain values of $\beta$ and \Qtwo, 
$\tilde{F}_2^{D}$ does not represent the diffractive fraction of the  
experimentally measured $F_2$. Although the errors are large, and the 
number of bins small, a relatively hard parton structure consistent with  
scaling in \Qtwo\ was found. This approximate scaling, as
well as the observation of jet structure  in diffractive 
events~\cite{HERA-Diff-jets}, supports the  
picture that the Pomeron contains point-like interacting constituents. 
 
More recently the two experiments have been able to make significant  
further progress. H1 has used the data taken in 1994, which gives almost a  
factor of ten improvement in integrated luminosity over that taken in  
1993, to make much more detailed measurements of the parton structure  
of diffraction. ZEUS has introduced a model independent method of 
subtracting the non-diffractive background by fits to data distributions and  
used this method to determine the diffractive structure function using  
1993 data. ZEUS has also used data from hard photoproduction to  
constrain the parton content of the Pomeron. 
 
\subsubsection{Preliminary H1 analysis of 1994 data} 
\label{sec-diff-exp-H1-94} 
Although the analysis discussed here remains preliminary at the time 
of writing, the information presented 
to the 1996 Warsaw Conference~\cite{H1-Warsaw-F2D} is of sufficient 
interest to warrant its inclusion in this review.  
 
The greatly  
increased statistics allow H1 to study diffraction over a significantly 
larger kinematic range. The variables used to describe the diffractive 
final state are functions of the variables $M_X, x, y$ and \Qtwo\ (see 
\eref{eq-LRG-variables}). Good reconstruction of $M_X$ is 
assured by using the hadronic measurement and  cutting on $\eta_{max}$  
in order to ensure that the hadronic system is well contained within the 
detector. The $\Sigma$ method (see section~\ref{sec-kin-var-recons}) is 
used to reconstruct the other kinematic variables.  
 
The data thus obtained are used to measure the diffractive cross-section 
in bins of {\xpom}, $\beta$ and \Qtwo\ in the kinematic range  
$2.5 < \Qtwo\ < 65$ GeV$^2$, 0.01 $< \beta <$ 0.9  
and $10^{-4} < \xpom <$ 0.05. The  
cross-section is then converted into the structure function $F_2^{D(3)}$ 
using equation~\ref{eq-f2D3-sigma}, correcting for backgrounds, losses,  
migration etc.\ bin by bin using Monte Carlo models.  
 
Figure~\ref{fig-H1-F2D3-94} shows \xpom\ $\cdot F_2^{D(3)}$ in bins of 
$\beta$ and $Q^2$. The data, particularly in 
the region $\beta < 0.2$, are inconsistent with the factorisation 
hypothesis, which predicts that $F_2^{D(3)}$ has the
same dependence on \xpom\
independent of $\beta$ and \Qtwo. The curves are fits to 
the form 
\begin{equation} 
A(\beta, Q^2) \xpom^{-n(\beta)} 
\end{equation} 
where 
\begin{equation} 
n(\beta) = b_0 + b_1\cdot \beta + b_2 \cdot \beta^2 
\end{equation} 
It can be seen that such a parameterisation gives a good description 
of the data.  
\ffigq{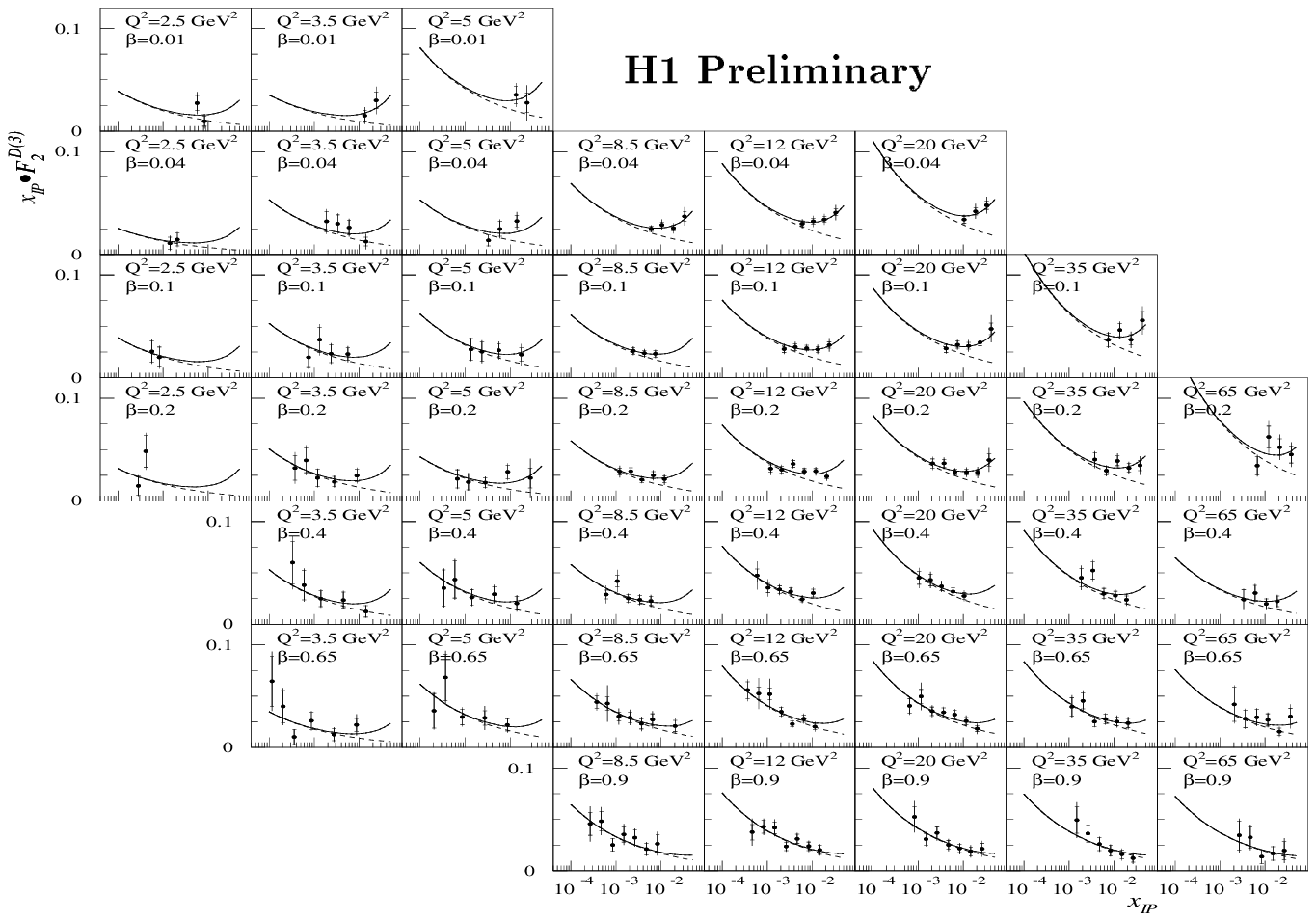} 
{140mm}{140mm}{Plot of \xpom $\cdot F_2^{D(3)}$ in bins of $\beta$ and 
\Qtwo. Statistical and systematic errors have been added in quadrature. 
The curves show fits to $F_2^{D(3)}$ of the form \xpom$^{-n(\beta)}$.} 
{1cm}{fig-H1-F2D3-94}  
There are several possible explanations for this departure from  
factorisation. The most straightforward is that factorisation is simply 
invalid and that therefore models in which factorisation is not 
assumed (see for example the review by McDermott and 
Briskin~\cite{McDermott-Briskin}), or in which factorisation
in soft diffraction processes is broken to satisfy unitarity 
(see for example the review by Goulianos~\cite{Goulianos-PIC}),
are preferred. It is also possible that the 
observed behaviour is indicative of the importance of non-leading 
Regge trajectories, such as the $f_2^{\,0}(1270)$, which must indeed 
contribute at some level if the Regge picture is valid. If the partonic 
structure of such exchanges were to be significantly softer than the  
Pomeron then this could explain the behaviour of the data. Since 
unfortunately there is no information on the structure function of 
the $f_2$, this is allowed to vary in a fit. H1 has carried out 
such a fit which does indeed give a good 
representation of the data and leads to values of intercepts of the 
trajectories which are consistent with what would be expected for the 
Pomeron and $f_2$.  
 
Despite the complication due to the  violation of factorisation, it is still  
possible to obtain information on the partonic structure of the exchanged 
trajectories from $\tilde{F}_2^D$. In this situation this quantity represents 
some sort of average partonic structure of all contributing exchanges.  
$\tilde{F}_2^D$ integrated from \xpom\ = 0.0003 to 0.05 is shown in  
\fref{fig-H1-F2tilde-94}.  
\ffig{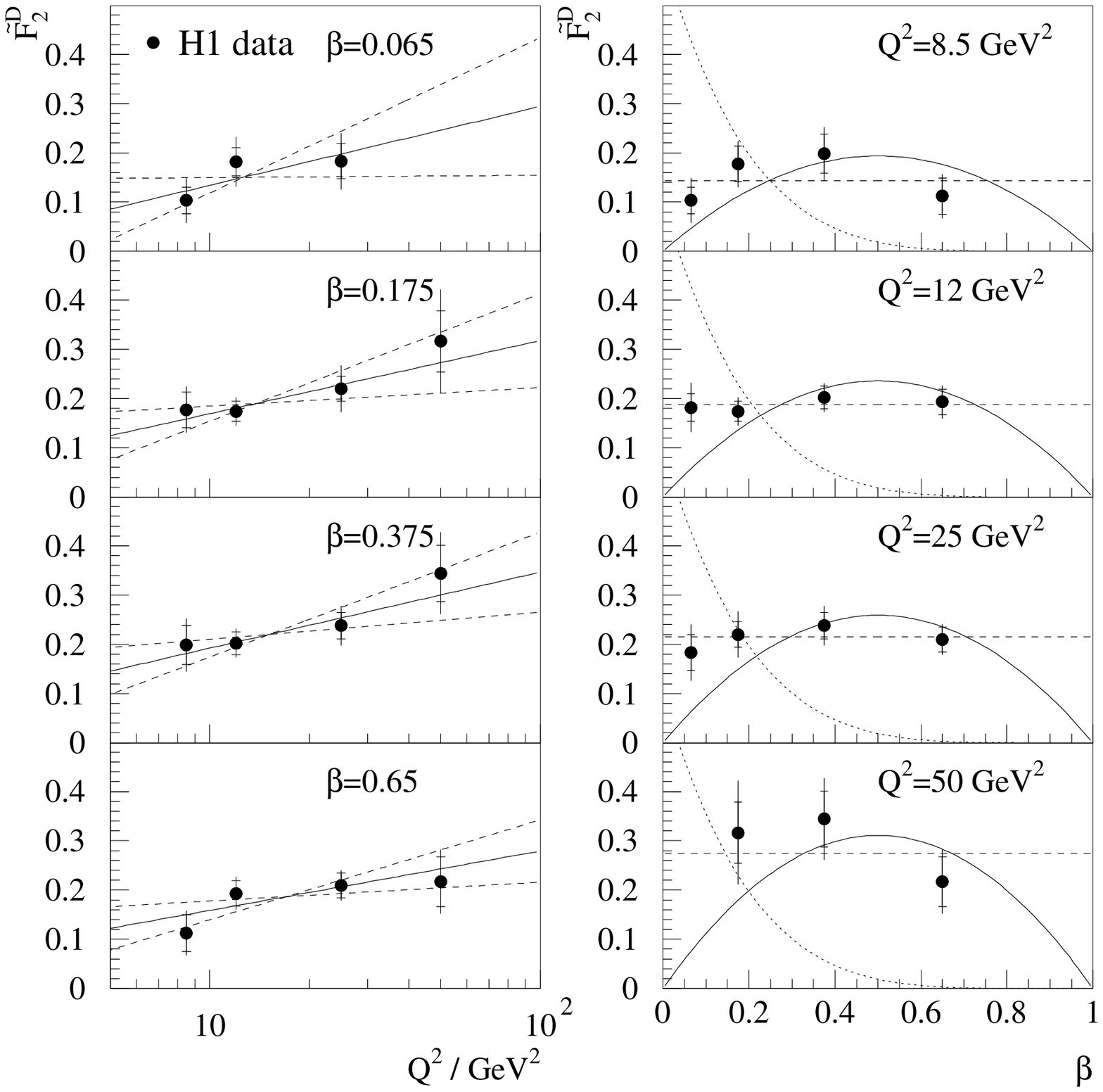} 
{120mm}{Plot of $\tilde{F}_2^{D}$ as a function of \Qtwo\ 
 for different values of  $\beta$ and vice-versa. The solid line on the 
left-hand plots corresponds to the best fit to a linear $\ln Q^2$ 
dependence with one sigma errors shown by the dashed lines.  
The dashed lines in the right-hand plots correspond to the best 
fits to a constant dependence with $\beta$.} 
{fig-H1-F2tilde-94}  
It is clear that $\tilde{F}_2^D$ is essentially independent of $\beta$ 
at fixed \Qtwo, whereas it depends logarithmically on \Qtwo\ at fixed 
$\beta$. The most striking aspect of the data is that this rise with  
$\ln Q^2$ persists even at large values of $\beta$, where, had there been 
a substantial valence quark contribution, the slope would have been 
negative. We are thus led to the qualitative conclusion that the exchanged 
trajectories consist dominantly of gluons.  
 
The above conclusion can be made more quantitative by a QCD analysis,  
assuming a singlet $u + d + s$ quark distribution and a gluon distribution 
at a starting scale $Q^2_0 =$ 2.5 GeV$^2$ and evolving these distributions 
to higher values of \Qtwo\ using the DGLAP equations. No acceptable fit 
to $\tilde{F}^D_2$ can be obtained if the initial distribution is assumed to 
be purely quark-like, whereas with a non-zero initial gluon component, 
excellent fits are obtained. \Fref{fig-H1-pom-qg-94} shows the results
of such a  
fit at two values of \Qtwo. At low \Qtwo\ the dominance of the 
gluon is clear; at higher \Qtwo\ the gluon distribution evolves down to 
lower $x$. However, in all cases the gluon is dominant in the parton 
structure of the exchange, with in excess of 80\% of the momentum 
being carried by gluons, in good agreement with models of ``leading'' 
gluons (see for example~\cite{Buchmuller-Hebecker}).  
\ffigp{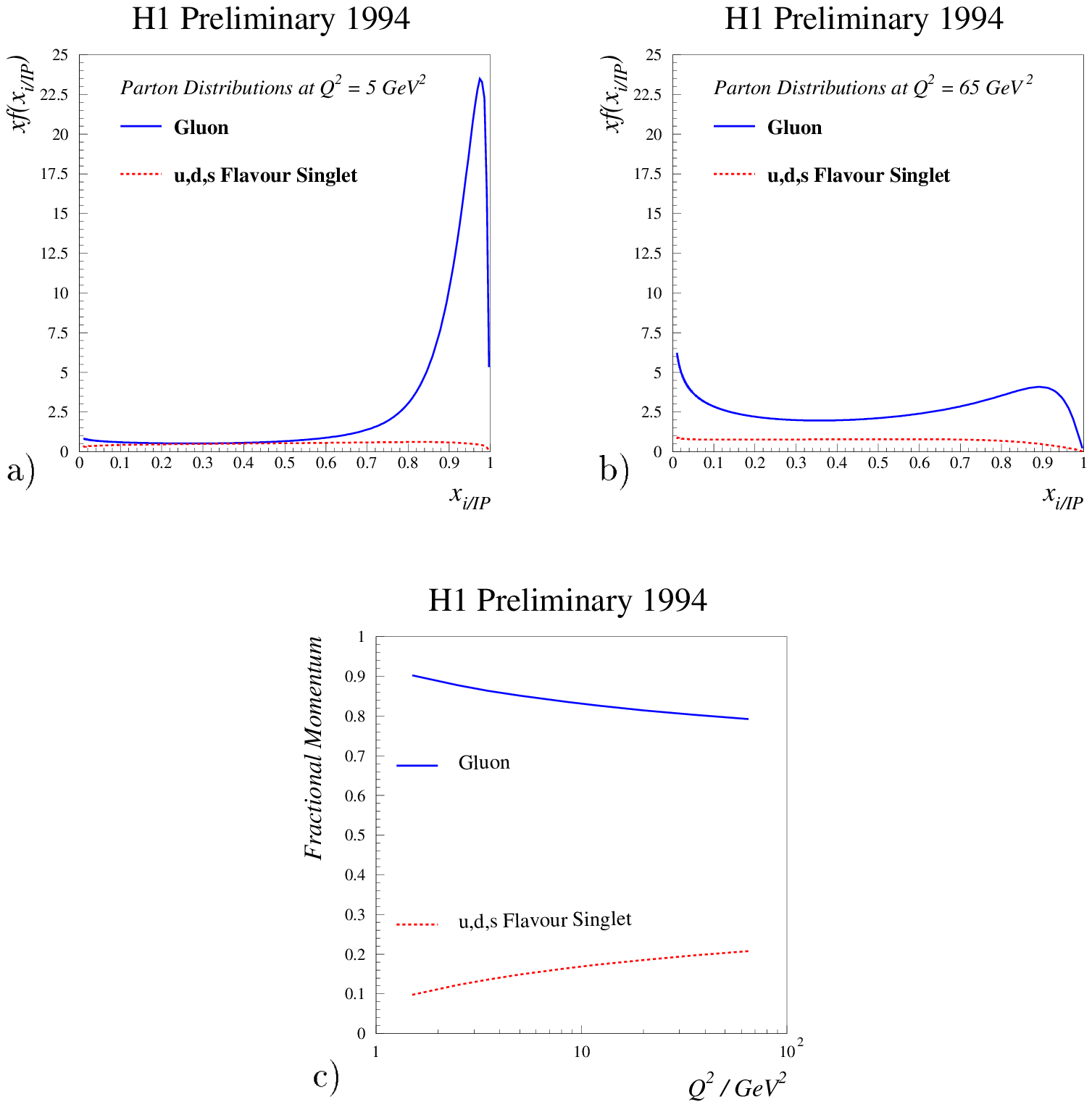} 
{110mm}{140mm}{Results of a DGLAP fit on the H1 $\tilde{F}^D_2$ data 
at a) \Qtwo\ = 5 GeV$^2$ and b) \Qtwo\ = 65 GeV$^2$; c) shows 
the fraction of the total momentum carried by quarks and gluons 
as a fraction of the total momentum transfer as a function of \Qtwo.  
The errors in these fits are not well determined and are therefore 
not shown.} 
{fig-H1-pom-qg-94}  
 
It can thus be seen that the H1 data give striking evidence for a ``leading'' 
gluon behaviour of the exchange in deep inelastic diffraction, as well as 
evidence that exchanges in addition to the Pomeron are important.  
However, it is important to stress that these conclusions are clearly 
model dependent. The theoretical situation is very confusing, and 
there is no consensus on many fundamental aspects of the  
phenomenology. We have already commented on many models  
which do not assume 
factorisation~\cite{Pom-DL,Goulianos-PL,Collins,Gehrmann-Stirling}. 
Some papers insist that the whole 
concept of a parton structure of diffraction is  
controversial~\cite{Ellis-Ross}. 
What is clear is that there are alternative models, for instance making 
different assumptions about the form factor appropriate to the  
Pomeron-proton coupling, which can lead to the observed behaviour of 
$\tilde{F}^D_2$ without implying QCD evolution from a ``leading'' 
gluon component. Furthermore, it is unclear whether or not a ``direct'' 
point-like coupling of the Pomeron to quarks is  
appropriate~\cite{Gehrmann-Stirling}. An excellent review of 
the different types of model can be found in~\cite{McDermott-Briskin}. 
\subsection{ZEUS results in DIS} 
\label{sec-diff-exp-ZEUS-DIS} 
ZEUS has developed a method of subtracting the non-diffractive  
background using fits to the distribution of $M_X$, which is 
independent of the model used for the diffractive process. It relies 
on the QCD prediction that in non-diffractive DIS events 
low $M_X$ values are exponentially suppressed~\cite{Andersson-83}. By 
fitting the data at large $M_X$ and assuming that the slope of the 
exponential is also valid for non-diffractive 
DIS events in the small $M_X$ region, a clean model-independent  
subtraction of the non-diffractive background 
to the diffractive signal can be made.  
 
This method has been applied to the 1993 data  
sample~\cite{ZEUS-MXfit}. The number of background-subtracted  
diffractive events is converted into a cross-section using kinematic  
variables calculated using the Double Angle (DA) Method  
(see section~\ref{sec-kin-var-recons}). The differential cross-section 
thus derived can be plotted as a function of $M_X$ and fitted 
to the form $(W^2)^{2\alpha_{\pom} -2}$ expected in the Regge picture. 
The value derived for 
$\alpha_{\pom}$ is around 1.2, somewhat larger 
than would be expected in a ``soft'' Pomeron  
process\footnote{Note that some of the results 
quoted here are smaller than those in the original  
paper~\cite{ZEUS-MXfit} due to 
a correction to the treatment of the radiative effects.}.  
 
The differential cross-section can also be used to determine the diffractive 
structure function $F^{D(3)}_2$ using equation~\ref{eq-f2D3-sigma}. Since 
the amount of data available is substantially smaller than the H1  
preliminary study discussed in the previous section (which used 1994  
data), only a relatively small number of bins was used. The data are 
consistent with the factorisation hypothesis, and when the data are 
fit to a form 
\begin{equation} 
F^{D(3)}_2 \propto \left(\frac{1}{\xpom}\right)^a 
\end{equation} 
gives a value of $a$ of approximately 1.4 $\pm 0.04 (stat.)\pm 0.08 (syst.)$, 
in good agreement with the previous ZEUS analysis. The 1993 H1 analysis 
produced a value of $n$ of $1.19 \pm 0.06 \pm 0.07$, which is on the limit 
of compatibility with ZEUS. The 1994 preliminary H1 analysis is, as we have seen, incompatible with factorisation, but the average value of $a$  
can be determined and is 
similar to that in the 1993 analysis. The source of  
any discrepancy between the two experiments is likely partially to be  
found in the method of background subtraction and the definition of what 
constitutes a diffractive event, as well
as differences in the range of $\beta$ covered. 
ZEUS estimated the value of $a$ obtained 
using a Monte Carlo model to subtract the background and obtained a 
substantially lower value. Nevertheless, both ZEUS and H1 agree that the 
value obtained for $\alpha_{\pom}$ is substantially less than would be  
expected were BFKL dynamics to play an important role. Similarly, both  
experiments agree that the Pomeron intercept obtained is 
somewhat larger than  
that normally obtained by analyses of ``soft'' hadronic processes.  
\subsection{ZEUS results from hard photoproduction} 
\label{sec-diff-exp-ZEUS-hpp} 
We once again slightly enlarge the scope of this review in 
order to consider some very relevant results on the partonic structure 
of diffraction from hard photoproduction events. Since the virtual photon  
in DIS couples only to electrically charged 
partons, further direct information on gluonic contributions 
to the Pomeron structure must be obtained by other 
means. We have already discussed H1's results using a QCD fit to 
determine the gluon fraction from an examination of the 
scaling violations with \Qtwo. An alternative and 
more direct probe of the gluon content of the Pomeron can be 
achieved using a ``resolved photon'' process, in which a gluon  
or quark inside the photon may interact with a quark or gluon 
inside the Pomeron.  
 
A clear signal for diffractive-like events in hard 
photoproduction can be obtained from a sample of events containing 
jets in the final state in the range $-1 < 
\eta^{jet} < 1$ and $E_T^{jet} > 8$ GeV where the most forward  
going hadron has 
\etamax $< 1.8$~\cite{ZEUS-hpp-93}. 
A possible contribution from non-diffractive 
processes to the measured cross-section is both too 
low and also has the wrong  
shape in $\eta$.. In addition, predictions 
from Monte Carlo models assuming the Pomeron consists of 
soft gluons or hard quarks fall substantially below the 
data.  
 
The cross-section was compared to predictions from the 
POMPYT Monte Carlo using the Donnachie-Landshoff~\cite{Pom-DL} flux factor and 
allowing for a mixture of hard gluons and quarks in the Pomeron, 
together with some non-diffractive background.   
\Sigmap\ is a free parameter in these fits and is defined as   
the sum of the Pomeron constituents such that if 
the momentum sum rule is satisfied \Sigmap = 1.   
If we assume that the Pomeron contains a fraction $c_g$ of gluons, 
then for each  
value of $c_g$ a fit can be made to the cross-sections 
by varying \Sigmap.  
The results are shown in \fref{fig-pomeron-qgfit}. For $c_g = 0$,  
\Sigmap = $2.5 \pm 0.9$, whereas for $c_g = 1$, \Sigmap = $0.5 \pm 0.2$ . 
 
On the assumption that the parton densities in the pomeron are universal, we 
can combine the results from the hard photoproduction and ZEUS DIS  
diffractive data~\cite{ZEUS-Pom-95}.  
Defining \Sigmapq\ as the 
contribution to \Sigmap\ coming from the quarks, 
\vskip0.3cm 
\begin{eqnarray} 
  \int_{\xpom_{min}}^{\xpom_{max}} d\xpom \int_{0}^{1} 
  d\beta \; F_2^{D(3)}(\beta,Q^2,\xpom) 
  =  
k_f \cdot  \Sigmapq(Q^2) \cdot I_{flux} 
\label{eq-F2D3int} 
\end{eqnarray} 
where $I_{flux}$ is the integral of the pomeron flux factor over $t$ 
and $k_f$ is either 5/18 if $u,d$ quarks contribute or $2/9$ if 
$u,d,s$ contribute. \Eref{eq-F2D3int} can be evaluated using the 
measured $F_2^{D(3)}$ to give \Sigmapq $= 0.32 \pm 0.05$ assuming two 
flavours or $0.40 \pm 0.07$ for three flavours; this is also plotted  
in \fref{fig-pomeron-qgfit}.  
\ffigp{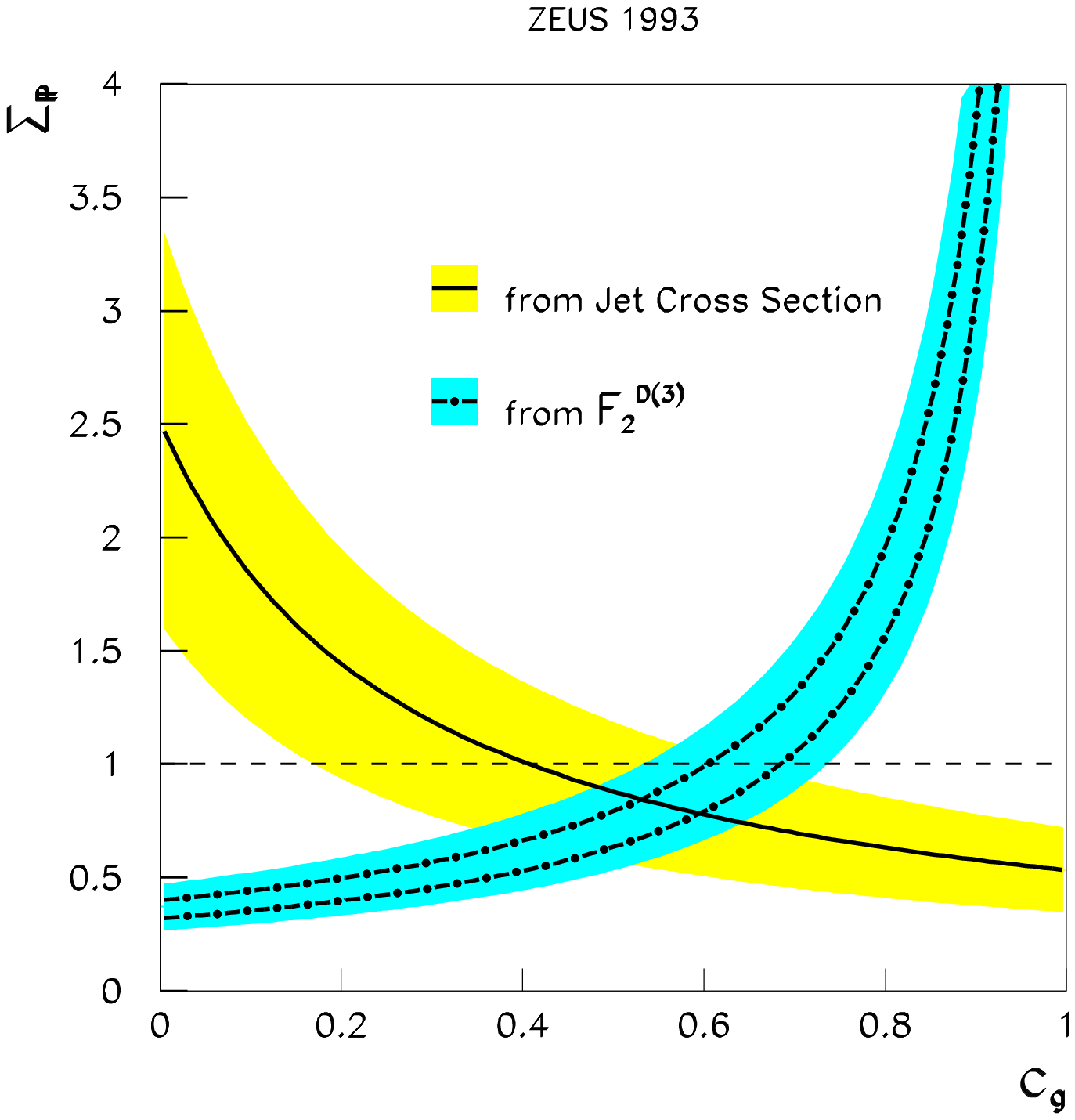} 
{100mm}{100mm}{\Sigmap\ against $c_g$. The solid line represents the 
$\chi^2$ 
minimum and the shaded band the $1\sigma$ limits for the fit to the 
hard photoproduction jet cross-section. The data from the diffractive 
DIS structure function is represented by the dashed-dotted lines for 
the two choices of the number of contributing quark flavours and the 
shaded area is the $1\sigma$ error band. } 
{fig-pomeron-qgfit} 
 
Taking into account the statistical errors, the result for  
\Sigmap ,  $ 0.5 <  \Sigmap < 1.1$,  
indicates that the data are consistent with the momentum 
sum rule for the pomeron being satisfied. Taking also into account  
those systematic errors which do not 
cancel in this comparison, the data indicate that approximately 1/3 of the  
momentum 
of the pomeron is carried by quarks and 2/3 by gluons. Furthermore,  
the fraction of gluons inside the pomeron can be limited to $0.3 < c_g 
< 0.8$ at the $1\sigma$ level, independent both of the 
validity of the momentum sum rule and of the definition of the flux 
factor, provided that it does not depend on \Qtwo. Thus there is very 
satisfactory agreement between H1 and ZEUS, based on very different 
assumptions, techniques and processes, that the pomeron is 
predominantly a gluonic object. 

Recent results from CDF~\cite{CDF-diff-W,CDF-diff-jets} give 
the gluon fraction inside the Pomeron as $c_g = 0.7 \pm 0.2$, in good
agreement with the ZEUS results. The total number of dijets observed
by CDF supports the idea of factorisation breakdown.  
\subsection{ZEUS results on energetic forward neutrons} 
\label{sec-diff-ZEUS-FNC} 
Although technically not a diffractive process, the production of leading  
forward neutrons is also a peripheral process and can also give  
information on the structure of an exchanged colourless particle, 
which in this case 
may be thought of as a charged pion. 
 
Both ZEUS and H1 have over the last few years installed forward neutron 
calorimeter (FNC) prototypes and latterly final devices. These are 
small sampling calorimeters situated at 0$^\circ$ from the interaction  
point and around 100 m downstream. Since the calorimeter is situated 
downstream of the HERA dipoles, the 820 GeV proton beam is bent 
upwards away from the FNC and the non-leading charged particles 
produced at the interaction point are swept away. Neutrons 
however are not affected by the magnets and leading neutrons interact in  
the calorimeter to produce a detectable shower. For details of
the final ZEUS FNC, see~\cite{ZEUS-FNC}. 
 
Events from the 1993 data taking period with signals in a
prototype  FNC, the so-called FNC-II~\cite{ZEUS-FNC-II}. 
The FNC-II
has a charged particle veto based on scintillation counters 
which ensures that  
the energetic particle entering the calorimeter is neutral. 
Events having signals from FNC-II thought to
correspond to leading neutrons were reconstructed inside the ZEUS  
detector using 
the Double Angle method. Those with \Qtwo\ $> 10$  
GeV$^2$ and 
with neutron energies greater than 400 GeV were used in the analysis.  
The data with leading neutrons have similar distributions in \Qtwo, 
$x$ and $W$ as other DIS events. The ratio of events containing 
a leading neutron is therefore independent of the kinematic variables 
and is around $0.45 \pm 0.02 \pm$ 0.02\%. While the 
distribution for large $\eta_{max}$ seems similar, there seems to be 
about a factor of two fewer events with a large rapidity gap compared 
to the average in DIS events. Monte Carlo models such as ARIADNE 
which do  
not contain peripheral processes reproduce neither the fraction of 
events with an energetic neutron nor the neutron energy distribution. 
However, reasonable agreement with the data is obtained from Monte 
Carlo models of one pion exchange.  
 
Thus it would seem that the fraction of events containing a leading  
neutron is independent of the kinematic variables in DIS. There 
is evidence that one pion exchange is required 
to explain the general features of the data. This, together with the  
installation of the final FNCs by ZEUS and H1, promise that future data  
taking will allow studies of the parton structure of the 
pion similar to those discussed above for the pomeron. 
\subsection{Conclusions and outlook for diffraction studies} 
\label{sec-diff-concl} 
It is clear that tremendous progress has been made in studies of diffraction  
since the discovery of large rapidity gap events in DIS at HERA in 1993.  
 ZEUS and H1 agree that there is strong evidence 
that diffraction is mediated by a colourless particle with the characteristics 
of the Pomeron. Furthermore, within specific models and with  
clearly stated assumptions, they agree that the Pomeron has a  
partonic structure 
which appears to be predominantly gluonic. It is gratifying to note 
that these conclusions have been reached with different techniques and 
differing assumptions. H1 in the preliminary 1994 data analysis 
has also shown that factorisation is broken and that a reasonable  
description of the data can be obtained with contributions from  
non-leading trajectories. Nevertheless, there remains much to be done.  
The theoretical situation and the 
basis for many of the assumptions made is still unclear. The availability 
of high quality experimental data has resulted in an explosion of 
theoretical work in response to the many open questions in this area. 
 
There are grounds for optimism not only in the greatly  
increased data sample to be analysed but also in the advent of new and  
powerful detectors. The larger data sample should open up 
the study of diffractive events in charged currents, which are now 
beginning to be seen in the two experiments~\cite{ZEUS-CC-1996}.  
This gives sensitivity in principle to the flavour content of the 
quarks in the Pomeron. With regard to new detectors, we have already  
mentioned the forward neutron 
calorimeters. Both experiments now also have leading proton  
spectrometers (LPS) either complete or nearing 
completion~\cite{ZEUS-LPS-rho,H1-LPS}. 

Leading proton spectrometers are complex devices consisting of silicon 
microstrip detectors inside ``Roman pots'' which can be inserted close 
to the beam line after injection. By using the HERA magnets as an 
analysing spectrometer, the trajectories of energetic protons as  
reconstructed in several stations of ``Roman pots'' placed downstream 
of the interaction point can be used to make an accurate measurement 
of the leading proton momentum. Since the detection of a leading 
proton is an unambiguous signal for a diffractive interaction, use of 
the LPS promises in principle the possibility of obtaining an 
essentially background free sample of diffractive events in DIS.  
 
First preliminary results from the ZEUS LPS are now becoming available. 
ZEUS has published a study of elastic $\rho^0$ events  
with a tagged leading proton~\cite{ZEUS-LPS-rho}. The published distributions 
and properties of these events demonstrate that the LPS can be used 
and calibrated accurately. In addition preliminary measurements of 
the $t$ distribution of diffractive events and  
of the diffractive structure function $F^{D(3)}_2$~\cite{Ela-Rome} 
using events tagged with the ZEUS LPS have been presented at  
conferences. Future publications and improvements in these 
measurements are to be expected, as well as the prospect for the 
measurement of the fully differential diffractive structure function, 
$F^{D(4)}_2$.  
 
A combination of these new detectors, improved knowledge of the 
current detectors and a large increase in integrated luminosity also 
opens up the possibility of the measurement of the ratio of the  
longitudinal to transverse diffractive structure functions,  
$R^D$. Since as we have seen, the pomeron seems to be  
predominantly gluonic, next to leading order QCD necessarily 
predicts a non-negligible value for $R^D$. Such a measurement 
would provide an important testing ground for QCD. Another 
topic which should become important is the measurement of 
the charm diffractive cross-section. Both experiments already have  
indications of charm production in diffractive DIS. A measurement 
of the charm diffractive structure function would be very important 
in constraining many of the theoretical models, since the relatively 
large mass of the $c$ quark provides a hard scale. This can not only 
allow perturbative calculations in some regions of phase space, but  
also strongly influence the predictions of the many models which are  
sensitive to the  
degree of virtuality of the quarks coupling to the virtual photon.  
A review of some of the topics in diffraction 
which can be investigated at HERA in the future can be found in 
\cite{Mehta-Phillips-Waugh}. It is clear that the study of diffraction 
at HERA holds much more excitement in store. 
\section{High \Qtwo\ processes} 
\label{sec-CC}  
\subsection{Introduction} 
\label{sec-CC-Intro} 
HERA provides an unique opportunity to study the electroweak 
interaction at \Qtwo\ sufficiently high that the charged and neutral 
currents are of similar strength.  
 
At high \Qtwo\ both charged and neutral current processes can be 
clearly identified and separated from backgrounds.  Neutral 
current processes can be isolated by identifying  the high energy 
scattered charged lepton, which at high \Qtwo\ tends to be scattered at 
large angles and therefore into 
the barrel or forward sections of the calorimeters. In charged current  
events  the fact that the outgoing and undetected $\nu$ carries away a  
large amount of \pt\ means 
that above a \Qtwo\ threshold, these events can be cleanly isolated in  
most regions of phase space. 
 
In section~\ref{sec-dsigdo-dis} we 
discussed the general form of the differential cross-section 
for deep inelastic  
scattering. \Eref{eq-general-sigma} shows the 
general form for the differential cross-section 
in terms of the structure functions ${ F}_1$, ${ F}_2$ and  
${ F}_3$. These structure functions are products of quark distribution  
functions and the couplings of the currents 
mediating the interaction, as discussed in detail in 
section~\ref{sec-dsigdo-dis}. 
 
\subsection{Experimental results} 
\label{sec-CC-exp} 
 
As remarked earlier, the identification of both neutral and charged current  
events at high \Qtwo\ is relatively straight-forward. H1 base their analysis  
and event selection of the kinematic variables reconstructed from the  
energy of the hadronic final state; the presence or absence of an electron is  
used only to classify an event as either a charged or a neutral current 
candidate. The requirement of transverse momentum from the hadronic  
jet of greater than 25 GeV, together with vertex and trigger selections, leads 
to a clean sample of high \Qtwo\ events with little background and an 
efficiency of $\sim 85$ \%~\cite{H1-CC-1996}. ZEUS in contrast select 
charged current candidates by selecting on missing transverse momentum  
greater than 11 GeV. Similar vertex and trigger cuts are made as for H1,  
although ZEUS uses a rather more complicated set of cuts to reduce  
backgrounds from neutral current and photoproduction events. Both 
experiments use a combination of pattern recognition algorithms and  
visual scanning to remove remaining cosmic ray interactions in the  
detectors. 
 
\ffigp{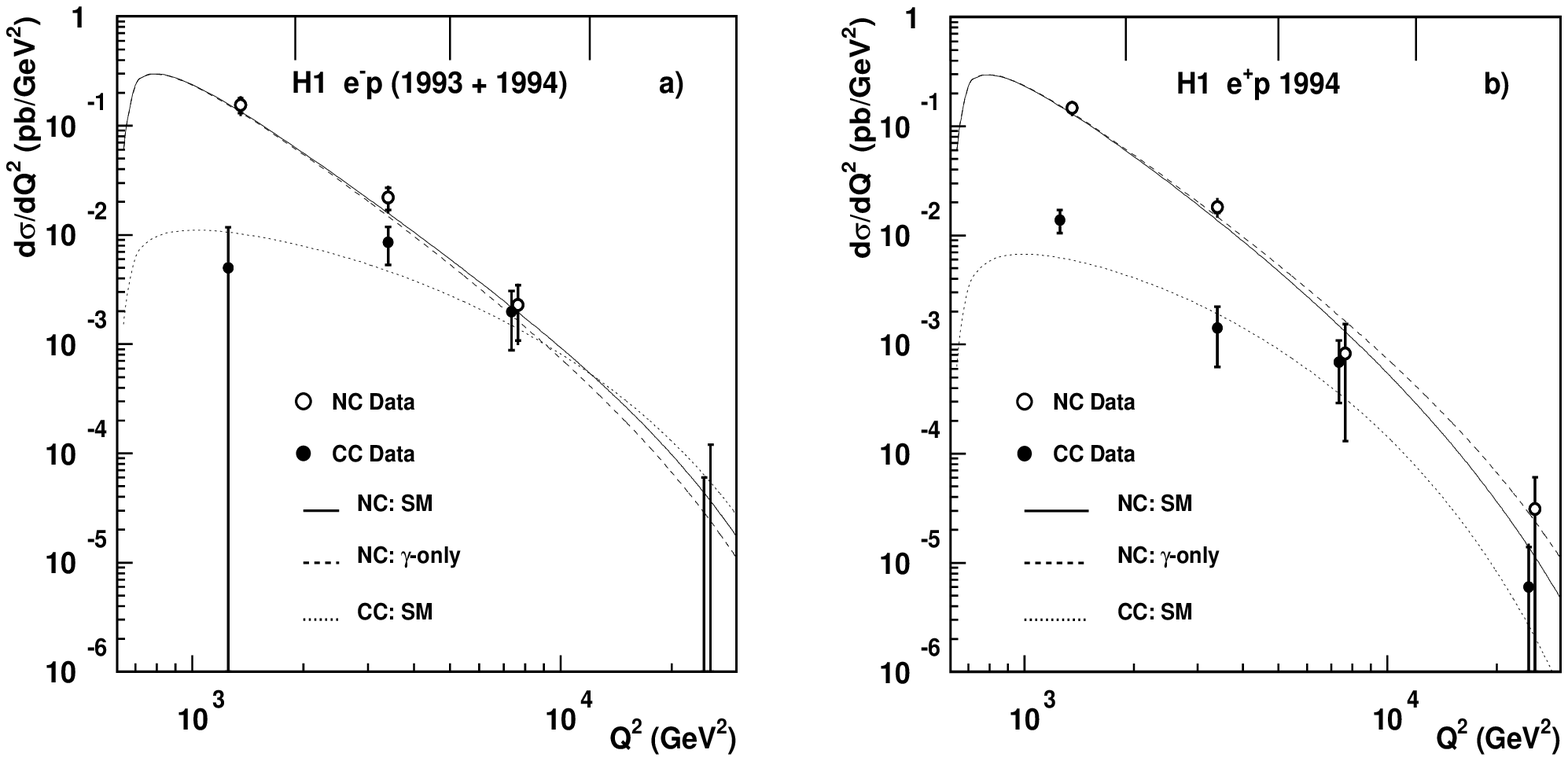} 
{70mm}{120mm}{Neutral current and charged current e$^-p$ and e$^+p$
cross-sections 
as a 
function of \Qtwo\ from the H1 1993 and 1994 data. The  
boundaries of each bin are denoted by the vertical lines at 
the top of the plot. The  
solid curve shows the Standard Model prediction 
for neutral currents, the dashed-dotted curve
the neutral current cross-section without the $Z$
contribution and the dashed curve the Standard Model prediction 
for charged currents.} 
{fig-H1-CCNC-sigmaQ} 
\ffigp{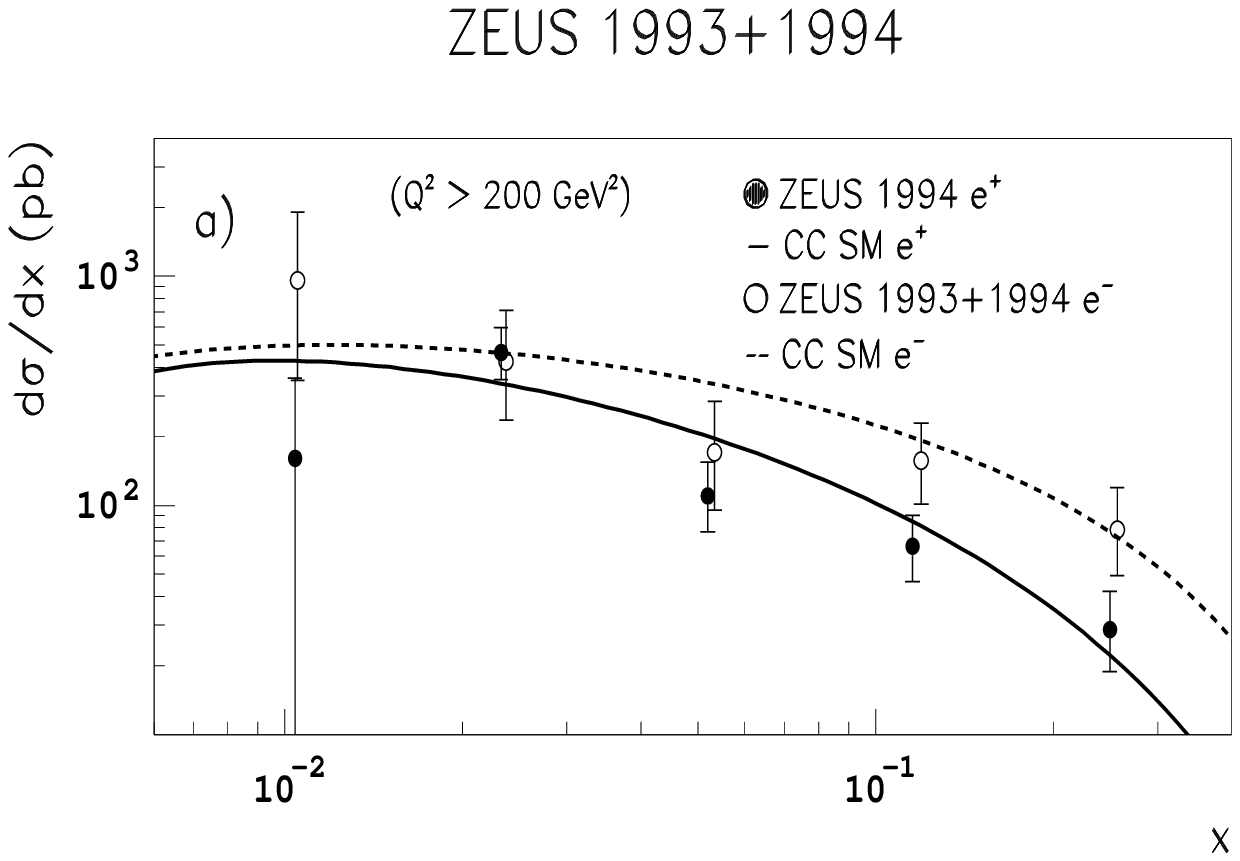} 
{80mm}{110mm}{Neutral current and charged current cross-sections as a 
function of $x$ from the ZEUS 1993 and 1994 data. The  
data points are centred at the generator level average $x$ of 
the points in that bin. The solid curve shows the Standard Model prediction 
for neutral currents and the dashed curve the Standard Model prediction 
for charged currents.} 
{fig-ZEUS-CCNC-sigmax}

\Fref{fig-H1-CCNC-sigmaQ} shows the differential cross-section for the 
charged and neutral currents as a function of \Qtwo\ from the 
H1 experiment~\cite{H1-CC-1996}, while \fref{fig-ZEUS-CCNC-sigmax}  
shows the ZEUS differential cross-section as a function of $x$. 
It can be seen that, for $e^-p$ interactions, 
these two processes become of equal strength at \Qtwo\ 
$\sim 10^4$ GeV$^2$. For $e^+p$ interactions, 
the charged current cross-section approaches the neutral current 
 cross-section, but remains below it. The reason for the smaller 
size of the $e^+p$ cross-section compared to $e^-p$ can be 
seen by inspection of  equations~\ref{eq-general-sigma}, 
\ref{eq-CCem} and \ref{eq-CCep}. Since at high \Qtwo\ both $ x, y \rightarrow  
1$, the $(1-y)^2$ terms in the cross-section, which 
arise from the $V - A$ helicity structure of the charged weak current, 
imply that 
quarks dominate  
the cross-section in the case of electrons and 
sea antiquarks in the case of positrons. Since 
the density of sea quarks tends to 0 as $x \rightarrow 1$, the 
ratio of positron to electron cross-sections will fall to 0. This 
behaviour is shown in \fref{fig-epemsigmas}, which shows 
data on the ratio of the positron and electron cross-sections 
as a function of \Qtwo\ from the ZEUS 1993  
data~\cite{ZEUS-ratios-1993}.  
\ffig{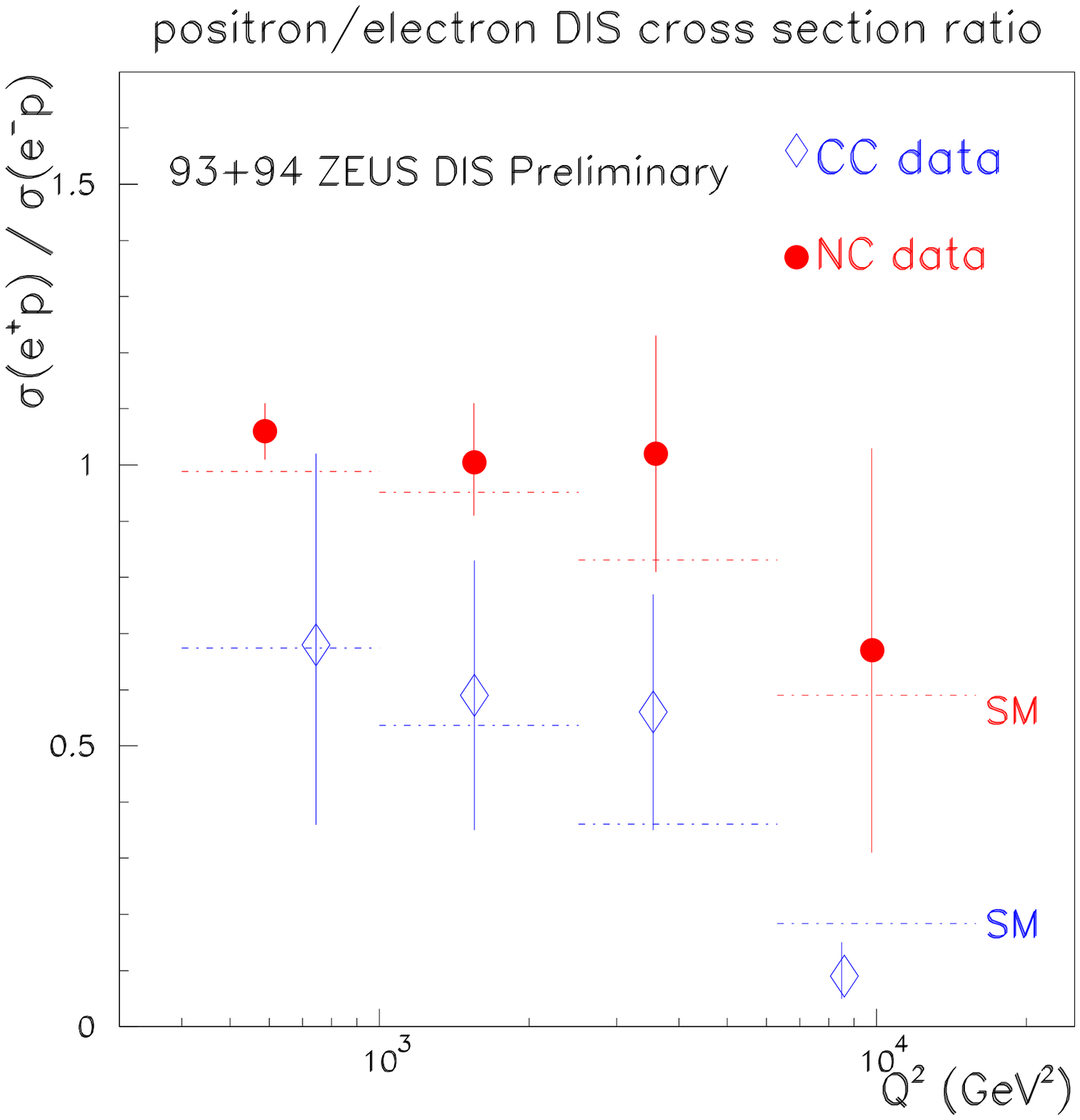} 
{70mm}{Ratio of positron and electron total cross-sections as a 
function of \Qtwo\ from ZEUS 1993 and 1994 data. The dotted lines show  
the Standard Model predictions.} 
{fig-epemsigmas}  
The ratio for neutral currents also falls as \Qtwo\ increases, again 
for reasons related to the current-quark coupling at high $y$, 
although the more complicated nature of the formula for the 
neutral current cross-section means that this is not as obvious as
in the charged current case.  
 
The fact that the charged current data are in good agreement with  
the Standard Model clearly excludes an infinite 
mass $W$ and allows in principle a determination of the $W$ mass. 
Both ZEUS and H1 have carried out such a determination. H1 uses 
the individual event characteristics to construct a likelihood function 
which is compared with the hypothesis of a varying $W$ mass. ZEUS 
performs a maximum likelihood fit to the binned differential 
cross-section data. H1 obtains 
$M_W = 84 ^{+9}_{-6}~^{+5}_{-4}$ GeV, while ZEUS obtains 
$M_W = 79 ^{+8}_{-7}~^{+4}_{-4}$ GeV. Both of these results 
are in good agreement with the world average of 
$80.22 \pm 0.26$~\cite{PDG}. It will clearly be some time before 
the mass determined in the space-like domain will approach the 
accuracy of the Tevatron and LEP-II in the time-like domain! 
\subsection{The highest attainable $x, Q^2$} 
\label{sec-CC-highestQ} 
Although we have seen in the previous section that the data taken up 
to 1994 showed excellent agreement with the Standard Model, the data  
taken in 1995 and 1996 represented an increase in luminosity of  
approximately a factor of 10. The analysis of this data  
by the two collaborations began to show very interesting possible  
discrepancies from the predictions of the Standard Model at the highest  
attainable \Qtwo\ and $x$.  
 
The data sample used by ZEUS~\cite{ZEUS-highqtwo} 
 corresponds to 20.1 pb$^{-1}$, while H1~\cite{H1-highqtwo} uses 
14.2 pb$^{-1}$. H1 uses a variety of methods to reconstruct the kinematic 
variables for neutral current events, preferring the Electron method and 
using the Double Angle method as a check. ZEUS prefers the DA method 
and uses the Electron method as a check. In contrast to ZEUS, H1 also  
presents results for charged currents; ZEUS charged current data have
now been presented at conferences~\cite{ZEUS-CC-highqtwo}.
 
The cause of the excitement can be seen when the data are 
plotted as a function of $x$ (or 
equivalently $M_e = \sqrt{sx}$) or \Qtwo\ and compared with the 
predictions of the Standard Model. \Fref{fig-ZEUS-highqtwo-x} 
and \fref{fig-H1-highqtwo-me} show the distributions of 
events as a function of $x (M_e)$ , while \fref{fig-ZEUS-highqtwo-qtwo} and 
\fref{fig-H1-highqtwo-qtwo} show the distributions as a function of 
\Qtwo. Both experiments show significant deviations 
from the expectations of the Standard Model, shown as a solid line, in
both $x (M_e)$ and \Qtwo. 
\ffig{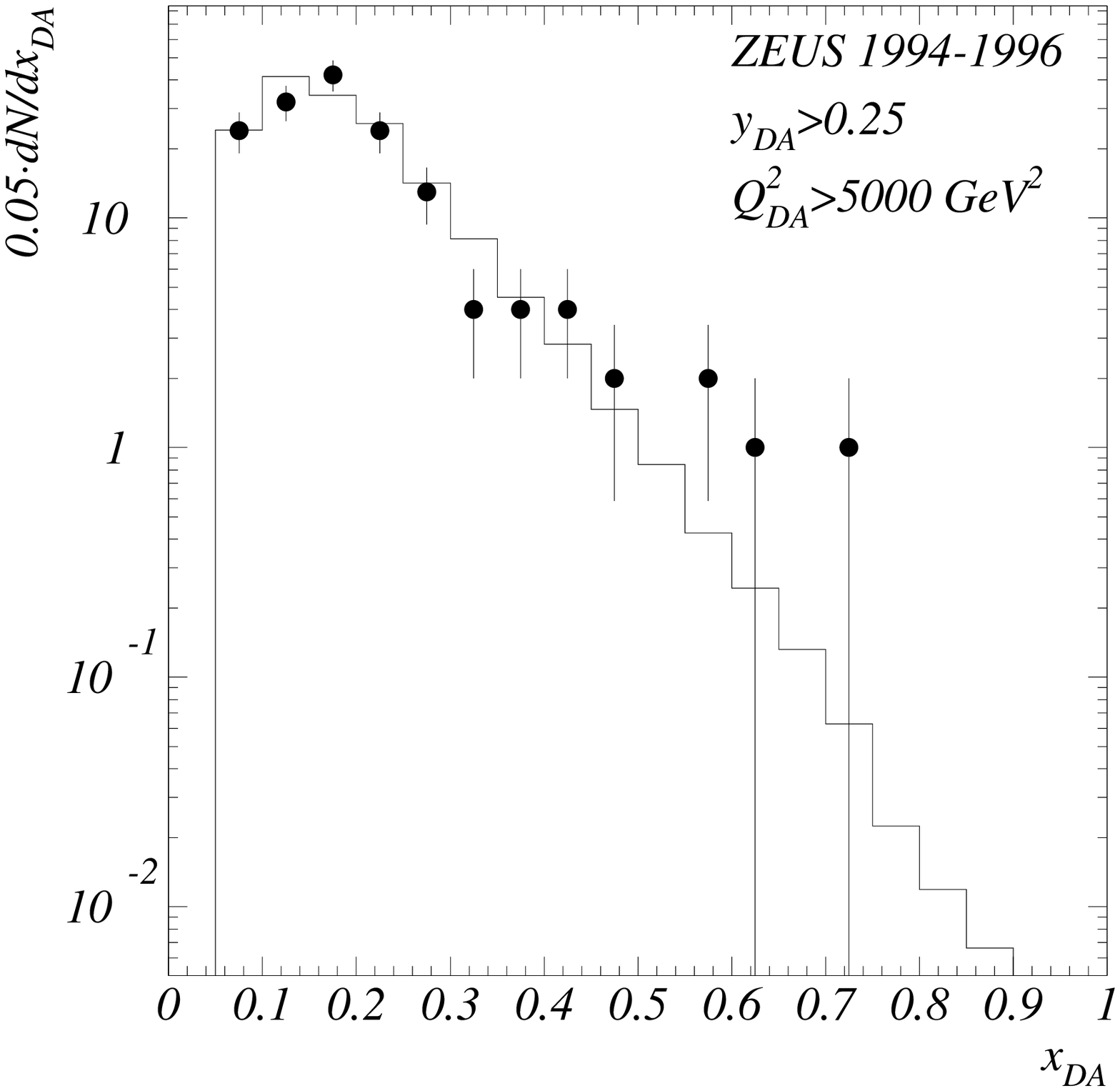} 
{80mm}{The $x_{DA}$ distribution of the observed events  
from ZEUS for the 
cuts in $y$ and \Qtwo\ indicated. The Standard Model prediction is 
shown as the histogram. The error bars correspond to 
the square root of the number of events in the bin.} 
{fig-ZEUS-highqtwo-x}  
\ffig{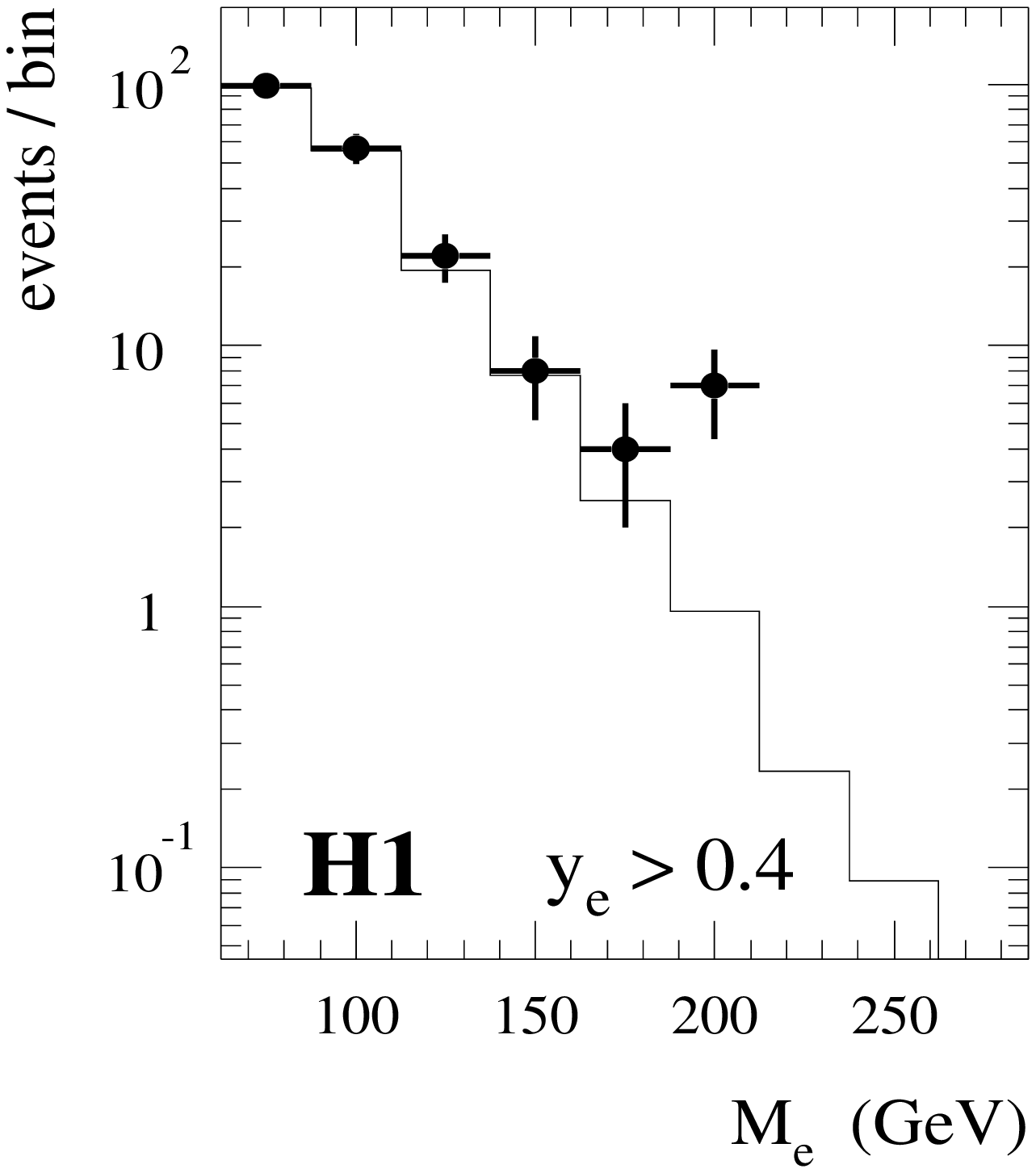} 
{100mm}{Distribution of $M_e$ for the observed events from H1 for  
$y_e > 0.4$ The error bars correspond to the statistical error 
in the bin.} 
{fig-H1-highqtwo-me}  
\ffig{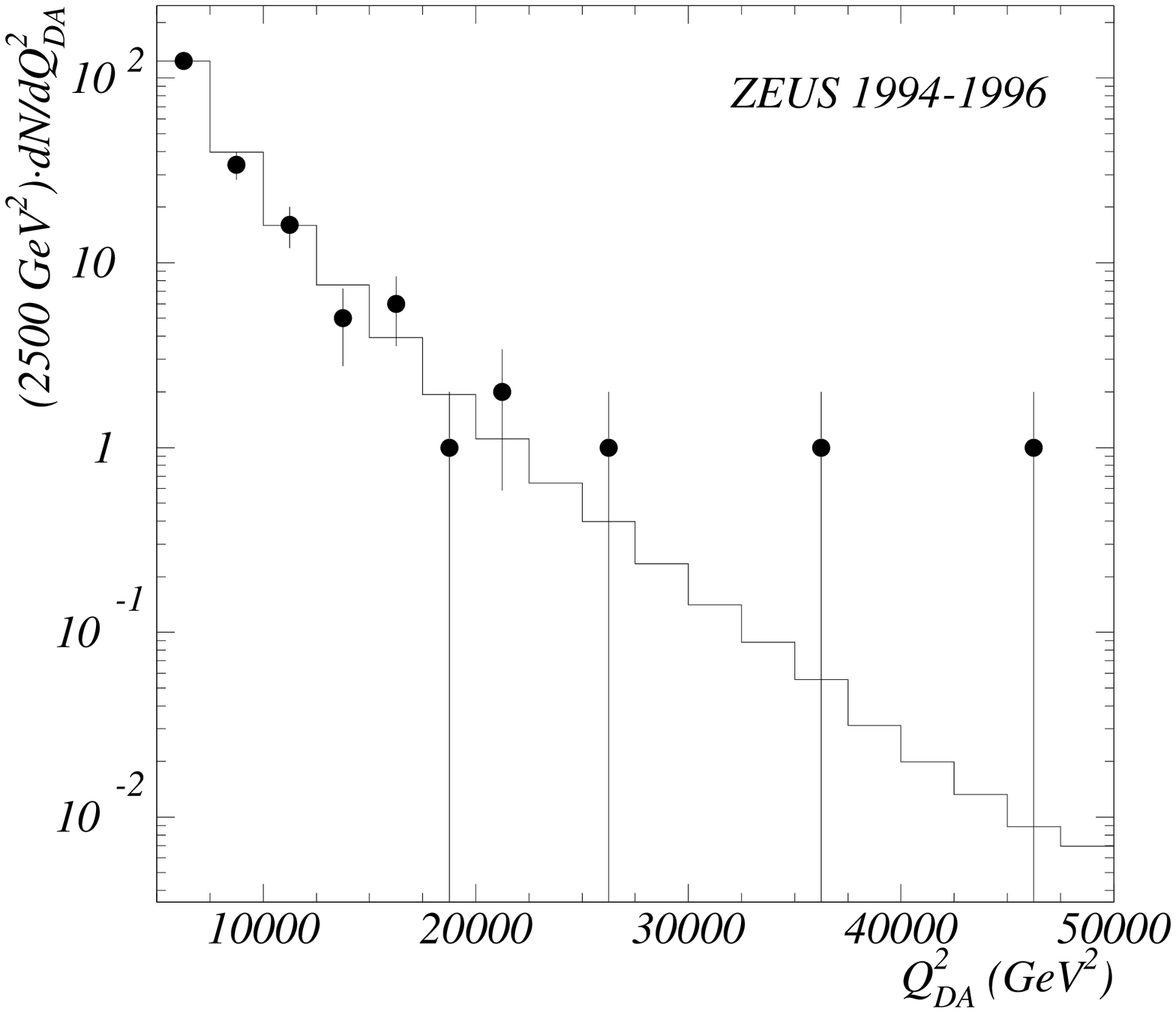} 
{80mm}{The \Qtwo\ distribution of the observed events from ZEUS.  
The Standard Model prediction is 
shown as the histogram. The error bars correspond to 
the square root of the number of events in the bin.} 
{fig-ZEUS-highqtwo-qtwo}  
\ffig{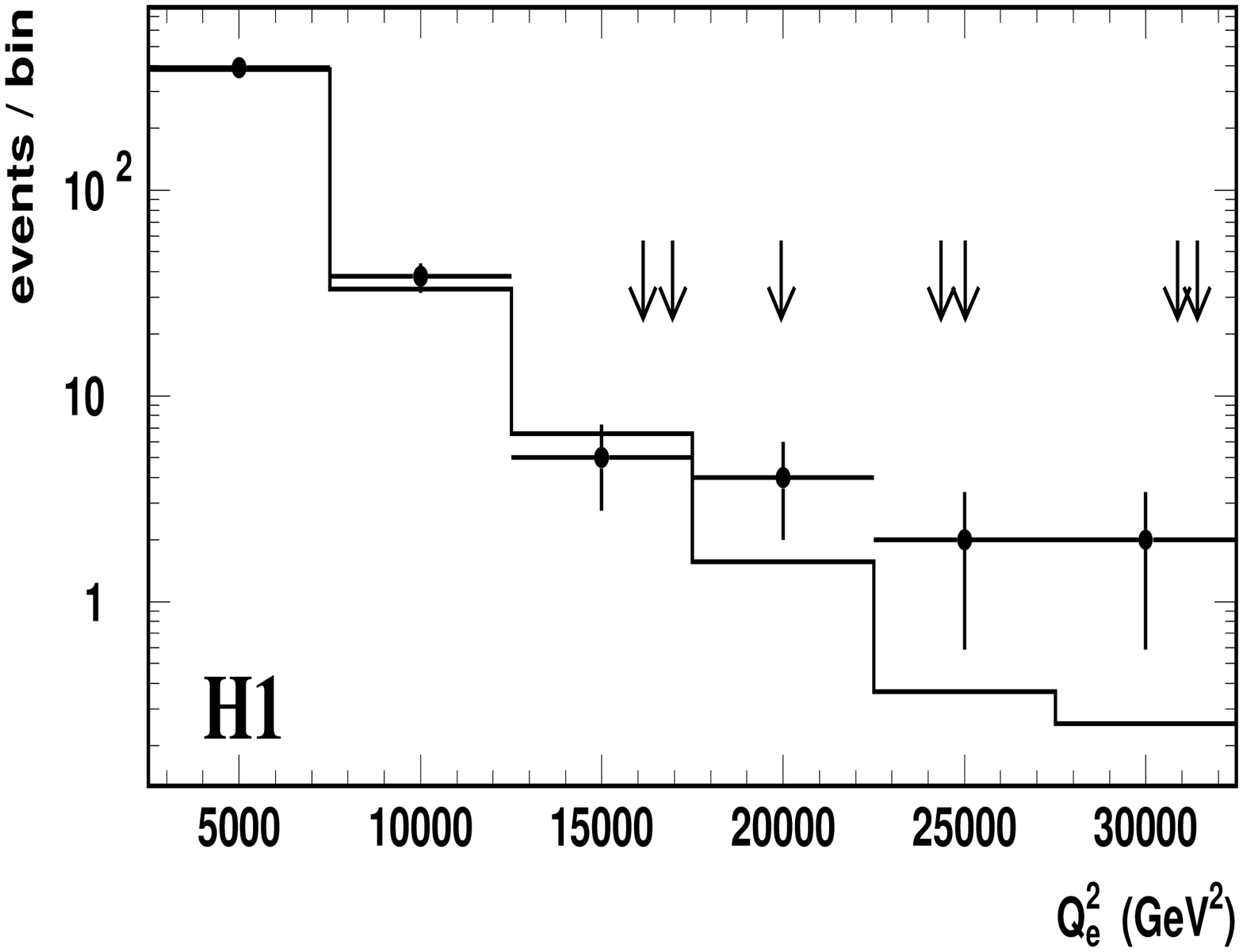} 
{100mm}{Distribution of $Q^2_e$ for the observed events from H1. 
The error bars correspond to the statistical error 
in the bin. The arrows indicate the $Q^2_e$ 
values for those events with $Q^2_e > 15000$ GeV$^2$ and $M_e > 180$ 
GeV.} 
{fig-H1-highqtwo-qtwo}  
 
H1 also see a small excess of 4 events compared to an expectation of  
$1.8 \pm 0.9$ for the region beyond \Qtwo\ $= 15000$ GeV$^2$ 
in the charged current sample. ZEUS also report a small excess,
though in neither case is this excess very statistically
significant. 

Of course the most natural interpretation of this signal, given the 
amazing resilience of the Standard Model over the past 30 years, is 
either an instrumental or background effect,  
possible freedom to vary the assumed 
parameters of the Standard Model fits, or a statistical fluctuation.  
Both experiments have carried out a thorough check for possible 
backgrounds and systematic effects, as well as investigating the  
uncertainty in the predictions of the Standard Model at these 
thus far unattainably high values of \Qtwo. Both experiments find 
essentially zero backgrounds from any known  
Standard Model processes. Neither could 
any systematic effects which could promote 
events from low to high \Qtwo\ be identified.  
 
Although there is some uncertainty from 
variations in parton distributions, $\alpha_s$ etc., the 
maximum effect which can be produced by varying the conventional  
parton distributions, of about 7 - 8\%, is much 
too small to explain the effect. However, Kuhlmann et 
al.~\cite{Kuhlmann}, have recently suggested that the effect could be 
explained without recourse to physics beyond the Standard Model by 
the addition of a small admixture of valence quark at very high $x$, 
beyond the region well-measured by fixed target experiments. This 
could also explain the excess observed in the jet cross-section 
at high $E_t$ by the CDF  
collaboration~\cite{CDF-highpt-jets}, although this can also 
be explained rather more plausibly by an increase in the poorly 
determined high $x$ gluon distribution~\cite{Houston-highpt-jets}.  
It is pointed out by Kuhlman {\it et al.} that such an admixture would 
actually {\it improve} the global fit to DIS data since there is 
a small excess in the BCDMS data compared to the QCD fits beyond $x$ 
of 0.75. The admixture at high $x$ could also be an  
intrinsic charm or 
bottom contribution. However, given that there is no 
longer any experimental motivation for intrinsic charm, to introduce it in 
this case would seem somewhat bizarre. Furthermore, although the 
addition of a small valence contribution at high $x$ would certainly 
not require physics beyond the Standard Model, such behaviour in 
hitherto well-behaved parton distributions seems scarcely less exotic. 
 
The final possibility not involving physics beyond the Standard Model 
is that the signal is a statistical fluctuation. Both experiments have 
done a very careful evaluation of the probability of the effect which 
is seen in its data arising by chance. ZEUS states that this 
probability is 0.72\% for $x > 0.55$ and $y > 0.25$, while H1 
consider the probability to be 0.6\%. Both observations are 
therefore rather unlikely to occur by chance. Furthermore, 
had these been fluctuations, they could have occurred anywhere in 
available phase space, rather than at the edge of phase space 
in a region hitherto unexplored.  
    
In the above experimental situation, it certainly seems reasonable to 
consider possible explanations for these effects in physics beyond 
the Standard Model. There are a variety of possible explanations for  
this effect. The publication of this data have lead to a veritable explosion  
of theoretical papers speculating on its origin; for example one of the most 
comprehensive is that by Altarelli {\it et al.}~\cite{Altarelli-leptoq}. Here 
we mention the possibilities only briefly.  
 
One possibility is an $s$ channel quark-lepton bound state, or  
resonance, which we can refer to generically as a ``leptoquark", 
although it could be one of several varieties of scalar or  
vector leptoquarks, or a 
supersymmetric $R$-parity violating squark.  H1's data 
is presented in the variables  $y$ and $M_e$, suggesting an explanation 
in terms of the production of such a particle. In contrast ZEUS works with 
\Qtwo\ and $x$. If the ZEUS data are expressed in terms of the H1  
variables, some sort of consistency check is possible. The mean value  
of $M_e$ in the H1 data is approximately $200 \pm 2 \pm 6$ GeV, 
where the first error corresponds to statistical and the second to 
a systematic effect due to the energy scale uncertainty.  
Using the preferred Double Angle reconstruction for  
ZEUS, $\overline{M}_{DA}$ is approximately $233 \pm 6$ GeV, where  
the error contains both statistical and systematic errors. At first glance 
these data differ by more than three standard deviations. However, 
some words of caution are in order here;  a Gaussian error in the 
contribution of the uncertainty of the calorimeter energy scale to 
the overall systematic error of the H1 data is scarcely justified,  
since their Monte Carlo studies show 
significant non-Gaussian tails (see figure 2d in~\cite{H1-highqtwo}). 
ZEUS uses more conservative RMS's. In addition, the overall 
$\chi^2$ between the $x_{DA}$ and $x_e$ calculations for ZEUS is not  
particularly good, implying the possibility that some additional 
systematic error is playing a role. The actual discrepancy 
could therefore be somewhat less. It is also true that possible 
initial state radiation of hard photons, which affects the
Electron and Double Angle determinations differently, could substantially improve the  compatibility of the data.

Several papers have discussed the compatibility of the ZEUS and
H1 data (see for example~\cite{Drees}). Although the analysis 
of Drees as to the compatibility of the ZEUS and H1 mass values is 
interesting, he does not consider the instrumental effects referred to above, 
nor the fact that ZEUS prefer to use DA variables for good reasons 
related to their understanding of their detector.  
However, Drees is certainly correct to point out that the 
number of events 
seen in H1 compared to ZEUS seems different what would be expected 
from the production 
of an $s$-channel resonance given the efficiencies and luminosities of 
the two experiments. This can only reasonably be ascribed to a 
statistical fluctuation. Whereas it would
seem premature to rule out the hypothesis that the signals seen in
the two experiments are compatible, there is clearly doubt on our 
ability to multiply the probabilities they quote for their compatibility 
with the Standard Model. Thus it would clearly be premature to assume that the Standard Model is excluded at the 99.99\% confidence level! 
 
Should the discrepancy between ZEUS and H1 turn out to be real, the 
explanation for the effect would then be unlikely to be an $s$-channel 
resonance. Some sort of contact interaction, implying quark and/or 
lepton substructure, could be possible. Such an effect would lead 
to an apparent excess in \Qtwo\, rather similar to the excess of large 
angle scatters observed by Rutherford when he resolved the atomic 
nucleus, and corresponding in this case to the resolution of substructure. 
Since there would be no $x$ or equivalently $M$ structure in such a 
case, the data of both experiments would be compatible, with the 
apparent clustering of the H1 data ascribed to a fluctuation.  
 
Whether any of the above explanations is correct, or whether the simple 
and disappointing possibility of an unlikely  
statistical fluctuation is the source of 
this interesting anomaly, must await the collection of more data 
over the next few years. 
\section{Future outlook and conclusions} 
\label{sec-future} 
HERA physics is poised at a fascinating stage. The ``high cross-section"  
physics of low $x$ DIS and photoproduction has proved to be much more 
interesting than had been anticipated when HERA was proposed. The 
steep rise in the gluon distribution inside the proton has been a catalyst 
to a great deal of theoretical work on the detailed properties of QCD.  
Nevertheless, as yet, no ``smoking gun" for BFKL behaviour has  
been observed. The 
study of exclusive vector meson production in DIS and the first 
observation of charm production in DIS has also opened up 
the possibility of further tying down the gluon distribution. 
A whole new field of study of the deep inelastic structure of 
diffraction has been opened up by HERA. This study has already 
produced much of interest and has the promise of truly revolutionising 
our understanding of the mechanism of diffraction. Finally, the 
data at high \Qtwo\ and $x$ are allowing detailed tests of the 
charged and neutral weak current, and at the edge of the kinematic 
region currently accessible there are indications of an excess of 
events which could indicate new physics. 
 
With such a wealth of interesting results it is easy to forget that the 
high \Qtwo\ and $x$ physics for which HERA was built has scarcely 
begun to be explored. The recent workshop on future physics at HERA 
has given an excellent overview of the detailed topics which HERA will 
be able to investigate over the next few years~\cite{HERA-future}. The 
potential of HERA will be greatly increased by the proposed luminosity 
upgrade; it is also likely that the energy of the proton beam
will be increased to at least 900 GeV. The changes in the machine 
configuration which are required to 
increase the luminosity will have consequences on the experiments,  
effectively excluding the study of low \Qtwo\ physics and introducing 
significant complications in the operation of detectors such as the 
leading proton spectrometers. It is particularly important therefore 
in the next few years to collect sufficient data with these unique detectors  
properly to exploit their physics potential. The prospect of easy use of both 
electrons and positrons and the existence of polarisation opens up the 
promise of detailed studies of the structure of the weak current in an  
analogous way to the detailed studies of the strong interaction which  
are currently being carried out. There is clearly a tremendous interest 
in pushing back the kinematic limit to as high values of \Qtwo\ as 
possible. Even if the current indications of possible new physics were not 
to persist with further luminosity, the potential of the future HERA 
programme is very great.  If the high \Qtwo\ - $x$ anomalies were 
to be confirmed, then it is difficult to overestimate the  
interest that physics at HERA would hold over the next few years.
\vskip1cm  
\noindent
{\large\bf Acknowledgements} 
\vskip1cm
I am grateful to my colleagues on ZEUS and H1 for many useful conversations. 
I am particularly grateful to M. Lancaster, J.D. McFall and A. Quadt for
carefully reading and commenting on the manuscript, and to J.D. McFall and
R. Hall-Wilton for help with the figures. 
\newpage
\noindent {\large\bf References} \\


\vskip1cm
\begin{thebibliography}{99} 
%
\bibitem{H1-detector} H1 Collaboration, I. Abt {\it et al.}, DESY  H1-96-01 
(1996).   
\bibitem{ZEUS-detector} The ZEUS Detector Status Report (DESY 1993).
\bibitem{H1-Si} H1 Collaboration, DESY Internal Report PRC 92/01; 
J. B\"{u}rger {\it et al.}, DESY 96-200, and references therein.
\bibitem{H1-CTD} J. B\"{u}rger {\it et al.}, Nucl. Instr. and Meth., A279 (1989) 217.
\bibitem{H1-FDET} G.A. Beck {\it et al.}, Nucl. Instr and Meth., A283 (1989)
471;
\newline J.M. Bailey {\it et al.}, Nucl. Instr and Meth., A323 (1992) 184;
\newline H. Gr\"{a}ssler {\it et al.}, Nucl. Instr and Meth., A323 (1992)
401, and references therein. 
\bibitem{H1-LA} B. Andrieu {\it et al.},   
    Nucl. Instr. and Meth., A336 (1993) 460.
\bibitem{H1-SPACAL} H1 Collaboration, DESY 96-171.
\bibitem{ZEUS-VXD} C. Alvisi {\it et al.}, Nucl. Inst. and Meth., A305 (1991) 30.
\bibitem{ZEUS-muVD} ZEUS Collaboration, DESY Internal Report PRC 97/01.
\bibitem{ZEUS-CTD} B. Foster {\it et al.}, Nucl. Instr. and Meth., A338
(1994) 254.
\bibitem{ZEUS-CTD-Electronics} N.~Harnew {\it et al.}.,
Nucl. Instr. and Meth., A279 (1989) 290;
\newline B.~Foster {\it et al.}, Nucl.~Phys.~B (Proc.~Suppl.) 32 (1993) 181;
\newline B.~Foster {\it et al.}, Nucl.~Instr. and Meth., A338 (1994) 254.
\bibitem{ZEUS-CAL} A. Andresen {\it et al.}, Nucl. Instr. and Meth., A305
 (1991) 101; A. Bernstein {\it et al.}, Nucl. Instr. and Meth., A336 (1993) 23.
\bibitem{Ellis-Stirling-Webber} R.K. Ellis, W.J. Sterling and B.R. Webber,
``QCD and Collider Physics'', Cambridge, ISBN 521 58189 3, (1996).
\bibitem{DGLAP}  V.N. Gribov and L.N. Lipatov, Sov. J. Nucl. Phys. 
15 (1972) 438, 675;
L.N. Lipatov, Sov. J. Nucl. Phys. 20 (1975) 94; 
Yu. L. Dokshitzer, Sov. Phys. JETP 46 (1977) 641;
G. Altarelli and G. Parisi, Nucl. Phys. B126 (1977) 298.
\bibitem{BFKL} E.A. Kuraev, L.N. Lipatov, and V.S. Fadin,
Sov. Phys. JETP 45 (1977) 199;
Y.Y. Bal$\breve{\rm i}$tsky and L.N. Lipatov, 
Sov. J. Nucl. Phys. 28 (1978) 822.
\bibitem{Callan-Gross} C.G. Callen and D.J. Gross, Phys. Rev. Lett. 22
(1969) 156.
\bibitem{DIS-scheme} G. Altarelli, R.K. Ellis and G. Martinelli,
Nucl. Phys. B143 (1978) 521, and 
\newline Nucl. Phys. B146 (1978) 544 (erratum).
\bibitem{MSbar-scheme} G. t'Hooft and M. Veltman, Nucl. Phys. B44
(1972) 189;
\newline G. t'Hooft, Nucl. Phys. B61 (1973) 455;
\newline W.A. Bardeen {\it et al.}, Phys. Rev D18 (1978) 3998.
\bibitem{rad-corr-refs}  See for example D. Bardin {\it et al.}, Proceedings
of the Workshop on Future Physics at HERA, Volume 1 (1996) 13, and
references therein. 
\bibitem{Jacquet-Blondel}F. Jacquet and A. Blondel, Proceedings
Study of an $ep$ facility in Europe, ed. U. Amaldi, DESY 79-48 (1979).
\bibitem{DA} S. Bentvelsen, J Engelen, and P. Kooijman, 
Proceedings of the Workshop on
Physics at HERA, DESY (1992) 23.
\bibitem{Sigma}  U. Bassler and G. Bernardi, Nucl. Instr. and Meth.,
A361 (1995) 197.
\bibitem{PT} ZEUS Collaboration, M. Derrick {\it et al.},
 Z. Phys. C72 (1996) 399.
\bibitem{H1-F2-1993} H1 Collaboration, T. Ahmed {\it et al.},
Nucl. Phys.  B439 (1995 )471. 
\bibitem{ZEUS-F2-1993} ZEUS Collaboration, M. Derrick {\it et al.}, 
Z. Phys. C65 (1995) 379.
\bibitem{GRV}  M. Gl\"{u}ck, E. Reya and A. Vogt, Z. Phys. C53 (1992) 127;
               Phys. Lett B306 (1993) 391; DESY-94-206 (1994).
\bibitem{MRSA} A.D. Martin, R.G. Roberts and W.J. Stirling,
Phys. Rev. D50 (1994) 6734. 
\bibitem{H1-1996-F2} H1 Collaboration, S. Aid {\it et al.}, Nucl. Phys. B470
(1996) 3.
\bibitem{ZEUS-1996-F2} ZEUS Collaboration, M. Derrick {\it et al.}, 
Z. Phys. C72 (1996) 399.
\bibitem{BCDMS} BCDMS Collaboration, A.C. Benvenuti {\it et al.},
Phys. Lett. B223 (1989) 485.
\bibitem{NMC-F2} NMC Collaboration, M. Arneodo {\it et al.},
Phys. Lett. B364 (1995) 107;
\newline Nucl. Phys. B483 (1997) 3.
\bibitem{E665-F2} E665 Collaboration, M.R. Adams {\it et al.},
Phys. Rev. D54 (1996) 3006..
\bibitem{H1-lowq2} H1 Collaboration, C. Adloff {\it et al.}, DESY 97-042
(1997).
\bibitem{ZEUS-BPC} ZEUS Collaboration, J. Breitweg {\it et al.},
DESY 97-135 (1997). 
\bibitem{F2-DL}  A. Donnachie and P.V. Landshoff, Z. Phys. C61 (1994)
139.
\bibitem{Capella} A. Capella {\it et al.}, Phys. Lett. B337 (1994) 358.
\bibitem{BK} B. Badelek and J. Kwiecinski, Phys. Lett. B295 (1992) 263.
\bibitem{GRV-various} 
M. Gl\"{u}ck, E. Reya and A. Vogt, 
Z. Phys. C67 (1995) 433;
As shown by A. Vogt, in Proceedings of Workshop
on Deep Inelastic Scattering and QCD, Paris (1995) 261.
\bibitem{Adel} K. Adel, F. Barreiro and F.J. Yndurain, 
Nucl. Phys. B495 (1997) 221.
\bibitem{DLL-Rujula} A. De R\'{u}jula {\it et al.}, Phys. Rev. D10 (1974) 1649.
\bibitem{Parton-recomb} L.V. Gribov, E.M. Levin and M.G. Ryskin, Phys. 
Rep 100 (1983) 1;
\newline A.H. Mueller and N. Quiu, Nucl. Phys. B268 (1986) 427
\bibitem{CCFM} M. Ciafaloni, Nucl. Phys. B296 (1988) 49;
\newline S. Catani, F. Fiorani and G. Marchesini, Phys. Lett. B234
(1990) 339; Nucl. Phys. B336 (1990) 18.
\bibitem{Kwiecinski-MS} J. Kwiecinski, A.D. Martin and P.J. Sutton,
Phys. Rev. D52 (1995) 1445.
\bibitem{MRS}  A.D. Martin, W.J. Stirling and R.G. Roberts, Phys. Lett. B306
 (1993) 145; Phys. Lett. B309 (1993) 492 (erratum).
\bibitem{CTEQ} CTEQ Collab., H.L Lai {\it et al.}, Phys. Rev. D51 (1995) 4763.
\bibitem{GRV-90} M. Gl\"{u}ck, E. Reya and A. Vogt, Z. Phys. C48
(1990) 471; 
\newline Nucl. Phys. B (Proc. Suppl.) 18C (1990) 49.
\bibitem{Brodsky-Schmidt} S.K. Brodsky and
I. Schmidt, Phys. Lett. B234 (1990) 144.
\bibitem{GRV-92} M. Gl\"{u}ck, E. Reya and A. Vogt, Z. Phys. C53
(1992) 127.
\bibitem{ZEUS-photoprod-SIGMA} ZEUS Collabation, M. Derrick {\it et al.},
Z. Phys. C63 (1994) 391.
\bibitem{H1-photoprod-SIGMA} H1 Collaboration, S. Aid {\it et al.},
Z. Phys. C69 (1995) 27.
\bibitem{GRV-F2} 
M. Gl\"{u}ck, E. Reya and A. Vogt, 
Z. Phys. C67 (1995) 433;
\bibitem{Tickner} ZEUS Collaboration, J. Tickner, private
communication.
\bibitem{Sch-Spi} D. Schildkneckt and H. Spiesberger, hep-ph/9707447.
\bibitem{Sak-Sch} J.J. Sakurai and D. Schildkneckt, Phys. Lett. B40
(1972) 121;
\newline B. Gorczyca and D. Schildkneckt, Phys. Lett. B47 (1973) 71. 
\bibitem{Ball-Forte} R.D. Ball and S. Forte, Phys. Lett B335 (1994) 77.
\bibitem{Lopez} C. L\'{o}pez and F.J. Yndur\'{a}in, Nucl. Phys. B171
(1980) 231.
\bibitem{Ellis-KL} R.K. Ellis, Z. Kunszt and E.M. Levin,
Nucl. Phys. B420 (1994) 517.
\bibitem{Ellis-HW} R.K. Ellis, F. Hautmann and B.R. Webber,
Phys. Lett B348 (1995) 582.
\bibitem{Forshaw-RT} J.R. Forshaw, R.G. Roberts, and R.S. Thorne,
Phys. Lett. B356 (1995) 79.
\bibitem{Askew} A.J. Askew, J. Kwiecinski, A.D. Martin and
P.J. Sutton, Phys. Rev. D47 (1993) 3775; 
\newline Phys. Rev. D49 (1994) 4402.
\bibitem{Thorne} R. Thorne, Phys. Lett. B392 (1997) 463; hep-ph/9701241
\bibitem{Catani} S. Catani, hep-ph/9609263, and Proceedings of DIS'96,
Roma, ed. G. D'Agostini and A. Nigro, World Scientific (1997) 165.
\bibitem{H1-gluon} H1 Collaboration, S. Aid {\it et al.}, 
Phys. Lett. B354 (1995) 494.
\bibitem{Bauerdick} L. Bauerdick, A. Glazov and M. Klein, Proceedings
of the Workshop on Future Physics at HERA, Volume 1 (1996) 77.
\bibitem{Fl-gamma} See for example A. Frey, Ph.D. Thesis, University
of Bonn (1996) (unpublished);
\newline A. Cassidy, Ph.D. Thesis, University of Bristol (1997) (unpublished). 
\bibitem{H1-Fl-sub} H1 Collaboration, S. Aid {\it et al.}, 
Phys. Lett. 393B (1997) 452.
\bibitem{ZEUS-gluon} ZEUS Collaboration, M. Derrick {\it et al.},
Phys. Lett. B345 (1995) 576.
\bibitem{NMC}  NMC Collaboration, M. Arneodo {\it et al.}, 
Phys. Lett. B309 (1993) 222.
\bibitem{GRV-charm1} M. Gl\"{u}ck, E. Hoffmann and E. Reya, 
Z. Phys. C13 (1982) 119.
\bibitem{GRV-charm2} M. Gl\"{u}ck, E. Reya and M. Stratmann, 
Nucl. Phys. B422 (1994) 37.
\bibitem{Laenen-charm} E. Laenen {\it et al.}, 
Nucl. Phys. B392 (1993) 162, 229;
\newline E. Laenen {\it et al.}, Phys. Lett. B291 (1992) 325;
\newline S.Riemersma {\it et al.}, Phys. Lett. B347 (1995) 143.
\bibitem{Marciano-charm} W.J. Marciano, Phys. Rev. D29 (1984) 580.
\bibitem{MRS-pin-glue} A. D. Martin, R. G. Roberts and W. J. Stirling,
Phys. Lett. B354 (1995) 155
\bibitem{Prytz} K. Prytz, Phys. Lett. B332 (1994) 393.
\bibitem{Ball-Forte-alphas} R.D. Ball and S. Forte, hep-ph/9607289,
and Proceedings of DIS'96,
Roma, ed. G. D'Agostini and A. Nigro, World Scientific (1997) 208.
\bibitem{NMC-g-Jpsi}
NMC Collaboration, D. Allasia {\it et al.},
Phys. Lett. B258 (1991) 493.
\bibitem{H1-gluon-jetrates} H1 Collaboration, S. Aid {\it et al.},
Nucl. Phys. B449 (1995) 3.
\bibitem{Mirkes-Zepp}
E. Mirkes and D. Zeppenfeld, Acta Phys. Polon. B27 (1996) 1393;
\newline Proceedings of the Zeuthen Workshop on Elementary 
Particle Theory, ``QCD
and QED in Higher Orders", Rheinsberg, Germany, April 1996; presented
by E. Mirkes, Nucl. Phys. (Proc. Suppl.) 51C (1996) 273.
\bibitem{Ryskin} M.G.~Ryskin, Z. Phys. C57 (1993) 89.
\bibitem{Brodsky} S.J.~Brodsky {\it et al.}, Phys. Rev. D50~(1994)~ 3134.
\bibitem{ZEUS-DIS-rho} ZEUS Collab., M. Derrick {\it et al.}, 
Z. Phys. C69 (1995) 39.
\bibitem{H1-DIS-rho} H1 Collaboration, S. Aid {\it et al.}, 
Nucl. Phys. B468 (1996) 3.
\bibitem{DL-rho} A. Donnachie and P.V.~Landshoff, Phys. Lett. 
B185 (1987) 403.
\bibitem{NMC-rho} NMC Collab., P. Amaudruz {\it et al.}, Z. Phys. C54~(1992)~ 
239; 
\newline  M. Arneodo {\it et al.}, Nucl. Phys. B429~(1994)~503.
\bibitem{H1-Jpsi-photo} H1 Collaboration, S. Aid {\it et al.},
Nucl. Phys. B472 (1996) 3.
\bibitem{ZEUS-Jpsi-photo} ZEUS Collab., M. Derrick {\it et al.}, 
Phys. Lett. B350 (1995) 120-134.
\bibitem{Jpsi-dl} A. Donnachie and P.V.~Landshoff, Phys. Lett. B348
(1995) 213.
\bibitem{ZEUS-Jpsi-photo-1997} ZEUS Collab., J. Breitweg {\it et al.}, 
Z. Phys. C75 (1997) 215. 
\bibitem{Jpsi-E516} FTPS Collaboration, B.H. Denby {\it et al.},
Phys. Rev. Lett. 52 (1984) 795.
\bibitem{Jpsi-E401} E401 Collaboration, M. Binklet {\it et al.},
Phys. Rev. Lett. 48 (1982) 73.
\bibitem{Jpsi-E687} E687 Collaboration, P.L. Frabetti {\it et al.},
Phys. Lett. B316 (1993) 197.  
\bibitem{ZEUS-Jpsi-photo-inelastic} ZEUS Collab., M. Derrick {\it et al.},
DESY 97-147 (1997) and hep-ex/9708010.
\bibitem{Jpsi-NA14} NA14 Collaboration, R. Barate {\it et al.}, 
Z. Phys. C33 (1987) 505.
\bibitem{Jpsi-EMC} EMC Collaboration, J. Ashman {\it et al.},
Z. Phys. C56 (1992) 21.
\bibitem{Jpsi-Cacciari} 
    M. Cacciari and M. Kr\"amer, Phys. Rev. Lett. 52 (1996) 4128.
\bibitem{Jpsi-CDF} CDF Collaboration, F. Abe {\it et al.},
Phys. Rev. Lett. 69 (1992) 3704;
\newline CDF Collaboration, A. Sansomi, FERMILAB-CONF-95/263-E.
\bibitem{Jpsi-Kramer}  
M.~Kr\"amer {\it et al.}, Phys. Lett. B348 (1995) 657;
\newline M.~Kr\"amer, Nucl. Phys. B459 (1996) 3.
\bibitem{Jpsi-DIS-Frankfurt} 
  L. Frankfurt, W. K\"opf and M. Strikman, 
 Phys. Rev. D54 (1996) 3194;
\newline see also: 
  H. Abramowicz, L. Frankfurt and M. Strikman,     
 DESY-95-047, hep-ph/9503437 (1995);
\newline   L. Frankfurt and M. Strikman, 
hep-ph/9510291 (1995).    
\bibitem{SLAC-D*} G. Feldman {\it et al.}, Phys. Rev. Lett. 38 (1977) 1313. 
\bibitem{ZEUS-photo-D*} ZEUS Collaboration, J. Breitweg {\it et al.},
Phys. Lett. B401 (1997) 192.
\bibitem{H1-photo-D*} H1 Collaboration, S. Aid {\it et al.},
Nucl. Phys. B472 (1996) 32. 
\bibitem{lowW-charm} CIF Collab., M.S. Atiya {\it et al.}, Phys. Rev. Lett
43 (1979) 414;
\newline BFP Collab., A.R. Clark {\it et al.}, Phys. Rev. Lett. 45 (1980)
682;
\newline SLAC HFP Collab., K. Abe at al., Phys. Rev. D30 (1984) 1;
\newline EMC Collab., M. Arneodo {\it et al.}, Z. Phys, C35 (1987) 1;
\newline PEC Collab., M. Adamovich {\it et al.}, Phys. Lett. B187 (1987) 437;
\newline E691 Collab., J.C. Anjos {\it et al.}, Phys. Rev. Lett. 65 (1990)
2503;
\newline NA-14$'$ Collab., M.P. Alvarez {\it et al.}, Z. Phys. C60 (1993) 53.
\bibitem{Frixione} S. Frixione {\it et al.}, Phys. Lett B348 (1995) 633;
\newline Nucl. Phys. B454 (1995) 3. 
\bibitem{GRV-photon} M. Gl\"{u}ck, E. Reya and A. Vogt,
Phys. Lett. B306 (1993) 391.
\bibitem{ZEUS-F2charm} ZEUS Collaboration, J. Breitweg {\it et al.},
DESY 97-089 (1997). 
\bibitem{H1-F2charm} H1 Collaboration, C. Adloff {\it et al.},
Z. Phys. C72(1996) 593.
\bibitem{LAC} H. Abramowicz, K. Charchula and A. Levy, Phys. Lett.
     B269 (1991) 458.
\bibitem{Laenen} E. Laenen {\it et al.}, Nucl. Phys. B392 (1993) 162;
\newline Nucl. Phys. B393 (1993) 229; Phys. Lett. B291 (1992) 325.
\bibitem{Riemersma} S. Riemersma, J. Smith and W.L. van Neerven,
Phys. Lett. B347 (1995) 143.
\bibitem{EMC-charm} EMC Collaboration, J.J. Aubert {\it et al.}, 
Nucl. Phys. B213 (1983) 31. 
\bibitem{H1-xfpt} H1 Collaboration, I. Abt {\it et al.},
Z. Phys. C63 (1994) 377.
\bibitem{ZEUS-xfpt} ZEUS Collaboration, M. Derrick {\it et al.},
Z. Phys. C70 (1996) 1.
\bibitem{DELPHI-xf} DELPHI Collab., P. Abreu {\it et al.}, Phys. Lett. B311
(1993) 408.
\bibitem{H1-nch} H1 Collaboration, S. Aid {\it et al.}, 
Z. Phys. C72 (1996) 573.
\bibitem{KNO} Z. Koba, H.B. Nielsen and P. Olesen, Nucl. Phys. B40
(1972) 317.
\bibitem{MEPS}  G. Ingelman, Proceedings of the Workshop 
         on Physics at HERA, DESY Vol.~3~(1992)~1366;
\newline M. Bengtsson, G. Ingelman, T. Sj\"{o}strand,
Nucl. Phys. B301 (1988) 554.
\bibitem{CDMBGF} T.~Sj\"{o}strand, Comp. Phys. Comm. 39 (1986) 347;
\newline T.~Sj\"{o}strand and M.~Bengtsson, {\it ibid.} 43 (1987) 367.
\bibitem{EMC-xf} EMC Collaboration, M. Arneodo {\it et al.},
Z. Phys. C35 (1986) 417.
\bibitem{E665-xf} E665 Collaboration, M.R. Adams {\it et al.}, FNAL-Pub
93/245-E (1993);
\newline Phys. Lett. B272 (1991) 163.
\bibitem{EMC-ptsq} EMC Collab., J. Ashman {\it et al.}, Z. Phys. C52 (1991) 361.
\bibitem{E665-ptsq} E665 Collab., M.R. Adams {\it et al.}, Phys. Lett. B272
(1991) 163.
\bibitem{H1-pt97} H1 Collaboration, C. Adloff {\it et al.}, 
Nucl. Phys. B485 (1997) 3.
\bibitem{HERWIG} G. Marchesini {\it et al.}, Comp. Phys. Comm. 67 (1992)
465;
\newline M. Seymour, Lund preprint LU-TP-94-12 (1994) and
Nucl. Phys. B436 (1995) 443;
\newline B.R. Webber, Nucl. Phys. B238 (1984) 492.
\bibitem{ZEUS-Breit} ZEUS Collaboration, M. Derrick {\it et al.},
Z. Phys. C67 (1995) 93.
\bibitem{H1-Breit} H1 Collaboration, S. Aid {\it et al.}, 
Nucl. Phys. B445 (1995) 3.
\bibitem{H1-Breit-97} H1 Collaboration, C. Adloff {\it et al.},
DESY 97-108 (1997).
\bibitem{MLLA-model} Yu.L. Dokshitzer {\it et al.}, ``Basics of Perturbative
QCD'', Editions Fronti\`{e}res (1991);
\newline Ya.I. Azimov {\it et al.}, Z. Phys. C31 (1986) 213; Z. Phys. C27
(1985) 65;
\newline  Yu.L. Dokshitzer, V.A. Khoze and S.I. Troyan,
Phys. Lett. 202B (1988) 276. 
\bibitem{Breit-ee} TASSO Collaboration, W. Braunschweig {\it et al.}, 
Z. Phys. 47 (1990) 187; Phys. Lett. B311 (1993) 408;
\newline DELPHI Collaboration, P. Abreu {\it et al.}, Phys. Lett B311 (1993)
408;
\newline OPAL Collaboration, G. Alexander {\it et al.}, Z. Phys. C72 (1996)
191; K.Z. Akrawy {\it et al.}, Phys. Lett. B247 (1990) 617.
\newline ALEPH Collaboration, R. Barate {\it et al.}, CERN PPE 96-186.
\bibitem{H1-K0} H1 Collaboration, S. Aid {\it et al.},
Nucl. Phys. B480 (1996) 3.
\bibitem{ZEUS-K0} ZEUS Collaboration, M. Derrick {\it et al.},
Z. Phys. C68 (1995) 29.
\bibitem{DELPHI-strangeness}
DELPHI Collab., P. Abreu {\it et al.}, Z. Phys. C65 (1995) 587
\bibitem{E665-strangeness}
E665 Collab., M.R. Adams {\it et al.}, Z. Phys. C61 (1994) 539
\bibitem{QCD-instanton}
A. Ringwald and F. Schremmp, DESY 94-147;
\newline M.J. Gibbs, A. Ringwald and F. Schremmp, DESY 95-119;
\newline M.J. Gibbs, A. Ringwald, B.R. Webber and J.T. Zadrozny, 
Z. Phys. C66 (1995) 285.
\bibitem{H1-alphas} H1 Collaboration, T. Ahmed {\it et al.}, 
Phys. Lett. B346 (1995) 415.
\bibitem{ZEUS-alphas} ZEUS Collaboration, M. Derrick {\it et al.},
Phys. Lett. B363 (1995) 201.
\bibitem{JADE-jet} JADE Collab., W. Bartel {\it et al.}, Z. Phys. C33,
(1986) 23;
\bibitem{PDG} Particle Data Group, R.M. Barnett {\it et al.}, 
Phys. Rev. D54, (1996) 1. 
\bibitem{H1-eflow} H1 Collaboration, S. Aid {\it et al.}, Z. Phys. C63 (1994) 377.
\bibitem{ZEUS-eflow} ZEUS Collaboration, paper 0391, submitted paper
to EPS Conference, Brussels, July 1995.
\bibitem{H1-eflow-warsaw} H1 Collaboration, pa02-073, submitted
paper to ICHEP'96, Warsaw Poland, July 1996.
\bibitem{H1-fjets} H1 Collaboration, S. Aid {\it et al.}, Phys. Lett.
B356 (1995) 118
\bibitem{H1-fjets-warsaw} H1 Collaboration, pa03-049, submitted
paper to ICHEP'96, Warsaw Poland, July 1996.
\bibitem{Bartels-BFKL-jets} J. Bartels {\it et al.},
Phys. Lett. B384 (1996) 300.
\bibitem{Goulianos-review} K. Goulianos, Phys. Rep. 101 (1983) 169.
\bibitem{pp-rapgap-disc} UA8 Collaboration, A. Brandt {\it et al.},
Phys. Lett. B297 (1992) 417;
\newline R. Bonino {\it et al.}, Phys. Lett. B211 (1988) 239. 
\bibitem{D0-rapgap} D0 Collaboration, S. Abachi {\it et al.},
Phys. Rev. Lett. 72 (1994) 2332.
\bibitem{CDF-rapgap} CDF Collaboration, F. Abe {\it et al.},
Phys. Rev. Lett. 74 (1995) 855.
\bibitem{ZEUS-rapgap-disc} ZEUS Collaboration., M. Derrick {\it et al.}, 
Phys. Lett. B315 (1993) 481.
\bibitem{H1-rapgap-disc} H1 Collaboration, T. Ahmed {\it et al.},
Nucl. Phys. B429 (1994) 477. 
\bibitem{Feynman-diffraction} R.P. Feynman, ``Photon-Hadron
 Interactions", p 344, Benjamin (1972).
\bibitem{Bj-diffraction} J.D. Bjorken, Proceedings of International 
Workshop on Deep Inelastic Scattering, Eilat, Israel, 1994, ed. A. Levy,
World Scientific (1995) 144.
\bibitem{DelDuca-95} V. Del Duca, hep-ph/9506355
in Proceedings of the $10^{th}$ Topical 
Workshop on
       Proton-Antiproton Collider Physics, Fermilab, 
Batavia, Illinois, May 9-13, 1995.
\bibitem{Pom-DL} A. Donnachie and P.V.~Landshoff, Nucl. Phys. B244
(1984) 322; Phys. Lett. B191 (1987) 309;
Nucl. Phys. B303 (1988) 634; Phys. Lett. B285 (1992) 172.
\bibitem{IS} G. Ingelman and P. Schlein, Phys. Lett. 152B (1985) 256.
\bibitem{H1-Pom-95} H1 Collaboration, T. Ahmed {\it et al.},
Phys. Lett. B348 (1995) 681.
\bibitem{ZEUS-Pom-95} ZEUS Collaboration, M. Derrick {\it et al.},
Z.  Phys. C68 (1995) 569.
\bibitem{HERA-Diff-jets} ZEUS Collaboration, M. Derrick {\it et al.},
Phys. Lett. B332 (1994) 228.
\bibitem{H1-Warsaw-F2D} H1 Collaboration, pa02-061, submitted
paper to ICHEP'96, Warsaw Poland, July 1996.
\bibitem{McDermott-Briskin} M.F. McDermott and G. Briskin, 
hep-ph/9610245, in Proceedings
of the Workshop on Future Physics at HERA, Volume 2 (1996) 691.
\bibitem{Goulianos-PIC} K. Goulianos, ``Results on diffraction'',
to be published in the Proceedings of the XVII International Conference
on Physics in Collision, Bristol (1997), ed. H.F. Heath, World 
Scientific.   
\bibitem{Buchmuller-Hebecker} W. Buchm\"{u}ller and A. Hebecker,
Phys. Lett. B355 (1995) 573.
\bibitem{Goulianos-PL} K. Goulianos, Phys. Lett. B358 (1995) 379; B363
(1995) 268.
\bibitem{Collins} J.C. Collins {\it et al.}, Phys. Rev. D51 (1995) 3182.
\bibitem{Gehrmann-Stirling} T. Gehrmann and W.J. Stirling, 
Z. Phys. C70 (1996) 89.
\bibitem{Ellis-Ross} J. Ellis and G. Ross, Phys. Lett. B384 (1996)
293.
\bibitem{Andersson-83} B. Andersson {\it et al.}, Phys. Rep. 97 (1983) 31.
\bibitem{ZEUS-MXfit} ZEUS Collaboration, M. Derrick {\it et al.}, 
Z. Phys. C70 (1996) 391.
\bibitem{ZEUS-hpp-93} ZEUS Collaboration, M. Derrick {\it et al.},
 Phys. Lett. B356 (1995) 129.
\bibitem{CDF-diff-W} CDF Collaboration, F. Abe {\it et al.},
Phys. Rev. Lett 78 (1997) 2698.
\bibitem{CDF-diff-jets} CDF Collaboration, F. Abe {\it et al.},
``Measurement of diffractive dijet production at the
Fermilab Tevatron'', submitted to Phys. Rev. Letters,
as reported in~\cite{Goulianos-PIC}. 
\bibitem{ZEUS-FNC} ZEUS Collaboration, ``Study of Events with an 
Energetic Forward Neutron in $ep$ collisions at HERA'', 
contributed paper N-641 to EPS-97 conference, Jerusalem, (1997). 
\newline ZEUS FNC group, S. Bhadra {\it et al.}, DESY 97-006 and
hep/ex 9701015. 
\bibitem{ZEUS-FNC-II} ZEUS Collaboration, M. Derrick {\it et al.},
Phys. Lett. B384 (1996) 388.
\bibitem{ZEUS-CC-1996} ZEUS Collaboration, M. Derrick {\it et al.}, 
Z. Phys. C72 (1996) 47.
\bibitem{ZEUS-LPS-rho} ZEUS Collaboration, M. Derrick {\it et al.}, 
Z. Phys. C73 (1997) 253.
\bibitem{H1-LPS} H1 Collaboration, 
     ``The H1 Forward Proton Spectrometer'', contributed paper
pa17-025 to ICHEP'96, Warsaw, Poland, July 1996. 
\bibitem{Ela-Rome} ZEUS Collaboration, presented by E. Barberis, 
Proceedings of DIS'96,
Roma, ed. G. D'Agostini and A. Nigro, World Scientific (1997) 343.
\bibitem{Mehta-Phillips-Waugh} A. Mehta, J. Phillips and B. Waugh, 
Proceedings
of the Workshop on Future Physics at HERA, Volume 1 (1996) 704.
\bibitem{H1-CC-1996} H1 Collaboration, S. Aid {\it et al.}, 
Phys. Lett. B379 (1996) 319. 
\bibitem{ZEUS-ratios-1993} ZEUS Collaboration, M. Derrick {\it et al.},
Phys. Rev. Lett. 75 (1995) 1006.
\bibitem{ZEUS-highqtwo} ZEUS Collaboration, J. Breitweg {\it et al.},
Z. Phys. C74 (1997) 207.
\bibitem{H1-highqtwo} H1 Collaboration, C. Adloff {\it et al.},
Z. Phys. C74 (1997) 191.
\bibitem{ZEUS-CC-highqtwo} ZEUS Collaboration, presented by
D. Krakauer,
to be published in Proceedings of Hadron Collider Physics XII,
State University of New York at Stony Brook, 1997.
\bibitem{Kuhlmann} S. Kuhlmann, H. L. Lai, W. K. Tung,
hep-ph/9704338.
\bibitem{CDF-highpt-jets} CDF Collaboration, F. Abe {\it et al.},
Phys. Rev. Lett. 77 (1996) 438.
\bibitem{Houston-highpt-jets}J. Huston {\it et al.},
Phys. Rev. Lett. 77 (1996) 444.
\bibitem{Altarelli-leptoq}G. Altarelli {\it et al.},
hep-ph/9703276.
\bibitem{Drees} M. Drees, Phys. Lett. B403 (1997) 353.
\bibitem{HERA-future} Proceedings
of the Workshop on Future Physics at HERA,
ed. G. Ingleman, A. De Roeck and R. Klanner, DESY,
Volumes 1 \& 2 (1996).
DESY, Hamburg, 1996.
\end{thebibliography}
\end{document}